\documentclass[fleqn,usenatbib]{mnras}

\usepackage{newtxtext,newtxmath}
\usepackage{threeparttable}

\usepackage[T1]{fontenc}

\DeclareRobustCommand{\VAN}[3]{#2}
\let\VANthebibliography\thebibliography
\def\thebibliography{\DeclareRobustCommand{\VAN}[3]{##3}\VANthebibliography}

\usepackage{graphicx}	
\usepackage{amsmath}	

\newcommand{\ocen}{$\omega\,Cen$}

\title[Chemical Taxonomy of $\omega$~Centauri]{Chemical Taxonomy of $\omega$~Centauri: Ten Populations Reveal a Multi-Phase Enrichment History}

\author[Akbaba et al.]{
Furkan Akbaba$^{1}$\thanks{E-mail: furkan.akbaba@ogr.iu.edu.tr},
Olcay Plevne$^{2}$,
Timur \c Sahin$^{3}$,
Sena Aleyna Şentürk$^{3}$
\\
$^{1}$Institute of Graduate Studies in Science, Istanbul University, Istanbul, Turkey\\
$^{2}$Faculty of Science, Department of Astronomy and Space Sciences, Istanbul University, Istanbul, Turkey\\
$^{3}$Department of Space Sciences and Technologies, Faculty of Science, Akdeniz University, 07058, Antalya, T\"{u}rkiye\\
}

\date{Accepted XXX. Received YYY; in original form ZZZ}

\pubyear{\the\year{}}

\begin{document}
\label{firstpage}
\pagerange{\pageref{firstpage}--\pageref{lastpage}}
\maketitle

\begin{abstract}
$\omega$~Centauri, the most massive globular cluster in the Milky Way, exhibits a level of stellar population complexity that has long resisted a unified chemical characterisation. We exploit high-resolution near-infrared spectroscopy from the Milky Way Mapper survey (MWM DR19) to construct one of the largest homogeneously analysed samples of $\omega$~Cen members to date. Applying Ward-linkage hierarchical clustering in a seven-dimensional chemical abundance space, without prior assumptions on population number or boundaries, we identify ten chemically distinct stellar populations. Their nucleosynthetic signatures trace four enrichment channels: iron-peak, $\alpha$-element, CNO-cycle, and high-temperature proton-capture processes. The populations organise into two dominant groups separated by a large light-element spread at a modest iron baseline, consistent with AGB-driven self-enrichment. This dichotomy reflects distinct enrichment pathways: core-collapse supernovae establish the iron baseline, while AGB stars dominate light-element and $s$-process enrichment. A decoupled rise in $s$-process abundances relative to iron-peak elements, together with sub-dominant Type~Ia contributions across all metallicities, indicates evolution on timescales shorter than the characteristic Type~Ia delay time. One intermediate-metallicity population retains a primordial composition, providing evidence for spatially segregated enrichment within the progenitor. The most metal-rich component may trace star formation continuing after accretion into the Milky Way halo. All populations lie in the accreted regime of the $[\mathrm{Al/Fe}]$--$[\mathrm{Mg/Mn}]$ plane, supporting an ex-situ origin. These results reinforce the interpretation of $\omega$~Cen as the remnant nucleus of an accreted dwarf galaxy and provide a framework for future chemo-dynamical studies.
\end{abstract}




\section{Introduction}
Among the approximately 150 globular clusters (GCs) of the Milky Way, NGC~5139 — more commonly known as $\omega$~Centauri — occupies a unique position. With a mass of approximately $4 \times 10^6\,M_\odot$ \citep{Baumgardt2018}, $\omega$~Cen is the most massive globular cluster in the Galaxy and has emerged as a powerful challenge to the classical notion of a simple stellar population (SSP) — an assembly of stars formed from a single molecular cloud, coeval and chemically homogeneous — from early observations onward \citep{FreemanRodgers1975, NorrisDaCosta1995}. This paradigm has proven so inadequate for $\omega$~Cen that the system is now widely regarded as the stripped nuclear star cluster (NSC) of an accreted dwarf galaxy \citep{BekkiFreeman2003, Massari2019, Haeberle2024}.
The first quantitative evidence for chemical heterogeneity came from photometric observations. \citet{Woolley1966} noticed an anomalously broad red giant branch (RGB) in early photographic data; \citet{CannonStobie1973} then demonstrated that this breadth arose not from observational errors but from a genuine spread in heavy-element abundances. Definitive spectroscopic evidence was provided by \citet{FreemanRodgers1975}, who showed that calcium line strengths measured in RR~Lyrae variables varied by more than 1.5~dex in $\mathrm{[Ca/H]}$ within the cluster. Over the following two decades, growing spectroscopic samples revealed that $\omega$~Cen's metallicity distribution function (MDF) is not only broad but also multi-modal, with prominent peaks near $\mathrm{[Fe/H]} \approx -1.75$, $-1.50$, $-1.20$, and $-0.80$~dex \citep{NorrisDaCosta1995, Pancino2000, Pancino2003, Johnson2010}. The total metallicity range reaches $\Delta\mathrm{[Fe/H]} \approx 1.5$~dex \citep{NorrisDaCosta1995, Johnson2010}, spanning nearly the full extent observed in the Galactic halo.
This complexity deepened in the early 2000s. High-resolution \textit{Hubble Space Telescope} (HST) imaging revealed that the main sequence of $\omega$~Cen splits into two distinct branches \citep{Bedin2004}. Under normal evolutionary expectations, more metal-rich stars should appear redder; consequently, the bluer branch being more metal-rich posed a striking paradox — one that can only be explained by invoking markedly enhanced helium abundances of $Y \approx 0.38$--$0.42$ \citep{Norris2004, Piotto2005, Milone2017}. Subsequent HST observations showed that the main sequence splits into at least 15 photometric sub-components \citep{Bellini2017}. This complexity became even more apparent with the oMEGACat project, which provided high-precision photometry and proper-motion measurements for 1.4 million stars \citep{Haberle2024b}. Recent MUSE integral-field spectroscopy has revealed the chemical and kinematic properties of more than 300,000 stars within the half-light radius, identifying at least 11 distinct groups \citep{Nitschai2023, Nitschai2024}. Accordingly, the reported number of sub-populations has continuously increased with observational precision — from three broad RGB groups \citep{Pancino2000}, to five metallicity peaks \citep{Johnson2010}, to ever finer classifications today \citep[e.g.,][]{Vernekar2025, Mason2026}.
The light-element abundances of $\omega$~Cen stars point to an equally complex formation history. Na--O and Mg--Al anti-correlations, chemical signatures of proton-capture nucleosynthesis at temperatures $T \gtrsim 7 \times 10^7$~K, are observed across the full metallicity range and indicate that successive stellar generations formed from gas processed by earlier generations \citep{Carretta2009a, Johnson2010, AlvarezGaray2024}. The more metal-rich populations also show marked enhancements in $s$-process elements such as Ba, La, and Ce, consistent with enrichment by intermediate-mass AGB stars on a timescale of a few $\times 10^8$~yr \citep{Meszaros2020, Pagnini2025}. Taken together, these findings point to a self-regulated star-formation history lasting at least 3--5~Gyr \citep{Romano2007}, a timescale more consistent with dwarf galaxy evolution than with that of an ordinary globular cluster.
Multiple independent observational lines of evidence support the interpretation that $\omega$~Cen is the nucleus of an accreted dwarf galaxy \citep{FreemanRodgers1975, BekkiFreeman2003}. The cluster's retrograde orbit with respect to the Milky Way disc and its measured proper motion strongly support an extragalactic origin \citep{Dinescu1999, Majewski2000, Zhao2001, Myeong2018}. Moreover, recent studies have revealed that $\omega$~Cen is surrounded by distant tidal tails known as the Fimbulthul stream \citep{Ibata2019}, direct evidence of ongoing stripping from its host galaxy. Its mass and structural properties are consistent with NSCs in dwarf elliptical galaxies \citep{Pfeffer2021}, and the merging of such systems with larger hosts is now recognised as a standard pathway for NSC formation \citep{Neumayer2020}. Earlier studies linked the progenitor galaxy of $\omega$~Cen to the Gaia-Enceladus/Sausage (GE) merger event \citep{Helmi2018, Belokurov2018}, though recent chemical and dynamic analyses suggest it may belong to a distinct accretion event, such as Sequoia or Nephele \citep{Massari2019, Pagnini2025}. A recent study analysing approximately 20~yr of HST archival data detected fast-moving stars at the cluster centre whose kinematics require an intermediate-mass black hole (IMBH) of $\approx 8200\,M_\odot$ \citep{Haeberle2024}. On Galactic scales, chemical tagging using \textsl{APOGEE}~DR17 data suggests that $\omega$~Cen may share a common origin with at least six other globular clusters under the Nephele hypothesis, pointing to a progenitor galaxy with an initial stellar mass of $\sim 10^8$--$10^9\,M_\odot$ \citep{Massari2019, Pagnini2025}. Phylogenetic analyses using chemical abundances as evolutionary characters have identified three main branches in $\omega$~Cen's chemical tree, interpreting them as the principal star-formation episodes \citep{Jofre2025}.

Despite all of this progress, the precise number of chemically distinct stellar populations in $\omega$~Cen and the abundance patterns that define them remain debated. The number of populations reported in the literature ranges from 3 to 15, depending on the photometric bands or chemical tracers adopted, the sample size, and the method of data partitioning. This wide range stems not from an intrinsic ambiguity in the system itself but from the heterogeneous nature of the datasets: different studies employ different element sets, spectral resolutions, and sample sizes, making direct comparison of population boundaries difficult and leading to non-homogeneous population definitions that cannot be directly cross-compared.
To overcome this limitation, the present study adopts a fully data-driven and homogeneous approach based on multi-dimensional chemical abundance space derived from a single spectroscopic pipeline. By applying hierarchical clustering without imposing any prior assumption on the number or shape of populations, we aim to reveal the intrinsic chemical structure of $\omega$~Cen in a reproducible and internally consistent manner.
To this end, we use high-resolution near-infrared spectroscopic data from the SDSS-V \textit{Milky Way Mapper} survey \citep[MWM DR19;][]{Kollmeier2017, Kollmeier2026} see also A. R. Casey et al., (in preparation). Recent studies based on large spectroscopic surveys such as \textsl{APOGEE} and \textsl{GALAH} have demonstrated the power of homogeneous multi-element abundance datasets in tracing the chemo-dynamical structure of $\omega$~Cen and identifying chemically related stars beyond the cluster itself \citep[e.g.,][]{Anguiano2025, Mason2026, Souza2026}. The expanded MWM DR19 dataset provides both increased sample size and improved homogeneity, enabling a more detailed and internally consistent reconstruction of the cluster’s multi-dimensional chemical structure. The MWM DR19 data were cross-matched with the \citet{Vasiliev2021} membership catalogue, yielding approximately 1000 high-quality $\omega$~Cen members. For these stars, homogeneous abundance measurements were obtained for seven elements: $\mathrm{[Fe/H]}$, $\mathrm{[C/H]}$, $\mathrm{[N/H]}$, $\mathrm{[O/H]}$, $\mathrm{[Al/H]}$, $\mathrm{[Na/H]}$, and $\mathrm{[Ca/H]}$. These elements represent four key nucleosynthetic channels relevant to multiple-population studies in globular clusters: iron-peak enrichment, $\alpha$-element production, CNO cycling, and proton-capture processes. Ward hierarchical clustering was applied in this seven-dimensional abundance space, and the dendrogram cut level was determined using silhouette and Davies--Bouldin criteria. The analysis yielded 10 chemically distinct sub-populations, whose abundance patterns, metallicity distributions, and spatial and kinematic properties are characterised.
The remainder of the paper is organised as follows. Section~\ref{sec:data_sample} describes the dataset, membership selection criteria, and quality cuts. Section~\ref{sec:method} explains the clustering methodology. Section~\ref{sec:results} presents the chemical properties of the 10 populations. Section~\ref{sec:discussion} discusses the results in the context of chemical evolution and the accretion-nucleus scenario, and Section~\ref{sec:conclusion} summarises the main findings.

\section{Data and Sample Selection}
\label{sec:data_sample}

To construct a spectroscopic sample of $\omega$ Centauri members, we cross-matched the membership catalogue of \citet{Vasiliev2021} with spectroscopic data from the Milky Way Mapper survey (MWM DR19). The MWM is one of the three core science programs of the Sloan Digital Sky Survey V (SDSS-V; \citep{Kollmeier2017, Kollmeier2026}) and is a near-infrared, high-resolution spectroscopic survey (R $\approx$ 22,500) designed to map stellar populations across the entire sky. Observations are carried out using twin \textsl{APOGEE} spectrographs mounted on the 2.5 m Sloan telescope at Apache Point Observatory (Northern Hemisphere) and the du Pont telescope at Las Campanas Observatory (Southern Hemisphere; \citep{Gunnetal2006, Wilson2019}). For MWM DR19, stellar atmospheric parameters and chemical abundance ratios were derived for approximately 1.4 million stars using the \textsl{astra} analysis framework (Casey et al., in prep), built on an updated version of the ASPCAP pipeline \citep{GarciaPerez2016} employing the FERE spectral synthesis code \citep{Allende2006, Allende2015}, together with the \textsl{APOGEE} atomic and molecular line lists \citep{Hasselquist2016, cunha2017, Smith2021}. Astrometric information---positions, parallaxes, and proper motions---was taken from \textsl{Gaia} Data Release 3 \citep{GaiaDR3}.

The membership catalogue of \citet{Vasiliev2021} assigns individual stellar membership probabilities based on \textsl{Gaia} DR3 proper motions and astrometry. We retained only stars with membership probability exceeding 90$\%$ to ensure a clean and highly reliable cluster sample, then cross-matched these against MWM DR19 using \textsl{Gaia} DR3 source identifiers. This procedure yielded 1807 stars with available MWM spectroscopic parameters.

\subsection{Sample Validation and Chemical Consistency}
\label{sec:sample_validation}

Before clustering, we checked whether the selected MWM sample broadly reproduces the established chemical and kinematic behavior of \ocen. In summary, the metallicity distribution and its metal-poor peak are compatible with high-resolution surveys \citep{Johnson2010, Marino2011}; the expected light-element trends (including O--Na/O--Al behavior) are visible in the MWM~DR19 abundance space \citep{Carretta2009a, Carretta2009b, AlvarezGaray2024}; and the mean radial velocity is consistent with published systemic kinematics for the cluster \citep{Meszaros2021}. A more detailed validation, including distribution-level comparisons and abundance-plane diagnostics, is provided in Appendix~\ref{sec:app_chem_validation}; see in particular the $[\mathrm{Fe/H}]$ distribution in Fig.~\ref{fig:feh_distribution}.

\begin{figure*}
    \centering
    \includegraphics[width=1\linewidth]{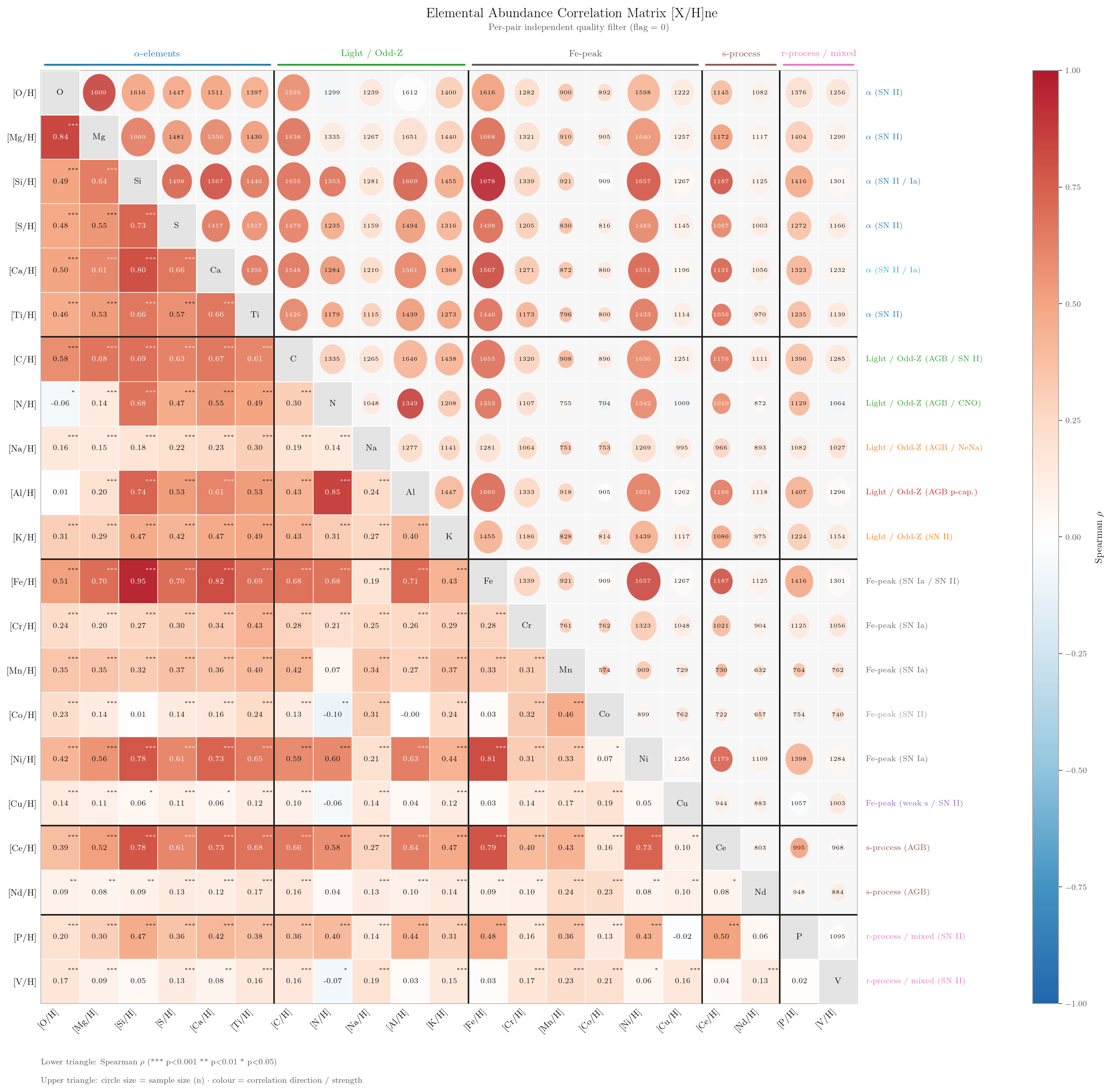}
    \caption{Spearman rank-correlation matrix for all available elemental abundances in the 1807-star MWM~DR19 \ocen\ sample, shown in the [X/H] frame. The dominant positive-correlation block among Fe-peak and $\alpha$-elements reflects the broad metallicity baseline, while weaker off-diagonal structure highlights comparatively less redundant dimensions for clustering.}
    \label{fig:corr_matrix_xh}
\end{figure*}

Before selecting clustering inputs, we explored the full elemental abundance space of this 1807-star sample by computing Spearman correlation matrices for all elements available in MWM DR19, in both the [X/H] and [X/Fe] frames (Fig.~\ref{fig:corr_matrix_xh} and Fig.~\ref{fig:corr_matrix}, respectively).
In [X/H], the matrix is largely shaped by the broad metallicity baseline of \ocen\ ($\Delta[\mathrm{Fe/H}] \approx 1.5$~dex; \citealt{Johnson2010, Marino2011}); in [X/Fe], light-element trends linked to proton-capture processing become comparatively more apparent.

Guided by these diagnostics, we adopted a compact and physically motivated feature set based on three practical criteria: nucleosynthetic coverage, limited pairwise redundancy, and sufficient completeness after quality filtering. We use seven abundances---$[\mathrm{Fe/H}], [\mathrm{C/H}], [\mathrm{N/H}], [\mathrm{O/H}], [\mathrm{Al/H}], [\mathrm{Na/H}]$, and $[\mathrm{Ca/H}]$---which sample the main enrichment channels relevant to multiple-population studies (SN~II/SN~Ia iron enrichment, CNO cycling, proton-capture processing, and $\alpha$-element production; \citealt{Gratton2004, Ventura2011, Bastian2018, Kobayashi2020}).

For these seven elements, we required per-element quality flag $=0$ and SNR $\geq 50$, resulting in a final working sample of 957 stars. The metallicity distribution of this 957-star clustering subsample closely matches that of the full 1807-star initial sample (Fig.~\ref{fig:feh_distribution}), demonstrating that the applied quality cuts do not introduce a significant metallicity bias. Additional details of the tracer-selection rationale are provided in Section~\ref{sec:element_selection}. In the following section we detail the Ward-linkage procedure, the dendrogram construction, and the objective threshold selection used to define the final chemical sub-populations.

\begin{figure*}
    \centering
    \includegraphics[width=1\linewidth]{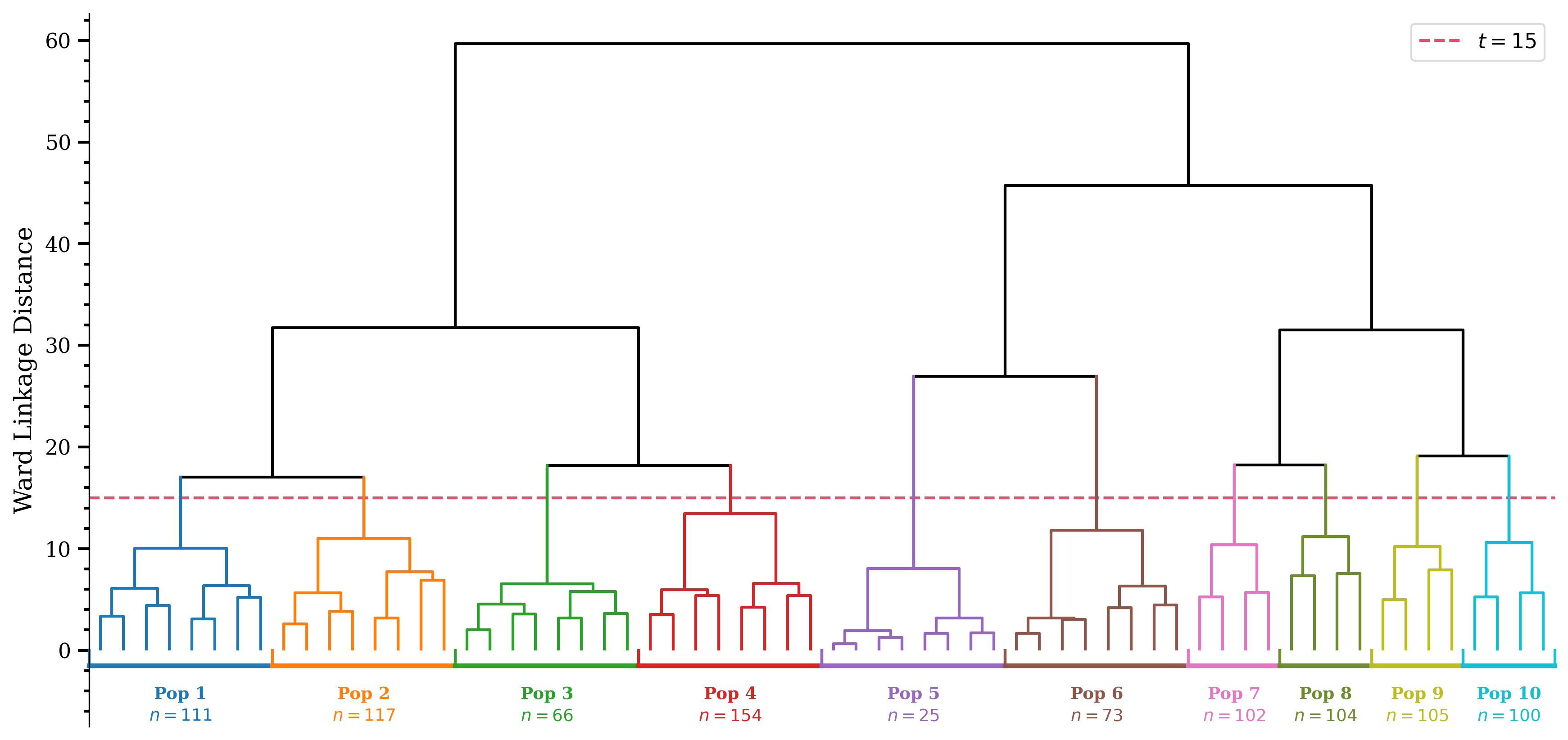}
    \caption{Ward-linkage hierarchical dendrogram for the 957 quality-selected \ocen\ members in the seven-dimensional abundance space ($[\mathrm{Fe/H}], [\mathrm{C/H}], [\mathrm{N/H}], [\mathrm{O/H}], [\mathrm{Al/H}], [\mathrm{Na/H}]$, and $[\mathrm{Ca/H}]$). The vertical axis shows linkage distance; the adopted cut yields 10 chemical sub-populations used throughout the analysis.}
    \label{fig:dendogram}
\end{figure*}

\section{Method}
\label{sec:method}


\subsection{Selection of Chemical Tracers for Hierarchical Clustering}
\label{sec:element_selection}

The selection of chemical abundance dimensions is critical for both the reproducibility and physical interpretability of the clustering results. Instead of using all available spectroscopic measurements, we adopt a targeted feature-selection strategy based on three criteria: (i) nucleosynthetic independence, (ii) limited redundancy without imposing a strict low-correlation criterion, and (iii) sufficient completeness after quality filtering.

The final clustering space consists of seven elemental abundances expressed in the $[\mathrm{X/H}]$ scale: \ion{Fe}{}, \ion{C}{}, \ion{N}{}, \ion{O}{}, \ion{Al}{}, \ion{Na}{}, and \ion{Ca}{}. These abundances are derived from the MWM~DR19 pipeline described in Section~\ref{sec:data_sample}.

The clustering is performed in the $[\mathrm{X/H}]$ frame rather than $[\mathrm{X/Fe}]$, since $[\mathrm{Fe/H}]$ is explicitly included as an independent dimension; removing it via ratio normalisation would discard the primary metallicity separation that motivates the analysis.

\subsubsection{Nucleosynthetic Basis of the Selected Elements}
\label{sec:nucleo}

The selected tracers collectively span the principal nucleosynthetic channels relevant to the chemical evolution of \ocen\ and are widely used in studies of globular cluster multiple populations \citep{Gratton2004, Bastian2018, Milone2022, Anguiano2025}.

Iron ($[\mathrm{Fe/H}]$) serves as the primary metallicity indicator, tracing the combined contributions of Type~II and Type~Ia supernovae and anchoring the broad metallicity baseline of \ocen\ ($\Delta[\mathrm{Fe/H}] \approx 1.5$~dex; \citealt{Johnson2010, Marino2011}).

Carbon and nitrogen probe CNO-cycle processing and stellar evolutionary mixing. While both elements are affected by internal processes in red giants, they retain partially independent information, as indicated by their modest correlation in the present dataset ($\rho = +0.30$).

Oxygen traces $\alpha$-element production in massive stars and plays a central role in identifying multiple populations through its well-known anti-correlations with proton-capture elements \citep{Denisenkov1990, Prantzos2007}. In the $[\mathrm{X/H}]$ space, it remains largely independent of $[\mathrm{N/H}]$ and $[\mathrm{Al/H}]$, ensuring its contribution as a distinct chemical dimension.

Aluminium is a key tracer of high-temperature proton-capture nucleosynthesis (Mg--Al cycle) and is among the most discriminating elements for population separation in globular clusters \citep{Carretta2009b, Bastian2018}. Its strong correlation with nitrogen ($\rho = +0.85$) reflects their shared origin, but both are retained due to their differing sensitivities to stellar evolutionary effects.

Sodium, produced via the NeNa cycle, exhibits relatively weak correlations with other selected elements ($|\rho| \leq 0.24$), making it one of the most orthogonal tracers in the dataset. It is also a key component of the classical Na--O anti-correlation observed in globular clusters \citep{Carretta2009a}.

Calcium, as an $\alpha$-element, traces chemical enrichment and provides important constraints on star formation timescales. Although it shows a strong correlation with iron in the $[\mathrm{X/H}]$ space ($\rho = +0.82$), this primarily reflects the global metallicity scaling shared by all elements. This correlation is therefore not interpreted as nucleosynthetic redundancy. Indeed, in the $[\mathrm{X/Fe}]$ space this dependence disappears, confirming that Ca contributes independent information as an $\alpha$-element tracer distinct from iron. Calcium is thus retained as a physically meaningful and independent dimension in the clustering space. We emphasise that correlation in the $[\mathrm{X/H}]$ space does not necessarily imply redundancy, as it may primarily arise from the shared metallicity scaling rather than identical nucleosynthetic origin. This distinction between correlation and redundancy is central to our feature-selection strategy.

Overall, this seven-element set provides a compact yet physically motivated representation of the dominant enrichment channels in \ocen, while avoiding excessive redundancy in the input feature space.

\subsubsection{Exclusion of Redundant and Low-Completeness Elements}
\label{sec:exclusion}

Fourteen additional elements available in the MWM~DR19 catalogue were evaluated but excluded from the clustering input based on redundancy, limited discriminating power, or insufficient completeness.

Several Fe-peak elements are nearly collinear with iron across the wide metallicity range of \ocen. In particular, silicon shows the strongest redundancy in the full abundance matrix ($\rho(\mathrm{[Fe/H],[Si/H]}) = +0.95$), indicating that it carries little additional information beyond $[\mathrm{Fe/H}]$. $[\mathrm{Ni/H}]$ exhibits a similarly strong correlation ($\rho = +0.81$). Including such elements would effectively overweight Fe-peak nucleosynthesis without increasing the dimensionality of the chemical space.

Magnesium shows strong correlation with oxygen ($\rho(\mathrm{[O/H],[Mg/H]}) = +0.84$), consistent with their common SN~II origin. In $[\mathrm{X/Fe}]$ space, this correlation strengthens further, indicating near-redundancy for clustering purposes. Magnesium is therefore excluded from the clustering input but retained as an independent validation tracer.

Other elements, including Mn, Cr, Co, Ni, Cu, P, V, K, Ce, and Nd, exhibit significantly weaker correlations with the Ward-cluster assignments compared to the selected tracers, as quantified by mutual information analysis (Appendix~\ref{sec:feature_selection}). While some of these elements trace specific enrichment channels (e.g. Mn and Cr as SN~Ia indicators; \citealt{Kobayashi2006, Nomoto2013}), their limited dynamic range reduces their effectiveness in separating closely spaced sub-populations within \ocen. The $s$-process elements Ce and Nd are retained for post-hoc chemical analysis but excluded from clustering to avoid sensitivity to stellar evolutionary state.

Finally, elements with fewer than $N < 800$ high-quality measurements were excluded to ensure adequate statistical sampling across all clusters. Sulfur ($[\mathrm{S/H}]$) was considered as an additional $\alpha$-element tracer but omitted due to its redundancy with oxygen and calcium. It is included in the extended correlation matrix for completeness. 
More generally, we note that imposing strict quality criteria (flag $=0$ and SNR cuts) on additional elements can significantly reduce the available sample size, leading to a loss of statistical robustness. 


These selection criteria ensure that the clustering is performed in a chemically informative yet minimally redundant feature space.

\subsubsection{Verification via the Extended Correlation Matrix}
\label{sec:corr_verify}

The element-selection rationale is summarised visually in the extended Spearman correlation matrices presented in Appendix~\ref{sec:feature_selection}, computed independently for both $[\mathrm{X/H}]$ and $[\mathrm{X/Fe}]$ representations. The $[\mathrm{X/H}]$ matrix (Fig.~\ref{fig:corr_matrix_xh}) shows that the strongest off-diagonal correlations involve element pairs that are either excluded from the clustering set (e.g.\ $[\mathrm{Fe/H}]$--$[\mathrm{Si/H}]$, $\rho = +0.95$; $[\mathrm{Fe/H}]$--$[\mathrm{Ni/H}]$, $\rho = +0.81$) or are retained as the sole representative of their nucleosynthetic channel (e.g.\ $[\mathrm{N/H}]$--$[\mathrm{Al/H}]$, $\rho = +0.85$, both tracers of the Mg--Al/NeNa proton-capture chain at high temperature). Within the selected seven-element set, the strongest pairwise correlation is $\rho(\mathrm{[Fe/H],[Ca/H]}) = +0.82$. As described in Section~\ref{sec:nucleo}, this correlation arises from the global metallicity scaling of $[\mathrm{X/H}]$ abundances across the wide baseline of \ocen, and is therefore not indicative of nucleosynthetic redundancy. No element pair exceeds $|\rho| = 0.85$, while the majority of off-diagonal terms lie below $|\rho| = 0.70$, confirming that the adopted feature set retains sufficient independence and discriminating power for Ward-linkage clustering.
The $[\mathrm{X/Fe}]$ matrix (Fig.~\ref{fig:corr_matrix}) provides complementary evidence: after removing the common metallicity trend, the anti-correlations expected from hot hydrogen burning become prominent ($\rho(\mathrm{[O/Fe],[N/Fe]}) = -0.77$, $\rho(\mathrm{[O/Fe],[Al/Fe]}) = -0.77$), confirming that the selected elements encode the physically distinct enrichment channels that define the multiple populations of \ocen\ \citep{NorrisDaCosta1995, Bedin2004, Villanova2014, Milone2017}. The Fe-peak elements excluded from the clustering basis (Cr, Mn, Ni) show only weak $[\mathrm{X/Fe}]$ correlations with $[\mathrm{Fe/H}]$ ($|\rho| < 0.20$), consistent with their limited discriminating power at the population level.

In summary, the chosen seven-element set provides a compact and physically meaningful representation that spans the four principal nucleosynthetic channels active in \ocen--iron enrichment (SN~II + SN~Ia), $\alpha$-element synthesis (SN~II), light-element proton-capture nucleosynthesis (AGB / massive star winds), and CNO cycling---while avoiding both numerical redundancy and the confounding effects of stellar evolutionary processing on surface abundances.

\subsubsection{Algorithm and Preprocessing}
\label{sec:ward_algorithm}

We identify chemically distinct stellar populations in \ocen\ using hierarchical agglomerative clustering with Ward's minimum-variance linkage in Euclidean space. This method iteratively merges clusters to minimise the increase in total within-cluster variance, naturally yielding compact and chemically homogeneous structures in the multi-dimensional abundance space.

A key advantage of hierarchical clustering is that it does not require specifying the number of clusters a priori. Instead, it constructs the full hierarchy of cluster mergers, allowing the final partition to be selected based on the structure of the dendrogram. This property is particularly important for \ocen, where stellar populations form a continuum rather than sharply separated groups.

The clustering is performed on the 957 stars that satisfy the membership, signal-to-noise, and abundance-quality criteria defined in Section~\ref{sec:data_sample}. The input feature space consists of the seven chemical abundances selected in Section~\ref{sec:element_selection}, chosen to capture the dominant nucleosynthetic channels and maximise the discriminatory power between stellar populations.

Prior to clustering, each abundance dimension is standardised to zero mean and unit variance using a \texttt{StandardScaler} transformation. This ensures that differences in the intrinsic dispersion of individual elements do not bias the Euclidean distance calculation, allowing each dimension to contribute equally to the clustering process. This step is essential when using variance-based linkage methods such as Ward’s algorithm. The clustering is performed exclusively in the chemical abundance space; no kinematic parameters or dimensionality-reduced representations are used at this stage.

\subsubsection{Dendrogram Construction and Partition Selection}
\label{sec:dendrogram}

\begin{figure}
    \centering
    \includegraphics[width=1\linewidth]{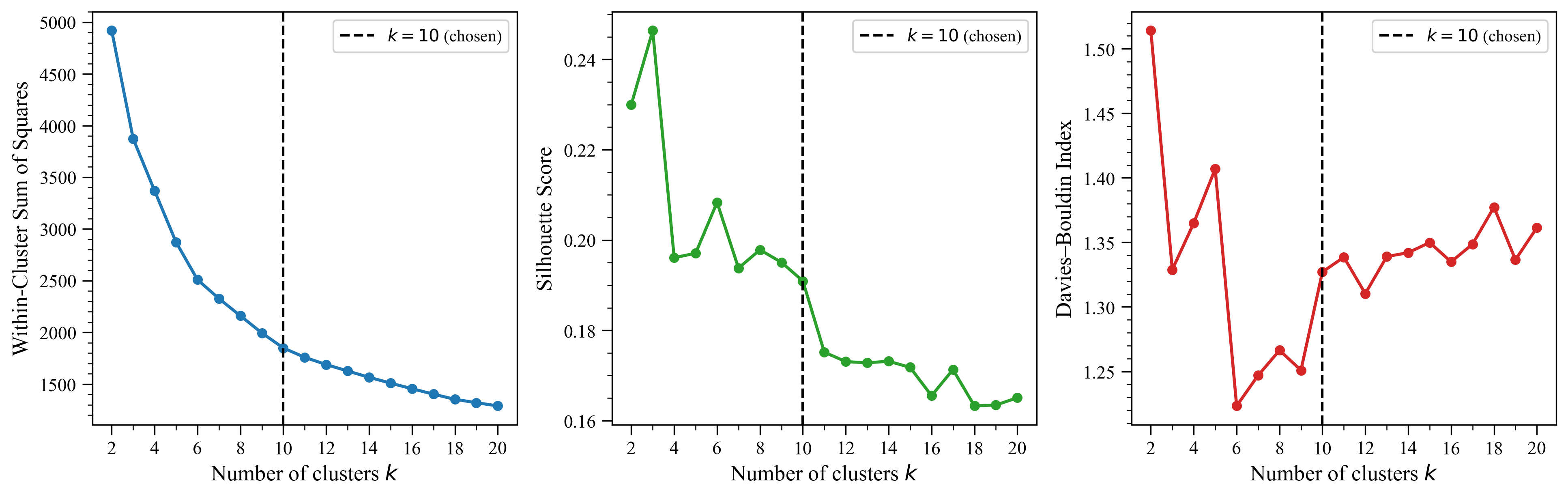}
    \caption{Validation of the Ward clustering solution in the seven-element abundance space ($[\mathrm{Fe/H}]$, $[\mathrm{C/H}]$, $[\mathrm{N/H}]$, $[\mathrm{O/H}]$, $[\mathrm{Al/H}]$, $[\mathrm{Na/H}]$, $[\mathrm{Ca/H}]$). Left: Within-cluster sum of squares (WCSS); the rate of decrease flattens near $k = 10$, indicating the elbow of the curve. Centre: Silhouette score; the peak at $k = 3$ is consistent with the three broad metallicity groups (metal-poor backbone, intermediate, metal-rich tail) present in $\omega$ Cen. Right: Davies–Bouldin index; values remain low and stable over $k = 6$--$10$ before rising, supporting $k = 10$ as the adopted number of populations (vertical dashed line).}
    \label{fig:cluster_metrics}
\end{figure}

The hierarchical structure of the data is visualised through the Ward dendrogram (Figure~\ref{fig:dendogram}), where the vertical axis represents the linkage distance between merged clusters.

To determine the optimal partition level, we evaluate the clustering structure as a function of the dendrogram cut threshold $t$. For each value of $t$, we compute the resulting number of clusters together with internal clustering diagnostics, including the silhouette coefficient, the Davies–Bouldin index, and the within-cluster sum of squares (WCSS) (Figure~\ref{fig:cluster_metrics}).

At small values of $t$, the data are fragmented into many small clusters, while increasing $t$ progressively merges them into larger structures. The silhouette score peaks at $k = 3$, broadly consistent with the classical metal-poor backbone, intermediate population, and metal-rich tail structure of $\omega$ Cen. However, this coarse partition does not capture the full chemical complexity of the system.

For larger numbers of clusters, the Davies–Bouldin index remains low and relatively stable over $k = 6$--$10$, while the WCSS curve exhibits diminishing returns beyond $k \approx 10$, indicating that additional clusters do not significantly improve compactness.

Around $t = 15$, the clustering diagnostics converge: the silhouette score reaches a plateau and the Davies–Bouldin index shows minimal improvement. This behavior, consistent with Figure~\ref{fig:cluster_metrics}, corresponds to a partition with $k = 10$ clusters.

We therefore adopt a threshold of $t = 15$, corresponding to $k = 10$ clusters. This choice provides a balance between capturing chemically distinct sub-structure and avoiding over-fragmentation, and is further supported by the stability of the Davies–Bouldin index and the reduced marginal gain in WCSS.

\subsubsection{Validation of the Clustering Solution}
\label{sec:validation}

Internal clustering diagnostics tend to favor coarser partitions, as expected for partially overlapping populations in multi-dimensional chemical space. We therefore use these metrics as guidance rather than as strict optimisation criteria, and adopt the 10-cluster solution as a compromise between resolution and statistical robustness. The robustness of the 10‑cluster solution against measurement uncertainties and sampling variance is quantitatively verified in Appendix \ref{sec:app_robustness}.

To assess the stability of this partition, we examine the separability of the clusters across the principal abundance dimensions and verify that the identified groups remain distinct in the selected chemical space.

\begin{figure}
    \centering
    \includegraphics[width=1\columnwidth]{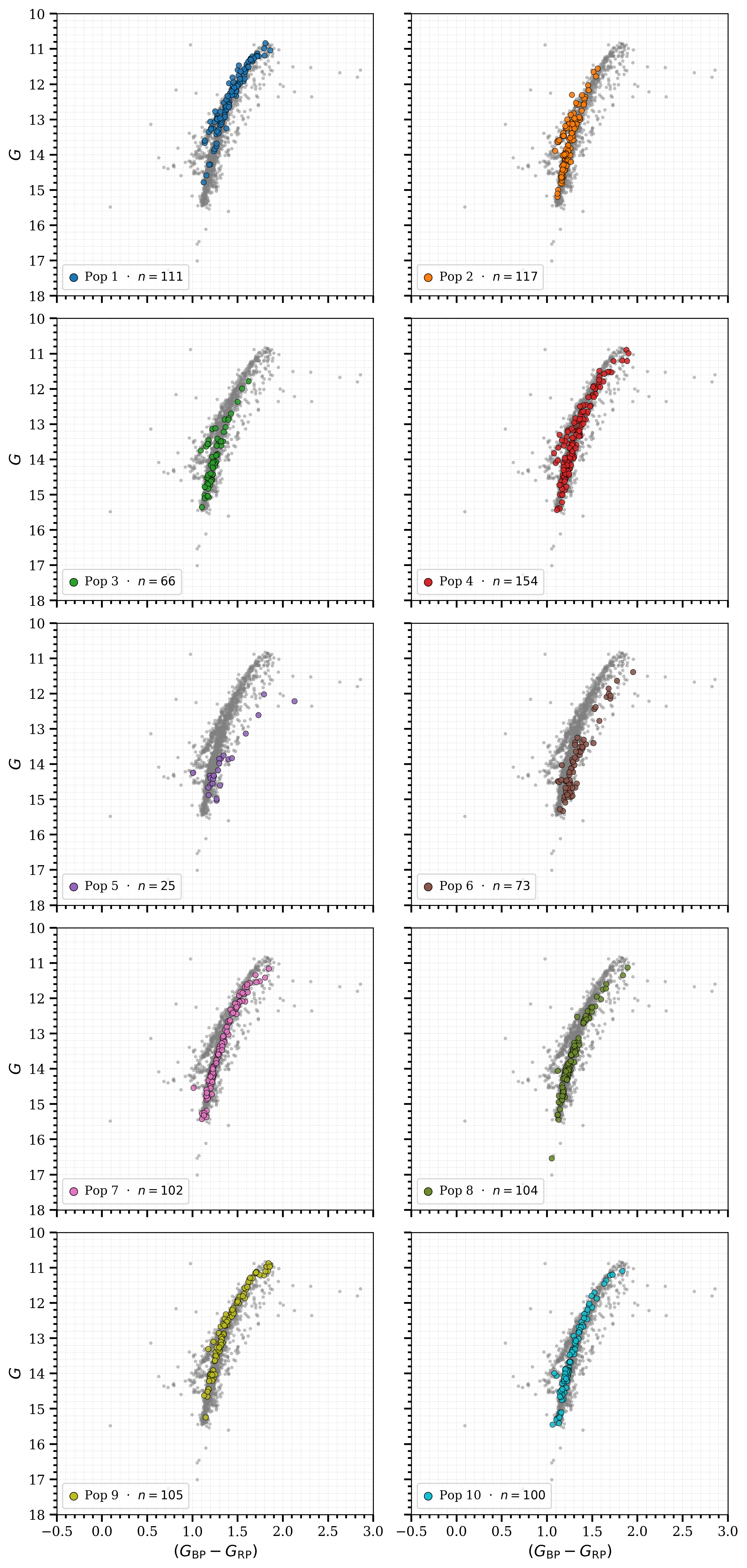}
    \caption{color--magnitude diagram of the 957 quality-selected \ocen\ members, color-coded by Ward-clustering assignment ($k=10$). The plot illustrates how the chemically defined groups are distributed across the observed photometric sequences.}
    \label{fig:color_mag_diagram}
\end{figure}

As an independent validation, we compare the chemically defined clusters with their distribution in the color--magnitude diagram (CMD) shown in Figure~\ref{fig:color_mag_diagram}. The CMD is not used as an input to the clustering; the partition is defined entirely in chemical abundance space. This comparison therefore provides an external consistency check on the clustering outcome.

The CMD-based validation also establishes the foundation for the subsequent analysis, where isochrone fitting is applied to each population in order to investigate their relative evolutionary properties.

\subsubsection{Population Summary}
\label{sec:pop_summary}

The ten Ward populations are discussed below, ordered by median $[\mathrm{Fe/H}]$.
Their sample sizes, age estimates, and median chemical properties are summarised in Table~\ref{tab:pop_summary}.

\begin{table*}[ht]
\centering
\small
\caption{Summary of stellar populations in $\omega$ Cen. Chemical abundances show median $\pm$ standard deviation for the clustering sample.}
\label{tab:pop_summary}
\begin{tabular}{lcccccccc}
\hline
Pop & $N$ & $[\mathrm{Fe}/\mathrm{H}]$ & $[\mathrm{C}/\mathrm{H}]$ & $[\mathrm{N}/\mathrm{H}]$ & $[\mathrm{O}/\mathrm{H}]$ & $[\mathrm{Al}/\mathrm{H}]$ & $[\mathrm{Na}/\mathrm{H}]$ & $[\mathrm{Ca}/\mathrm{H}]$ \\
\hline
Pop 1  & 111 & $-1.75 \pm 0.07$ & $-2.32 \pm 0.28$ & $-1.37 \pm 0.22$ & $-1.43 \pm 0.09$ & $-2.06 \pm 0.18$ & $-1.53 \pm 0.32$ & $-1.67 \pm 0.21$ \\
Pop 2  & 117 & $-1.75 \pm 0.07$ & $-1.91 \pm 0.29$ & $-1.32 \pm 0.27$ & $-1.41 \pm 0.18$ & $-1.70 \pm 0.38$ & $-0.92 \pm 0.25$ & $-1.72 \pm 0.25$ \\
Pop 3  & 66  & $-1.58 \pm 0.10$ & $-1.42 \pm 0.21$ & $-1.17 \pm 0.30$ & $-1.16 \pm 0.15$ & $-1.59 \pm 0.23$ & $-0.90 \pm 0.28$ & $-1.31 \pm 0.15$ \\
Pop 4  & 154 & $-1.61 \pm 0.09$ & $-1.50 \pm 0.27$ & $-1.14 \pm 0.23$ & $-1.21 \pm 0.17$ & $-1.67 \pm 0.19$ & $-1.78 \pm 0.24$ & $-1.42 \pm 0.24$ \\
Pop 5  & 25  & $-0.93 \pm 0.18$ & $-0.88 \pm 0.17$ & $0.47 \pm 0.22$  & $-0.63 \pm 0.22$ & $-0.22 \pm 0.11$ & $-0.41 \pm 0.18$ & $-0.61 \pm 0.16$ \\
Pop 6  & 73  & $-1.30 \pm 0.14$ & $-0.98 \pm 0.27$ & $-0.96 \pm 0.22$ & $-0.77 \pm 0.20$ & $-1.36 \pm 0.19$ & $-1.43 \pm 0.33$ & $-0.94 \pm 0.16$ \\
Pop 7  & 102 & $-1.35 \pm 0.09$ & $-1.31 \pm 0.27$ & $-0.15 \pm 0.18$ & $-1.36 \pm 0.17$ & $-0.30 \pm 0.11$ & $-1.01 \pm 0.29$ & $-1.06 \pm 0.12$ \\
Pop 8  & 104 & $-1.47 \pm 0.09$ & $-1.43 \pm 0.18$ & $-0.40 \pm 0.21$ & $-1.35 \pm 0.20$ & $-0.71 \pm 0.32$ & $-1.66 \pm 0.42$ & $-1.19 \pm 0.14$ \\
Pop 9  & 105 & $-1.66 \pm 0.08$ & $-1.94 \pm 0.28$ & $-0.81 \pm 0.30$ & $-1.55 \pm 0.17$ & $-1.14 \pm 0.26$ & $-1.71 \pm 0.33$ & $-1.50 \pm 0.16$ \\
Pop 10 & 100 & $-1.59 \pm 0.07$ & $-1.66 \pm 0.21$ & $-0.59 \pm 0.25$ & $-1.63 \pm 0.20$ & $-0.83 \pm 0.31$ & $-0.87 \pm 0.27$ & $-1.38 \pm 0.16$ \\ \hline
\end{tabular}
\end{table*}

Populations~1--4 comprise the four most metal-poor, proton-capture-poor populations.
Population~4 ($N = 154$) is the most numerous single group, together with Populations~1 ($N = 111$) and 2 ($N = 117$) forming the dominant metal-poor backbone of \ocen\
($-1.75 < [\mathrm{Fe/H}]_{\mathrm{med}} < -1.58$), analogous to the classical ``main'' population identified by \citet{NorrisDaCosta1995} and \citet{Johnson2010}.
Population~2 shows a marginally elevated $[\mathrm{Al/Fe}] = +0.028$ relative to Populations~1, 3, and 4 ($-0.31 < [\mathrm{Al/Fe}] < -0.05$), suggesting an incipient proton-capture contamination at the boundary between these population groups.

Populations~5--10 are chemically heterogeneous. Populations~7, 8, 9, and 10 are proton-capture enriched ($[\mathrm{Al/Fe}] = +0.56$ to $+1.01$), with Population~7 ($[\mathrm{Al/Fe}] = +1.007$) representing the most extreme nucleosynthetic enrichment in the sample. Population~6 ($N = 73$, $[\mathrm{Fe/H}]_{\mathrm{med}} = -1.30$) has primordial $[\mathrm{Al/Fe}] = -0.025$ but falls within the second group owing to its elevated metallicity, which displaces it from Populations~1--4 in the standardised 7-element space. Population~5 ($N = 25$, $[\mathrm{Fe/H}]_{\mathrm{med}} = -0.93$)
is the anomalous metal-rich tail of \ocen, corresponding to the ``RGB-a'' component of \citet{Pancino2000} and ``SGB-a'' component of \citet{Villanova2014}.

\subsection{Isochrone Fitting Framework}
\label{sec:age_estimation}

Having chemically partitioned the \ocen\ sample into ten distinct populations, we now turn to their relative evolutionary properties. To this end, we estimate stellar population ages by fitting \textsl{Gaia}~DR3 color--magnitude diagrams (CMDs) to theoretical isochrones from the BaSTI-IAC library \citep{Pietrinferni2021}. We adopt the standard $\alpha$-enhanced grid with $[\alpha/\mathrm{Fe}]=+0.4$ and initial helium abundance $Y=0.30$, in the \textsl{Gaia}~DR3 photometric system ($G$, $G_{\mathrm{BP}}$, $G_{\mathrm{RP}}$), as publicly available through the BaSTI-IAC web interface. The grid covers $\log(\mathrm{Age}/\mathrm{yr}) = 9.85$--$10.18$ ($\approx 7$--$15$~Gyr) in steps of $0.1$~Gyr, and ten metallicity nodes spanning $Z_{\mathrm{ini}} = 0.0001$--$0.004$ ($[\mathrm{Fe/H}] \approx -2.5$ to $-0.9$ on the BaSTI $[\alpha/\mathrm{Fe}]=+0.4$ scale), with a typical node spacing of $0.1$--$0.2$~dex in $[\mathrm{Fe/H}]$.

Spectroscopic metallicities are converted to the BaSTI-IAC $Z_{\mathrm{ini}}$ scale using the \citet{Salaris1993} prescription for $\alpha$-enhanced mixtures.  The total metallicity is
\begin{equation}
  [\mathrm{M/H}] = [\mathrm{Fe/H}]
    + \log\!\bigl(0.638\times10^{[\alpha/\mathrm{Fe}]}+0.362\bigr),
  \label{eq:salaris}
\end{equation}
which accounts for the fractional contribution of $\alpha$ elements
to the total metal budget.  For the adopted value
$[\alpha/\mathrm{Fe}]=+0.4$ this reduces to
\begin{equation}
  [\mathrm{M/H}] = [\mathrm{Fe/H}]
    + \log(0.638\times10^{0.4}+0.362)
    = [\mathrm{Fe/H}] + 0.2934,
  \label{eq:mh_numeric}
\end{equation}
and the corresponding initial metal fraction is
\begin{equation}
  Z_{\mathrm{ini}} = Z_{\odot}\times 10^{[\mathrm{M/H}]}
                   = 0.01508\times 10^{[\mathrm{M/H}]},
  \label{eq:zini}
\end{equation}
where $Z_{\odot}=0.01508$ is the solar metal fraction of the BaSTI-IAC models \citep{Pietrinferni2021}.

Likelihood evaluation at each nested-sampling live point requires predicting model photometry at arbitrary $(\log\mathrm{Age}, Z_{\mathrm{ini}}, M_{\mathrm{ini}})$, which is expensive if done by interpolating the discrete grid at runtime.  We instead train a \textsc{DecisionTreeRegressor} interpolator \citep{scikit-learn} on the full BaSTI-IAC grid restricted to the \ocen\ parameter space ($1.94\times10^6$ grid points after removing the lower main sequence, $\log L/L_\odot < -1.5$).  The interpolator maps $(\log\mathrm{Age},\,Z_{\mathrm{ini}},\,M_{\mathrm{ini}}) \rightarrow (G,\,G_{\mathrm{BP}}-G_{\mathrm{RP}})$ and is grown to full depth (\texttt{max\_depth=None}).

Decision-tree interpolation is well suited to isochrone grids because age and metallicity are discretely sampled at fixed nodes while $M_{\mathrm{ini}}$ varies continuously along each isochrone. The resulting axis-aligned partition structure reproduces the sharp morphological transitions --- MSTO, subgiant hook, RGB base --- without imposing global smoothness assumptions.  Grid-point accuracy is sub-millimagnitude:
$\mathrm{MAE}(G) = 0.17$~mmag and
$\mathrm{MAE}(G_{\mathrm{BP}}-G_{\mathrm{RP}}) = 0.06$~mmag,
less than typical \textsl{Gaia} photometric uncertainties for \ocen\ members.

\subsubsection{Log-Likelihood Function}
\label{sec:likelihood}

Each population CMD is described by four free parameters:
\begin{equation}
    \boldsymbol{\theta} = \bigl(\log\mathrm{Age},\; Z_{\mathrm{ini}},\;
    d_{\mathrm{pc}},\; E(G_{\mathrm{BP}}-G_{\mathrm{RP}})\bigr).
    \label{eq:theta}
\end{equation}
At each likelihood call the interpolator isochrone is shifted to the observed frame by applying the distance modulus and extinction corrections before comparison with the data.

We adopt a nearest-neighbour Gaussian likelihood in CMD space. For a given parameter vector $\boldsymbol{\theta}$, the interpolator produces a set of model isochrone points $\{(C_j^{\rm mod}, G_j^{\rm mod})\}_{j=1}^{N_{\rm iso}}$,
where $C = G_{\rm BP} - G_{\rm RP}$. Each observed star $i$ is assigned to the isochrone point closest to it in the
two-dimensional CMD, and the squared residuals in colour and magnitude are evaluated independently, normalised by the
respective \textsl{Gaia} photometric uncertainties. The log-likelihood is then:

\begin{equation}
    \ln \mathcal{L}(\boldsymbol{\theta}) =
    -\frac{1}{2}\sum_{i=1}^{N_\star}
    \left\{
    \min_j \left[
    \frac{\bigl(\Delta C_j^{(i)}\bigr)^2}{\sigma_{C,i}^2}
    +
    \frac{\bigl(\Delta G_j^{(i)}\bigr)^2}{\sigma_{G,i}^2}
    \right]
    + \ln\!\bigl(2\pi\,\sigma_{C,i}\,\sigma_{G,i}\bigr)
    \right\},
    \label{eq:logL}
\end{equation}
where $\Delta C_j^{(i)} = C_i^{\rm obs} - C_j^{\rm mod}$ and $\Delta G_j^{(i)} = G_i^{\rm obs} - G_j^{\rm mod}$ are the
colour and magnitude residuals between observed star $i$ and model point $j$, and the minimum is taken over all $N_{\rm iso}$ model points.

The physical motivation for this form is the following. If the photometric noise on each star is well described by
independent Gaussians in colour and magnitude then the probability of observing star $i$ at its measured CMD position, given that it lies on the isochrone at position $j$, is a two-dimensional Gaussian centred on $(C_j^{\rm mod}, G_j^{\rm mod})$ with widths $(\sigma_{C,i}, \sigma_{G,i})$.  Marginalising over the unknown position along the isochrone and approximating the marginal by its maximum (nearest-point assignment) yields exactly Equation~(\ref{eq:logL}). The normalisation term $\ln(2\pi\sigma_{C,i}\sigma_{G,i})$ ensures that the likelihood value is meaningful as an absolute quantity, in particular it enters the Bayesian evidence integral correctly, and penalises stars with small photometric uncertainties more strongly when the model fails to reproduce their CMD positions.

\subsubsection{Priors}
\label{sec:priors}

Prior distributions are listed in Table~\ref{tab:priors}.  We adopt uninformative uniform priors on age, distance, and reddening, allowing the likelihood to determine these parameters without imposing external anchors.  The only informative prior is placed on metallicity, where the APOGEE spectroscopy provides a direct constraint: a Gaussian centred on each population's spectroscopic median $[\mathrm{Fe/H}]_{\rm med}$ (converted to $Z_{\rm ini}$ via Equation~\ref{eq:mh_numeric}--\ref{eq:zini}) with $\sigma_{[\mathrm{Fe/H}]} = 0.15$~dex prevents the photometric fit from drifting to metallicities inconsistent with the measured stellar chemistry.

\begin{table}
\centering
\caption{Prior distributions for the nested-sampling fit.}
\label{tab:priors}
\begin{tabular}{lll}
\hline\hline
Parameter & Type & Range / ($\mu$, $\sigma$) \\
\hline
$\log\mathrm{Age}$            & Uniform           & $[9.90,\,10.18]$ \\
$Z_{\mathrm{ini}}$            & Gaussian  & ($\mathrm{[Fe/H]}_{\rm spec}$,\ $0.15$~dex) \\
$d_{\rm pc}$                  & Uniform           & $[5300,\,5700]$~pc \\
$E(B_{\rm P}{-}R_{\rm P})$   & Uniform           & $[0.0,\,0.5]$~mag \\
\hline
\end{tabular}
\end{table}

The distance range $[5300, 5700]$~pc encompasses the full spread of published heliocentric distances to \ocen\ \citep{Baumgardt2021, Vasiliev2021, Haberle2025} and includes a conservative allowance for the physical depth of the cluster ($\sim$100~pc line-of-sight extent). The reddening range $[0.0, 0.5]$~mag generously covers the total Schlegel \citep{Schlegel1998} dust column towards \ocen\ ($E(B - V) \approx 0.18$~mag) and accommodates any differential reddening across the $\sim$1\degr\ field. 

\subsubsection{Posterior Sampling}
\label{sec:dynesty}

The posterior is sampled with the \textsc{dynesty} \citep{Speagle2020} \textsc{DynamicNestedSampler} \citep{dynamicsampling}.  Sampling proceeds in two phases.  Phase~1 uses $N_{\rm live}=400$ initial
live points and runs until the evidence increment satisfies $\Delta\ln\mathcal{Z} < 0.01$.  Phase~2 adds dynamic batches
(up to five) targeted at the posterior mass until the effective sample size (ESS) reaches $N_{\rm eff} \geq 1500$. This two-phase strategy ensures both reliable evidence estimates and well-resolved posterior shapes, including any multi-modal structure in age. Posterior chains are stored per population in HDF5 format and parameter estimates are reported as posterior means with $68\%$ credible intervals (16th--84th percentiles).

Two convergence diagnostics accompany every run and are reported in Table~\ref{tab:nested_results}: the log-evidence $\ln\mathcal{Z}$ and the effective sample size ESS.

\textit{Log-evidence $\ln\mathcal{Z}$.}
The Bayesian evidence is the prior-weighted average of the likelihood over the entire parameter space,

\begin{equation}
  \mathcal{Z} = \int \mathcal{L}(\boldsymbol{\theta})\,
                     \pi(\boldsymbol{\theta})\,
                     \mathrm{d}\boldsymbol{\theta},
  \label{eq:evidence}
\end{equation}

and its logarithm is the primary output of nested sampling. Absolute values of $\ln\mathcal{Z}$ depend on both the number of
stars $N_\star$ and their photometric quality: populations with more members or tighter photometric uncertainties produce larger (less negative) evidence values simply because there is more data constraining the isochrone position.  In our application, $\ln\mathcal{Z}$ therefore scales roughly with $N_\star$, as can be verified in Table~\ref{tab:nested_results} (e.g.\ Pop.~4 with $N=154$ yields $\ln\mathcal{Z} = 284.2$, while Pop.~5 with $N=25$ yields $\ln\mathcal{Z} = 10.4$).
The estimated uncertainty $\sigma_{\ln\mathcal{Z}} \leq 0.13$ for all populations confirms that the evidence integral has
converged: with 400 live points the random error on $\ln\mathcal{Z}$ from nested sampling is well below the
astrophysically relevant scale of $\ln(10)/2 \approx 1.15$ used in Bayes factor comparisons.  In future work,
$\ln\mathcal{Z}$ differences between competing isochrone models (e.g.\ different $[\alpha/\mathrm{Fe}]$ or $Y$ prescriptions) will provide a rigorous model-selection criterion.

\textit{Effective sample size.}
The posterior samples returned by nested sampling are not independent draws: each sample carries a weight $w_k \propto
\mathcal{L}_k\,\Delta\mathcal{Z}_k$ inherited from the prior volume compression at that iteration. The ESS measures
the number of equivalent independent samples represented by the weighted chain,

\begin{equation}
  \mathrm{ESS} = \frac{\bigl(\sum_k w_k\bigr)^2}{\sum_k w_k^2},
  \label{eq:ess}
\end{equation}

and quantifies the effective resolution of the posterior. All ten populations reach $\mathrm{ESS} \geq 1658$ after Phase~2,
with a median of $\approx 2000$. An ESS of this magnitude is sufficient to resolve the posterior means, standard deviations, and 16th--84th percentile credible intervals to better than $1\%$ statistical precision, and to faithfully represent any multi-modal or skewed structure in the age posterior. The high ESS values for populations with tightly constrained parameters (e.g.\ Pop.~5 with ESS $= 2781$) reflect the Phase~2 dynamic batching: additional live points are targeted at the peak posterior mass, producing a denser sampling of the well-localised posterior regions.

\begin{figure}
\centering
\includegraphics[width=\columnwidth]{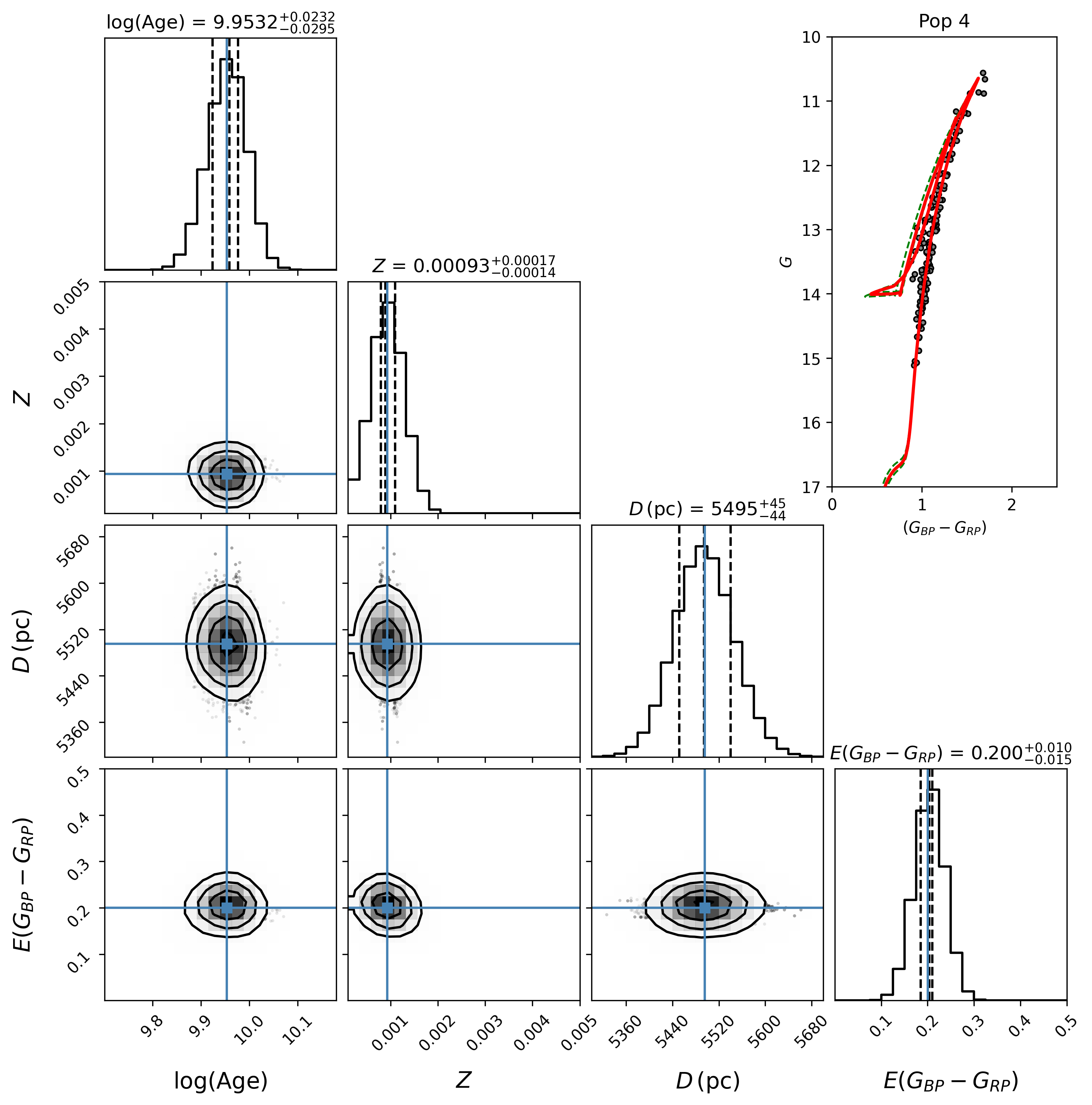}
\caption{%
  Nested-sampling result for Population~4 ($N=154$).
  \textit{Top right:} \textsl{Gaia}~DR3 CMD of Population~4
  members (black points) with the best-fitting BaSTI-IAC isochrone
  ($[\alpha/\mathrm{Fe}]=+0.4$, $Y=0.30$; red curve) drawn at the
  posterior mean parameters.
  \textit{Corner panels:} Marginalised posterior distributions for
  the four free parameters $(\log\mathrm{Age},\,Z,\,D\,[\mathrm{pc}],\,
  E(G_{\rm BP}-G_{\rm RP}))$.  Diagonal panels show 1D histograms
  with vertical dashed lines at the 16th, 50th, and 84th percentiles;
  quoted values above each histogram give the posterior mean and
  asymmetric $1\sigma$ credible interval.  Off-diagonal panels show
  2D marginal posteriors with $1\sigma$ and $2\sigma$ contours.
}
\label{fig:cmd_pop4}
\end{figure}

Figure~\ref{fig:cmd_pop4} shows the nested-sampling result for Population~4 ($N = 154$ stars), the most populous population in the spectroscopic clustering sample.  The top-right panel displays the observed \textsl{Gaia}~DR3 CMD with the best-fitting BaSTI-IAC isochrone overlaid; the other panels show the marginalised one- and two-dimensional posterior distributions for all four parameters.  The isochrone passes through the upper RGB and red clump simultaneously, confirming that the fit is driven by the global CMD morphology rather than by any single feature. The log(Age) posterior peaks at $9.95^{+0.02}_{-0.03}$ and is significantly narrower than the uniform prior, indicating that the RGB color and luminosity distribution carries non-negligible age information for this high-membership population. The distance ($d = 5495^{+45}_{-44}$~pc) and reddening ($E(G_{\rm BP}-G_{\rm RP}) = 0.200^{+0.010}_{-0.015}$~mag) are tightly constrained and consistent across all ten populations (Table~\ref{tab:nested_results}), validating the internal coherence of the fits.

\begin{table*}
\centering
\caption{%
  BaSTI-IAC nested-sampling results for the ten Ward populations of
  $\omega$~Cen ($[\alpha/\mathrm{Fe}]=+0.4$, $Y=0.30$).  $N$ is the
  number of stars in the spectroscopic clustering sample
  (flag-clean, $p_{\rm mem}\geq0.9$).  All parameters are posterior
  means with $68\%$ credible intervals (16th--84th percentiles).
  $\ln\mathcal{Z}$ is the log-evidence with its $1\sigma$ sampling
  uncertainty; ESS is the posterior effective sample size after
  Phase~2 dynamic batching (see Section~\ref{sec:dynesty}).
}
\label{tab:nested_results}
\setlength{\tabcolsep}{4pt}
\renewcommand{\arraystretch}{1.3}
\begin{tabular}{c c c c c c c c}
\hline\hline
Pop. & $N$ & Age (Gyr) & $\mathrm{[Fe/H]}$ & $d$ (pc) & $E(B_{\rm P}{-}R_{\rm P})$ (mag) & $\ln\mathcal{Z}$ & ESS \\
\hline
  1 & 111 & $9.07^{+0.62}_{-0.66}$ & $-1.71^{+0.04}_{-0.04}$ & $5450^{+46}_{-47}$ & $0.197^{+0.006}_{-0.006}$ & $203.5\pm0.12$ & 2258 \\
  2 & 117 & $8.91^{+0.92}_{-0.78}$ & $-1.77^{+0.07}_{-0.07}$ & $5505^{+54}_{-53}$ & $0.206^{+0.010}_{-0.011}$ & $222.1\pm0.11$ & 1823 \\
  3 &  66 & $9.76^{+1.42}_{-1.38}$ & $-1.56^{+0.09}_{-0.10}$ & $5504^{+48}_{-48}$ & $0.202^{+0.014}_{-0.013}$ & $125.9\pm0.10$ & 1658 \\
  4 & 154 & $8.98^{+0.49}_{-0.59}$ & $-1.50^{+0.07}_{-0.07}$ & $5495^{+45}_{-44}$ & $0.200^{+0.010}_{-0.015}$ & $284.2\pm0.13$ & 2152 \\
  5 &  25 & $11.26^{+0.85}_{-0.46}$ & $-1.13^{+0.03}_{-0.06}$ & $5523^{+49}_{-50}$ & $0.321^{+0.020}_{-0.022}$ & $10.4\pm0.13$  & 2781 \\
  6 &  73 & $12.28^{+0.64}_{-0.06}$ & $-1.17^{+0.04}_{-0.04}$ & $5521^{+53}_{-53}$ & $0.195^{+0.005}_{-0.008}$ & $124.2\pm0.13$ & 1806 \\
  7 & 102 & $9.43^{+1.24}_{-1.14}$ & $-1.30^{+0.06}_{-0.08}$ & $5506^{+48}_{-48}$ & $0.159^{+0.012}_{-0.012}$ & $188.5\pm0.10$ & 1911 \\
  8 & 104 & $11.19^{+2.81}_{-2.62}$ & $-1.39^{+0.05}_{-0.05}$ & $5510^{+51}_{-50}$ & $0.163^{+0.014}_{-0.014}$ & $189.6\pm0.10$ & 1854 \\
  9 & 105 & $9.86^{+1.36}_{-1.20}$ & $-1.58^{+0.05}_{-0.07}$ & $5482^{+44}_{-45}$ & $0.174^{+0.013}_{-0.012}$ & $198.8\pm0.12$ & 2358 \\
 10 & 100 & $8.91^{+0.64}_{-0.71}$ & $-1.58^{+0.06}_{-0.07}$ & $5511^{+49}_{-50}$ & $0.186^{+0.012}_{-0.010}$ & $189.8\pm0.11$ & 1777 \\
\hline
\end{tabular}
\end{table*}

\section{Results}
\label{sec:results}

\subsection{Robustness of the Clustering Solution}
\label{sec:robustness_sol}

The stability of the clustering solution is assessed using a combination of internal diagnostics, hierarchical consistency checks, and independent validation tests.

First, internal validation metrics (Figure~\ref{fig:cluster_metrics}) show a coherent behavior as a function of cluster number. The silhouette score peaks at low values ($k \approx 3$), reflecting the dominant global metallicity structure of $\omega$~Cen. However, this coarse partition does not capture the known chemical complexity of the system. For higher cluster numbers, the Davies--Bouldin index remains low and approximately constant over $k \approx 6$--$10$, while the WCSS curve exhibits diminishing returns beyond $k \approx 10$. The adopted solution ($k = 10$) lies within this stable regime, indicating that additional clusters do not significantly improve compactness.

Second, the hierarchical structure of the Ward dendrogram shows that the main population groupings persist across a range of linkage thresholds. Moderate variations in the cut level do not alter the separation into metal-poor, intermediate, and metal-rich components, supporting the robustness of the global structure.

Third, the clustering is performed in a standardised seven-dimensional abundance space, ensuring that no single element dominates the Euclidean distance metric. This confirms that the identified populations arise from multi-dimensional chemical structure rather than a single abundance trend.

Finally, an external consistency check is provided by the color--magnitude diagram (Figure~\ref{fig:color_mag_diagram}), where the chemically defined populations remain coherent in photometric space despite the CMD not being used as an input. This independent agreement supports the physical relevance of the clustering solution.

In addition, the robustness of the clustering solution is explicitly tested against both measurement uncertainties and sampling variance using Monte Carlo perturbation and bootstrap resampling experiments (Appendix~\ref{sec:app_robustness}). These tests show that the identified populations remain stable under perturbations, with the lowest stability ($P \sim 0.42$) consistent with the chemically continuous nature of the metal-poor backbone, while the more discrete enriched populations reach $P \sim 0.95$. This variation reflects the intrinsic chemical structure of $\omega$~Cen, where the most chemically extreme populations exhibit the highest coherence, whereas the lower stability of the metal-poor component naturally arises from its continuous distribution in abundance space. Cross-contamination between distinct groups remains low ($P \leq 0.27$).

Overall, these results demonstrate that the adopted ten-population partition is statistically robust, reproducible, and physically meaningful, capturing the intrinsic chemo-dynamical structure of $\omega$~Cen without over-fragmenting the sample.

\subsection{Validation of the Nested Sampling Fits}
\label{sec:nested_validation}

\begin{figure}
    \centering
    \includegraphics[width=1\linewidth]{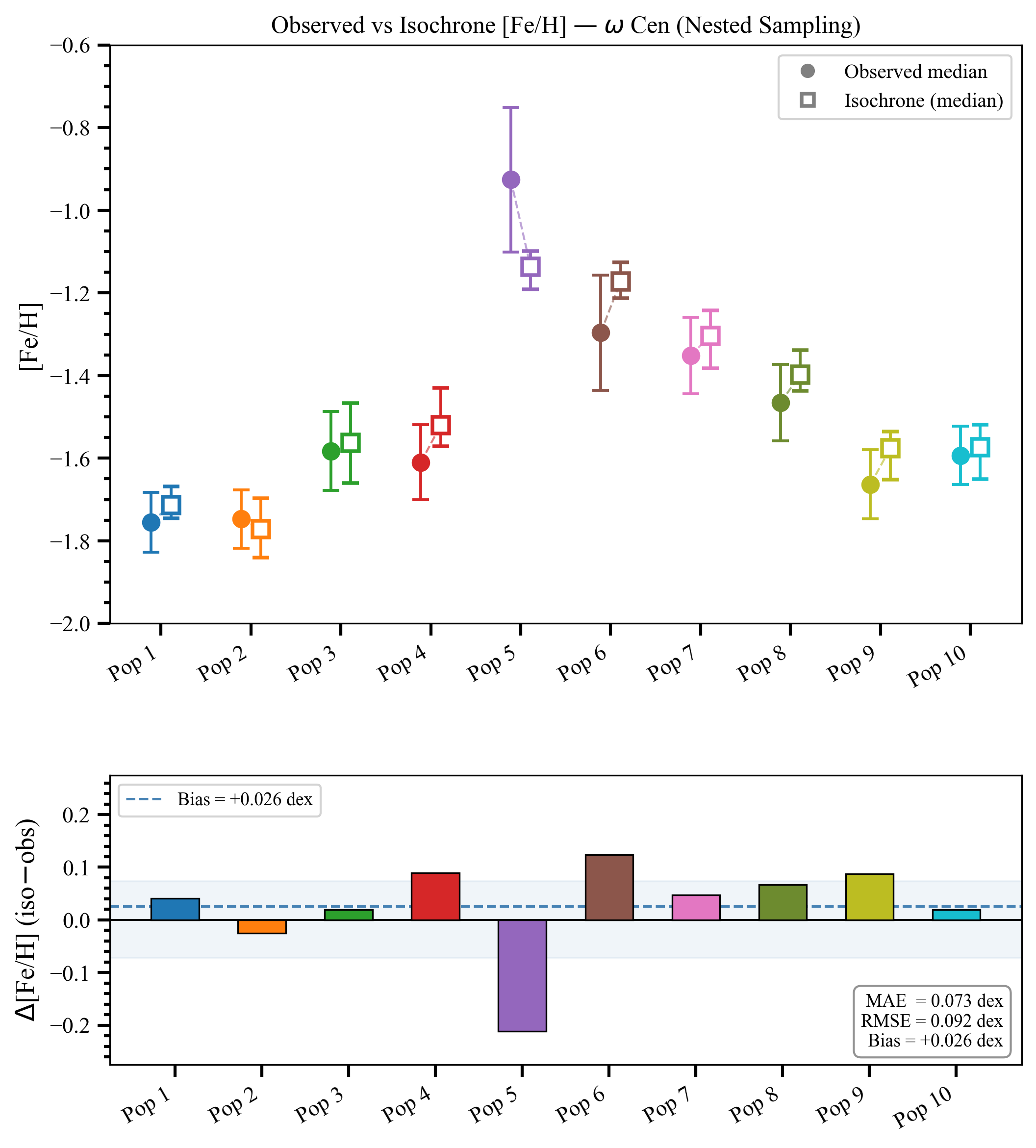}
    \caption{Comparison of spectroscopic metallicities from MWM~DR19 (filled circles, observed median per population) with values from isochrone fitting (open squares) for the ten Ward-defined populations. Error bars represent the 16th--84th percentile range of the posterior. The residual panel (bottom) shows $\Delta[\mathrm{Fe/H}] = [\mathrm{Fe/H}]_\mathrm{iso} - [\mathrm{Fe/H}]_\mathrm{obs}$; the dashed line marks the global bias of $+0.026$~dex. The mean absolute error of $0.073$~dex and root-mean-square error of $0.092$~dex indicate broad agreement, with Population~5 contributing the largest residual.}
    \label{fig:feh_comparison}
\end{figure}

\begin{figure}
    \centering
    \includegraphics[width=\columnwidth]{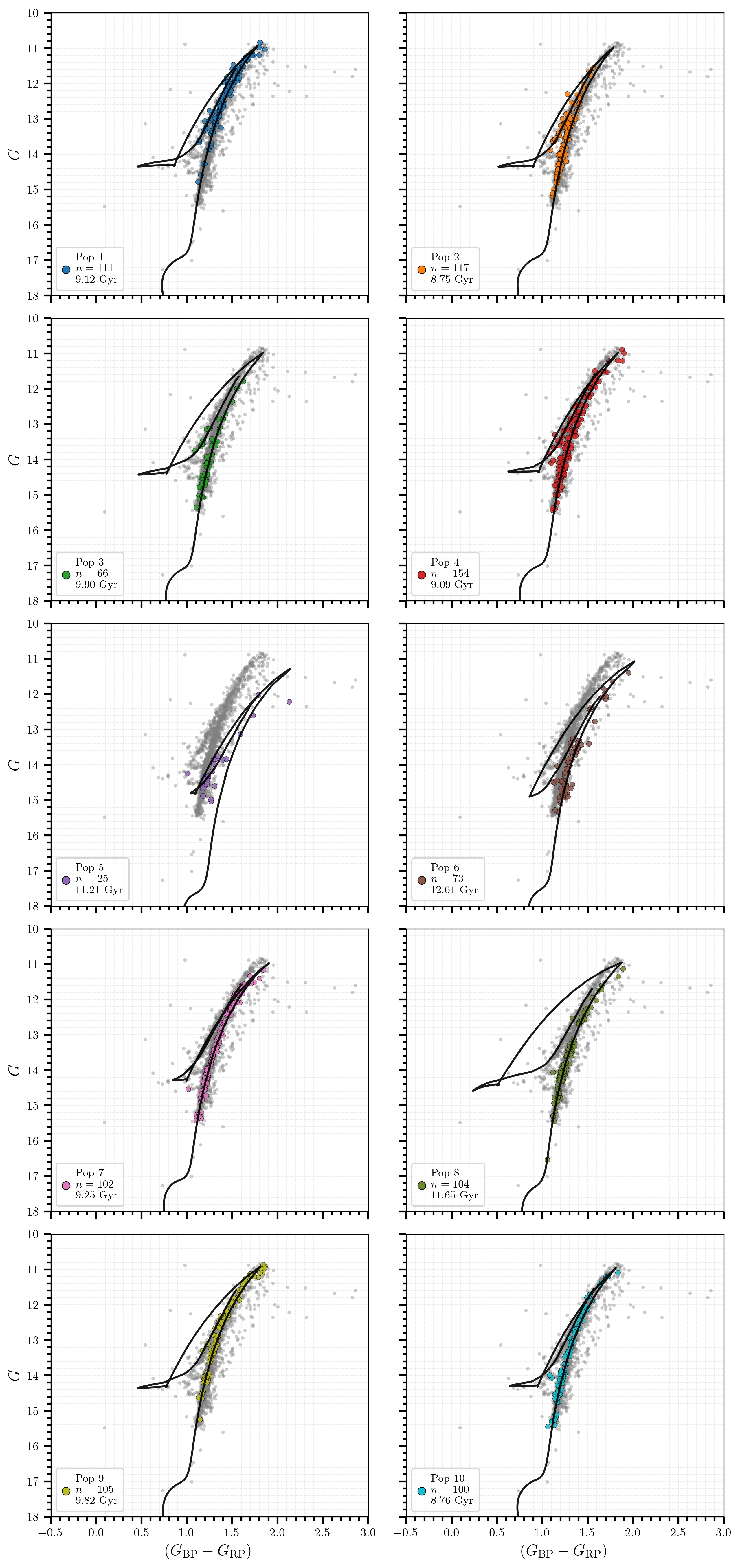}
    \caption{CMD distributions of the ten chemically defined populations, overlaid with the best-fitting isochrones derived from the nested sampling. The fitted models reproduce the observed RGB loci consistently across all populations, demonstrating the internal coherence of the fitting framework.}
    \label{fig:cmd_iso}
\end{figure}

The reliability of our age-determination framework is first assessed by comparing the spectroscopic metallicities from MWM~DR19 with the best-fit values derived from the \textit{\textsl{Gaia}} CMD isochrone fitting. We demonstrate the nested fitting results in Figure~\ref{fig:feh_comparison}, Figure~\ref{fig:cmd_iso}, and summarize them in Table~\ref{tab:nested_results}.

The three remaining parameters are tightly constrained and highly consistent across all populations:

\begin{itemize}

  \item \textbf{Distance.} 
  The fitted distances span $5450$--$5523$~pc across all ten populations (Table~\ref{tab:nested_results}), with a median of $5505$~pc and an inter-population scatter of only $\sim 20$~pc. This is consistent with the Gaia EDR3-based distance from \citet{Baumgardt2021} and \citet{Vasiliev2021} and in particularly close agreement with the kinematic distance of \citet{Haberle2025}, $d = 5494 \pm 61$~pc, derived from $1.4\times10^6$ proper motions and $3\times10^5$ radial velocities in the oMEGACat survey. Our median deviates from the \citeauthor{Haberle2025} value by only $11$~pc ($0.2\sigma$), providing a strong external validation of the nested sampling distance posterior. The sub-percent level of inter-population scatter confirms that the ten Ward populations form a single, spatially compact system and that the nested sampler is not compensating for CMD morphology differences by artificially shifting the distance modulus.

  \item \textbf{Extinction.}
  For nine of the ten populations the posterior mean extinction lies in the range $E(B_{\mathrm{P}}-R_{\mathrm{P}}) = 0.159$--$0.206$~mag, with a median of $0.195$~mag (Table~\ref{tab:nested_results}). This is consistent at the $0.01$~mag level with the canonical catalogue value of $E(B{-}V) = 0.12$ toward $\omega$\,Cen \citep{Harris1996}, which translates to $E(B_{\mathrm{P}}-R_{\mathrm{P}}) = 0.185$~mag via the \citet{Wang2019} Gaia-band coefficients, and with the \citet{Schlegel1998} dust map. Population~5 is a notable exception, with $E(B_{\mathrm{P}}-R_{\mathrm{P}}) = 0.321 \pm 0.021$~mag --- roughly $0.13$~mag above the median of the remaining populations. This elevated value likely arises from the well-known age--metallicity--extinction degeneracy in photometric fitting. Because Population~5 is both extreme in metallicity and sparsely sampled ($N=25$), the nested sampler has limited CMD leverage along the sparsely populated RGB, leading the algorithm to accommodate the morphology via a slight artificial shift in the reddening posterior rather than reflecting true spatial differential extinction.

  \item \textbf{Metallicity.}
  The ten Ward populations span a spectroscopic metallicity range of $[\mathrm{Fe/H}] = -1.76$ to $-0.93$~dex (MWM medians per population), a total spread of $0.83$~dex that reflects the well-known chemical complexity of $\omega$\,Cen. The internal dispersions within each group are small ($\sigma_{[\mathrm{Fe/H}]} \approx 0.07$--$0.14$~dex), confirming that the Ward clustering has produced chemically coherent groups rather than arbitrary mixes of the metallicity distribution. The BaSTI isochrone priors for each population are centred on these spectroscopic medians; the resulting posterior metallicities show moderate agreement with the spectroscopic values, with a mean absolute error (MAE) of $0.073$~dex, a root-mean-square error of $0.092$~dex, and a global bias of $+0.026$~dex. The dominant contribution to this mismatch comes from Population~5 (Figure~\ref{fig:feh_comparison}).

\end{itemize}

\subsection{Chemical Structure of $\omega$ Cen}
\label{sec:res_structure}

The ten chemically defined populations, which are shown to be robust against measurement uncertainties and sampling variations (Appendix~\ref{sec:app_robustness}), reveal a structured and multi-phase enrichment history in $\omega$~Cen, characterised by both discrete components and partially decoupled evolutionary pathways.

The dominant separation occurs between a metal-poor backbone (Populations~1--4; $\sim 46$\%) spanning $-1.75 \leq [\mathrm{Fe/H}] \leq -1.58$, and an enriched intermediate component (Populations~7--10; $\sim 43$\%) extending up to $[\mathrm{Fe/H}] = -1.35$. These two regimes define the primary chemical structure of the system. The remaining $\sim 10$\% of the sample is divided between Population~6 ($N = 73$; $\sim 8$\%) at intermediate metallicity with a primordial chemical composition, and the metal-rich Population~5 ($N = 25$; $\sim 3$\%), which together define the chemically distinct outlier components of the system.

\begin{figure}
    \centering
    \includegraphics[width=1\linewidth]{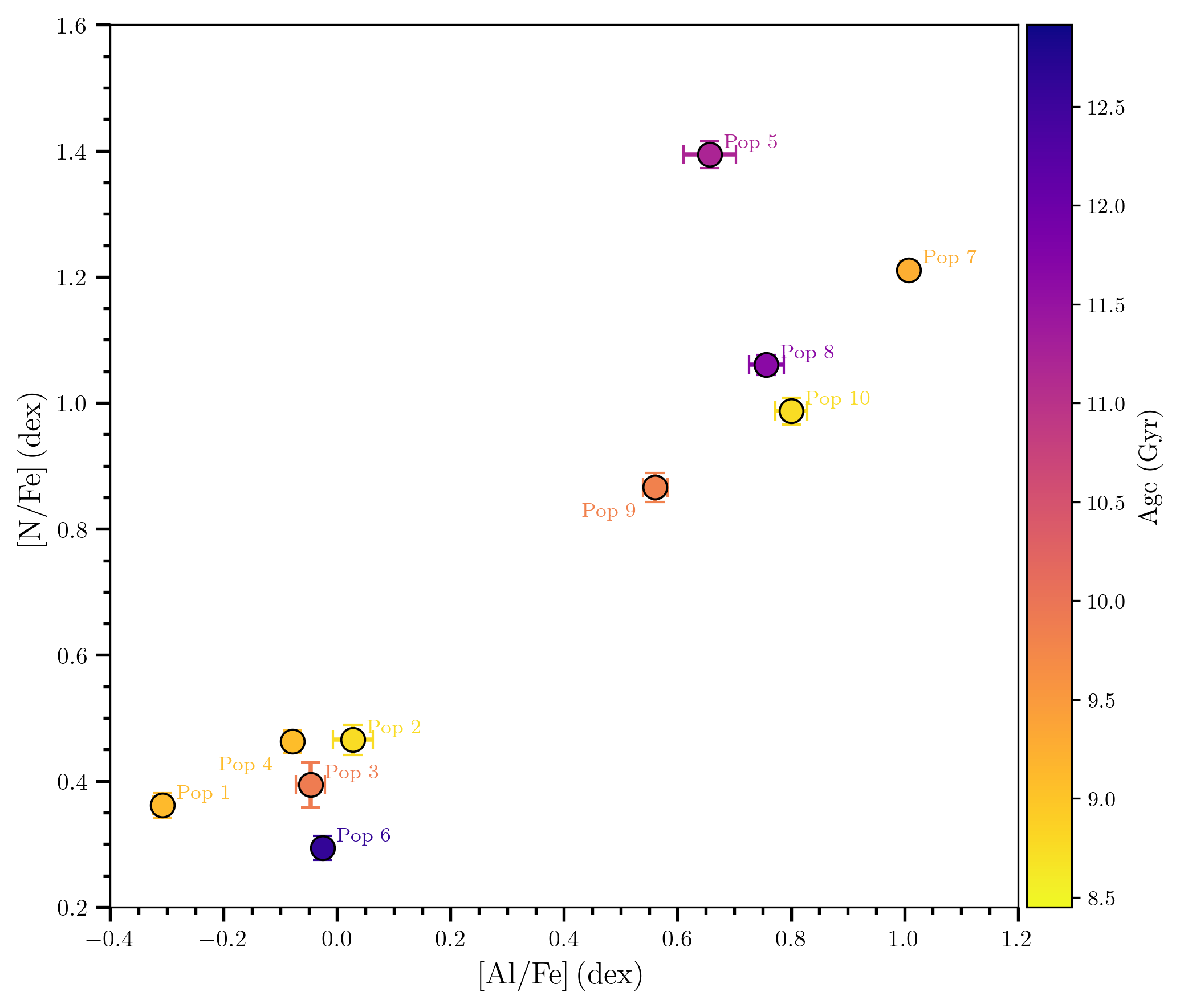}
    \caption{Distribution of the ten chemically defined populations in the $[\mathrm{Al/Fe}]$--$[\mathrm{N/Fe}]$ plane. Each point represents the median abundance of a population, with error bars indicating the standard error in the corresponding abundance ratios. The color coding reflects the median age of each population derived from isochrone fitting.}
    \label{fig:al_n_plane}
\end{figure}

This separation is most clearly illustrated in the $[\mathrm{Al/Fe}]$--$[\mathrm{N/Fe}]$ plane (Figure~\ref{fig:al_n_plane}), where the metal-poor populations cluster at low $[\mathrm{Al/Fe}]$ and $[\mathrm{N/Fe}]$, while the intermediate populations trace a well-defined sequence toward strong proton-capture enrichment. This behavior reflects the progressive activation of high-temperature hydrogen burning in subsequent stellar generations.

Superimposed on this main sequence, two chemically distinct populations emerge. Population~6 occupies the intermediate metallicity regime ($[\mathrm{Fe/H}] = -1.30 \pm 0.14$) while retaining a primordial light-element composition, indicating that not all gas followed the same enrichment pathway.

At the high-metallicity end, Population~5 ($[\mathrm{Fe/H}]_{\rm spec} = -0.93 \pm 0.18$) forms a clear outlier, representing the most chemically evolved component. Its strong proton-capture and $s$-process signatures suggest formation from the most enriched gas reservoir prior to the cessation of star formation.

Taken together, these results support a scenario in which $\omega$~Cen experienced both sequential enrichment and partially decoupled star formation episodes, rather than evolving along a single homogeneous chemical track.

\subsubsection{The Metal-Poor Backbone: Populations 1--4}
\label{sec:res_metal_poor}

Populations~1--4 form the metal-poor backbone of $\omega$~Cen, comprising 448 stars (47\% of the sample; Table~\ref{tab:pop_summary}). These groups occupy a narrow metallicity range of $-1.75 \leq [\mathrm{Fe/H}] \leq -1.58$ and define the baseline chemical composition of the cluster.

These populations exhibit predominantly primordial light-element compositions, with median $[\mathrm{Al/Fe}]$ ranging from $-0.30$ to $-0.05$~dex, indicating minimal proton-capture processing. Their enhanced $\alpha$-element ratios (with $[\mathrm{O/Fe}]$ values of $+0.3$ to $+0.5$~dex) are consistent with enrichment dominated by core-collapse supernovae.

Population~2 represents a transitional case within this group. While sharing a similar metallicity to Population~1, it shows slightly elevated $[\mathrm{Al/Fe}]$ ($\sim +0.03$~dex) and enhanced $[\mathrm{N/Fe}]$, indicating the earliest detectable onset of internal chemical processing.

Overall, Populations~1--4 are consistent with the classical ``main'' population of $\omega$~Cen \citep{NorrisDaCosta1995, Pancino2000, Johnson2010} and define the initial chemical baseline from which subsequent enrichment pathways developed.
Their comparatively lower clustering stability (Appendix~\ref{sec:app_robustness}) further supports the interpretation that this metal-poor component is intrinsically continuous in chemical abundance space, rather than composed of sharply separated sub-populations.

\subsubsection{The Intermediate Enriched Group: Populations 7--10}
\label{sec:res_intermediate}

Populations~7--10 form a chemically enriched intermediate-metallicity component comprising 411 stars (43\% of the sample; Table~\ref{tab:pop_summary}), spanning $-1.66 \leq [\mathrm{Fe/H}] \leq -1.35$.

These populations display strong proton-capture signatures, with extreme enhancements in $[\mathrm{Al/Fe}]$ ($+0.56$ to $+1.01$~dex) and $[\mathrm{N/Fe}]$, accompanied by oxygen depletion. Population~7 represents the most extreme case, reaching $[\mathrm{Al/Fe}] = +1.01$~dex and exhibiting a pronounced O--Al anti-correlation, tracing high-temperature hydrogen burning.

In addition, these populations show significant $s$-process enrichment, consistent with enrichment from intermediate-mass AGB stars. The simultaneous presence of proton-capture and $s$-process signatures is consistent with extended star formation and self-enrichment over timescales of several $10^8$~yr.

The chemical diversity of this group defines a chemically distinct enrichment sequence within the system and is consistent with previous interpretations of $\omega$~Cen as the remnant nucleus of a disrupted dwarf galaxy.

This partial continuity is also reflected in the clustering stability analysis, where moderate cross-assignment between adjacent populations (particularly Pop~7 and Pop~8) indicates partially overlapping abundance regimes, consistent with a continuous chemical enrichment sequence rather than strictly discrete population boundaries.

\subsubsection{The Intermediate-Primordial Population: Population 6}
\label{sec:res_pop6}

Population~6 ($N=73$; Table~\ref{tab:pop_summary}) occupies a distinct position at intermediate metallicity ($[\mathrm{Fe/H}] = -1.30$) but exhibits a predominantly primordial chemical signature.

Unlike the neighbouring enriched populations (Populations~7--10), Population~6 shows low $[\mathrm{Al/Fe}] = -0.05$~dex and suppressed $[\mathrm{N/Fe}]$, indicating minimal proton-capture processing. At the same time, it retains strong $\alpha$-enhancement ($[\mathrm{O/Fe}] = +0.44$~dex), consistent with enrichment dominated by core-collapse supernovae.

This combination is consistent with formation from gas that had reached higher metallicity while remaining largely unaffected by AGB-driven pollution. Population~6 therefore highlights the presence of partially decoupled chemical enrichment pathways within the progenitor system.

\subsubsection{The Metal-Rich Extreme: Population~5}
\label{sec:res_pop5}

Population~5 ($N = 25$; Table~\ref{tab:pop_summary}) represents the most metal-rich and chemically evolved component of \ocen, with $[\mathrm{Fe/H}]_{\rm spec} \approx -0.93$, forming the high-metallicity tail of the distribution. Its iron abundance represents a $\sim 0.4$--$0.8$~dex jump relative to all other populations. In combination with the persistently low $[\mathrm{Mn/Fe}]$ values discussed in Section~\ref{sec:discussion_ncap}, this offset is more consistent with rapid, high-efficiency enrichment dominated by core-collapse supernovae, while any Type~Ia contribution remained sub-dominant.

This population exhibits significant proton-capture processing, with enhanced $[\mathrm{Al/Fe}] \sim +0.6$~dex and extremely high $[\mathrm{N/Fe}] \approx +1.39$~dex — the highest in the sample. It is further characterised by strong CNO-cycle signatures, reflected in the lowest $[\mathrm{C/N}]$ ratio among all populations ($-1.38$~dex; Figure~\ref{fig:cn_violin}), consistent with near-complete conversion of carbon to nitrogen. Substantial $s$-process enrichment is also observed, with $[\mathrm{Ce/Mg}] \approx +0.45$~dex, consistent with late-stage AGB contributions \citep{Ventura2013, Cristallo2011}. Notably, Pop~5 lies outside the main Al--N sequence defined by 
Populations~7--10 in Figure~\ref{fig:al_n_plane}, occupying a position of extreme $[\mathrm{N/Fe}]$ at comparatively moderate $[\mathrm{Al/Fe}]$, which suggests a different nucleosynthetic balance rather than a simple continuation of the same enrichment pathway.

The chemical properties of Population~5 are consistent with its identification as the ``RGB-a/SGB-a'' component of \ocen\ \citep{Pancino2000, Villanova2014}, placing it among the most chemically evolved populations in the system and likely representing the endpoint of its self-regulated star-formation history in a regime where AGB enrichment became prominent while Type~Ia supernovae remained sub-dominant.

\section{Discussion}
\label{sec:discussion}

\subsection{Comparison with Previous Population Classifications}
\label{sec:discussion_matching}

The ten chemically defined populations identified in this work (summarised in Tables~\ref{tab:pop_summary} and \ref{tab:xfe_populations}) can be interpreted within the broader context of existing sub-population classifications from the literature. Different studies employ distinct observational tracers, spectral resolutions, and analysis techniques. 
As a result, the correspondence is not always one-to-one and is primarily guided by metallicity ranges and characteristic abundance patterns.

At the global level, our clustering solution recovers the classical three-component structure of $\omega$~Cen --- metal-poor, intermediate, and metal-rich --- consistent with earlier studies. In particular, low-dimensional analyses such as \citet{Mason2026} capture this broad structure, while our results demonstrate that each of these components is itself composed of multiple chemically distinct sub-populations. In this sense, our results provide a direct bridge between traditional one-dimensional metallicity classifications and fully data-driven multi-dimensional population decompositions.

The metal-poor backbone (Pop~1--4) corresponds most directly to the dominant metal-poor peak identified in virtually all previous surveys \citep{NorrisDaCosta1995, Pancino2000, Johnson2010}. These populations are characterised by primordial light-element abundances and enhanced $\alpha$-elements, consistent with enrichment dominated by core-collapse 
supernovae. Within this backbone, Pop~2 shows enhanced $[\mathrm{Na/Fe}]$ ($+0.83$~dex; Table~\ref{tab:xfe_populations}) and mildly elevated $[\mathrm{Al/Fe}]$. This combination closely matches the ``intermediate'' population identified by \citet{Carretta2009a} in the Na--O plane and likely traces the earliest onset of proton-capture processing.

A direct comparison with the fourteen chemically defined populations of \citet{Wang2026} reveals strong correspondence across the Al--N--O abundance space, despite the different number of recovered components. Their most metal-poor, chemically primordial component (Pop~4; $[\mathrm{Al/Fe}] = -0.23$, $[\mathrm{O/Fe}] = +0.36$, $[\mathrm{Fe/H}] = -1.78$) is in good agreement with our Pop~1 ($[\mathrm{Al/Fe}] = -0.31$, $[\mathrm{O/Fe}] = +0.32$), both representing a first-generation population dominated by core-collapse supernova ejecta. At the other extreme, their most strongly proton-capture-enriched component (Pop~6; $[\mathrm{Al/Fe}] = +1.08$, $[\mathrm{O/Fe}] = -0.11$) closely matches our Pop~7 ($[\mathrm{Al/Fe}] = +1.01$, $[\mathrm{O/Fe}] = 0.00$), both representing the most extreme processed material at $[\mathrm{Fe/H}] \approx -1.7$. The difference in the total number of recovered components --- fourteen versus ten --- is most naturally attributed to the finer resolution achieved by \citet{Wang2026} in the most metal-poor regime ($[\mathrm{Fe/H}] < -1.87$), where their sample resolves five distinct groups (Pop~0--4) that our near-infrared MWM~DR19 dataset treats as a single broad metal-poor backbone. Conversely, our multi-dimensional approach more cleanly isolates the chemically primordial intermediate component (Pop~6; discussed further below) that \citet{Wang2026} subsume into their Pop~8.

The chromosome map classifications assigned by \citet{Wang2026} (P1, P2, Im) provide an independent validation of our chemical decomposition. Their P2-classified populations (Pop~2, 6, 9, 10, 12, 13) uniformly exhibit $[\mathrm{Al/Fe}] > +0.75$, consistent with our proton-capture-enriched groups (Pop~7--10). Their intermediate Im class (Pop~1, 5, 7; $[\mathrm{Al/Fe}] = +0.42$--$+0.64$) corresponds to our transitional populations (Pop~2, 3). This chromosome-space anchoring confirms that our purely spectroscopic clustering recovers physically meaningful chemical sequences that are also photometrically distinguishable.

Population~6 represents the most challenging component to cross-identify, precisely because its chemical properties --- intermediate metallicity combined with predominantly primordial light-element abundances --- do not fit naturally into any single previous classification. Its high $[\mathrm{O/Fe}]$ ($+0.53$~dex; Table~\ref{tab:xfe_populations}) suggests 
an association with the metal-intermediate, oxygen-rich component reported by \citet{Marino2012}, while its metallicity places it within the RGB-MInt2 regime of \citet{Pancino2000}. However, its lack of proton-capture enrichment at this metallicity sets it apart from all other intermediate-metallicity populations and suggests that it corresponds to a previously identified, but rarely isolated, fraction of chemically primordial stars \citep{Johnson2010, Marino2011}. In the 
classification of \citet{Wang2026}, this component finds its closest counterpart in their Pop~8 ($[\mathrm{Al/Fe}] = 0.00$, 
$[\mathrm{O/Fe}] = +0.32$, $[\mathrm{Fe/H}] = -1.63$), which similarly stands out as a chemically unenriched island within the intermediate-metallicity regime.

Importantly, our clustering separates into two chemically distinct groups what previous metallicity-based classifications (e.g.\ \citealt{Johnson2010}) treated as a single population: Pop~6 and Pop~7. This distinction is not visible in one-dimensional metallicity space and demonstrates the key advantage of a multi-dimensional abundance-space approach over traditional metallicity-based classifications.

The proton-capture-enriched intermediate populations (Pop~7--10) span the metallicity range $-1.65 \leq [\mathrm{Fe/H}] \leq -1.35$ and exhibit strong enhancements in $[\mathrm{Al/Fe}]$ and $[\mathrm{N/Fe}]$, together with oxygen depletion (Table~\ref{tab:xfe_populations}). These trends are fully consistent with canonical Mg--Al and Na--O patterns reported in 
previous studies \citep{Johnson2010, Nitschai2024}. Moreover, they are in good agreement with the large-scale spectroscopic compilation of \citet{Wang2026}, whose populations at comparable metallicities (Pop~7, 9, 10) show $[\mathrm{Al/Fe}]$ values of $+0.42$, $+1.10$, and $+1.10$~dex, respectively, bracketing the range we recover in Pop~7--10 ($[\mathrm{Al/Fe}] = +0.56$--$+1.01$~dex). This consistency across independent datasets further supports the role of intermediate-mass 
AGB stars in driving the chemical evolution of the cluster at these metallicities.

\subsection{A Multi-Phase Chemical Evolution Scenario}
\label{sec:discussion_multiphase}

\begin{table*}
    \centering
    \caption{Median $[\mathrm{X/Fe}]$ and $[\mathrm{Fe/H}]$ abundances with standard
    deviations for each identified population. Values are given as median $\pm \sigma$.}
    \label{tab:xfe_populations}
    \resizebox{\textwidth}{!}{%
    \begin{tabular}{lcccccccccc}
    \hline\hline
    Element & Pop~1 & Pop~2 & Pop~3 & Pop~4 & Pop~5 & Pop~6 & Pop~7 & Pop~8 & Pop~9 & Pop~10 \\
    \hline
    $[\mathrm{Fe/H}]$ & $-1.75 \pm 0.07$ & $-1.75 \pm 0.07$ & $-1.58 \pm 0.10$ & $-1.61 \pm 0.09$ &
    $-0.93 \pm 0.18$ & $-1.30 \pm 0.14$ & $-1.35 \pm 0.09$ & $-1.47 \pm 0.09$ & $-1.66 \pm 0.08$ &
    $-1.59 \pm 0.07$ \\
    \hline
    $[\mathrm{C/Fe}]$  & $-0.59 \pm 0.25$ & $-0.16 \pm 0.28$ & $+0.16 \pm 0.21$ & $+0.14 \pm 0.24$ &
    $+0.04 \pm 0.14$ & $+0.36 \pm 0.19$ & $-0.00 \pm 0.22$ & $+0.05 \pm 0.19$ & $-0.31 \pm 0.28$ &
    $-0.07 \pm 0.21$ \\
    $[\mathrm{N/Fe}]$  & $+0.36 \pm 0.21$ & $+0.47 \pm 0.26$ & $+0.39 \pm 0.29$ & $+0.46 \pm 0.22$ &
    $+1.39 \pm 0.11$ & $+0.29 \pm 0.16$ & $+1.21 \pm 0.14$ & $+1.06 \pm 0.16$ & $+0.87 \pm 0.23$ &
    $+0.99 \pm 0.21$ \\
    $[\mathrm{O/Fe}]$  & $+0.32 \pm 0.08$ & $+0.33 \pm 0.17$ & $+0.44 \pm 0.10$ & $+0.39 \pm 0.12$ &
    $+0.29 \pm 0.06$ & $+0.53 \pm 0.11$ & $+0.00 \pm 0.10$ & $+0.13 \pm 0.17$ & $+0.13 \pm 0.18$ &
    $-0.08 \pm 0.22$ \\
    $[\mathrm{Mg/Fe}]$ & $+0.20 \pm 0.07$ & $+0.20 \pm 0.13$ & $+0.29 \pm 0.07$ & $+0.28 \pm 0.09$ &
    $+0.31 \pm 0.08$ & $+0.41 \pm 0.08$ & $-0.04 \pm 0.12$ & $+0.14 \pm 0.17$ & $+0.08 \pm 0.20$ &
    $-0.07 \pm 0.23$ \\
    $[\mathrm{Al/Fe}]$ & $-0.31 \pm 0.16$ & $+0.03 \pm 0.38$ & $-0.05 \pm 0.21$ & $-0.08 \pm 0.16$ &
    $+0.66 \pm 0.23$ & $-0.03 \pm 0.08$ & $+1.01 \pm 0.12$ & $+0.76 \pm 0.31$ & $+0.56 \pm 0.22$ &
    $+0.80 \pm 0.28$ \\
    $[\mathrm{Ca/Fe}]$ & $+0.10 \pm 0.19$ & $+0.02 \pm 0.25$ & $+0.28 \pm 0.14$ & $+0.18 \pm 0.19$ &
    $+0.35 \pm 0.05$ & $+0.33 \pm 0.08$ & $+0.28 \pm 0.07$ & $+0.28 \pm 0.10$ & $+0.17 \pm 0.15$ &
    $+0.19 \pm 0.15$ \\
    $[\mathrm{Na/Fe}]$ & $+0.26 \pm 0.33$ & $+0.83 \pm 0.25$ & $+0.67 \pm 0.29$ & $-0.14 \pm 0.27$ &
    $+0.48 \pm 0.20$ & $-0.17 \pm 0.33$ & $+0.36 \pm 0.30$ & $-0.21 \pm 0.43$ & $-0.06 \pm 0.34$ &
    $+0.71 \pm 0.29$ \\
    $[\mathrm{Ni/Fe}]$ & $+0.00 \pm 0.08$ & $+0.02 \pm 0.14$ & $-0.03 \pm 0.16$ & $-0.01 \pm 0.10$ &
    $-0.07 \pm 0.08$ & $-0.05 \pm 0.08$ & $+0.01 \pm 0.07$ & $-0.01 \pm 0.11$ & $+0.01 \pm 0.12$ &
    $+0.02 \pm 0.11$ \\
    $[\mathrm{Mn/Fe}]$ & $-0.46 \pm 0.21$ & $-0.51 \pm 0.29$ & $-0.42 \pm 0.29$ & $-0.50 \pm 0.26$ &
    $-0.36 \pm 0.15$ & $-0.43 \pm 0.26$ & $-0.52 \pm 0.22$ & $-0.47 \pm 0.24$ & $-0.42 \pm 0.25$ &
    $-0.54 \pm 0.26$ \\
    $[\mathrm{Ce/Fe}]$ & $-0.41 \pm 0.24$ & $-0.37 \pm 0.35$ & $+0.20 \pm 0.37$ & $+0.17 \pm 0.37$ &
    $+0.79 \pm 0.29$ & $+0.42 \pm 0.23$ & $+0.73 \pm 0.31$ & $+0.42 \pm 0.42$ & $+0.03 \pm 0.39$ &
    $+0.19 \pm 0.48$ \\
    \hline
    $N$ & 111 & 117 & 66 & 154 & 25 & 73 & 102 & 104 & 105 & 100 \\
    \hline
    \end{tabular}}
\end{table*}

The chemically defined populations identified in this work reveal a complex and multi-phase enrichment history in \ocen. The separation between the metal-poor backbone (Populations~1--4) and the enriched intermediate populations (Populations~7--10) is most clearly visible in the $[\mathrm{Al/Fe}]$--$[\mathrm{N/Fe}]$ plane (Figure~\ref{fig:al_n_plane}), where the two groups occupy distinct regions of chemical space.

The magnitude of this separation is substantial. From Population~1 to Population~7, $[\mathrm{Al/H}]$ increases by $\approx +1.76$~dex (from $-2.06$ to $-0.30$) and $[\mathrm{N/H}]$ by $\approx +1.22$~dex (from $-1.37$ to $-0.15$), while $[\mathrm{Fe/H}]$ rises by only $+0.40$~dex. This asymmetry — a large light-element spread over a comparatively modest iron baseline — is the defining hallmark of self-enrichment driven by proton-capture nucleosynthesis rather than by supernovae alone, and is quantitatively consistent with the $[\mathrm{Al/Fe}]$ excesses reported for the most enriched giants 
in high-resolution optical surveys of \ocen\ \citep{Johnson2010, Marino2011, AlvarezGaray2024}.

This separation indicates that the system did not evolve along a single, continuous chemical track, but rather experienced at least two distinct enrichment phases. In the first phase, the metal-poor backbone (Populations~1--4) formed from gas whose composition reflects enrichment dominated by core-collapse supernovae, with $[\mathrm{O/Fe}] \sim +0.3$--$+0.4$~dex and minimal proton-capture processing ($[\mathrm{Al/Fe}] \lesssim -0.05$~dex; Table~\ref{tab:pop_summary}). In the second phase, gas pre-processed by an earlier stellar generation was retained within the potential well of the progenitor system and incorporated into the intermediate populations (Populations~7--10), which show extreme enhancements in $[\mathrm{Al/Fe}]$ (up to $+1.0$~dex) and strong oxygen depletion \citep[cf.][]{Prantzos2007, Ventura2011, Bastian2018}.

\begin{figure}
    \centering
    \includegraphics[width=1\linewidth]{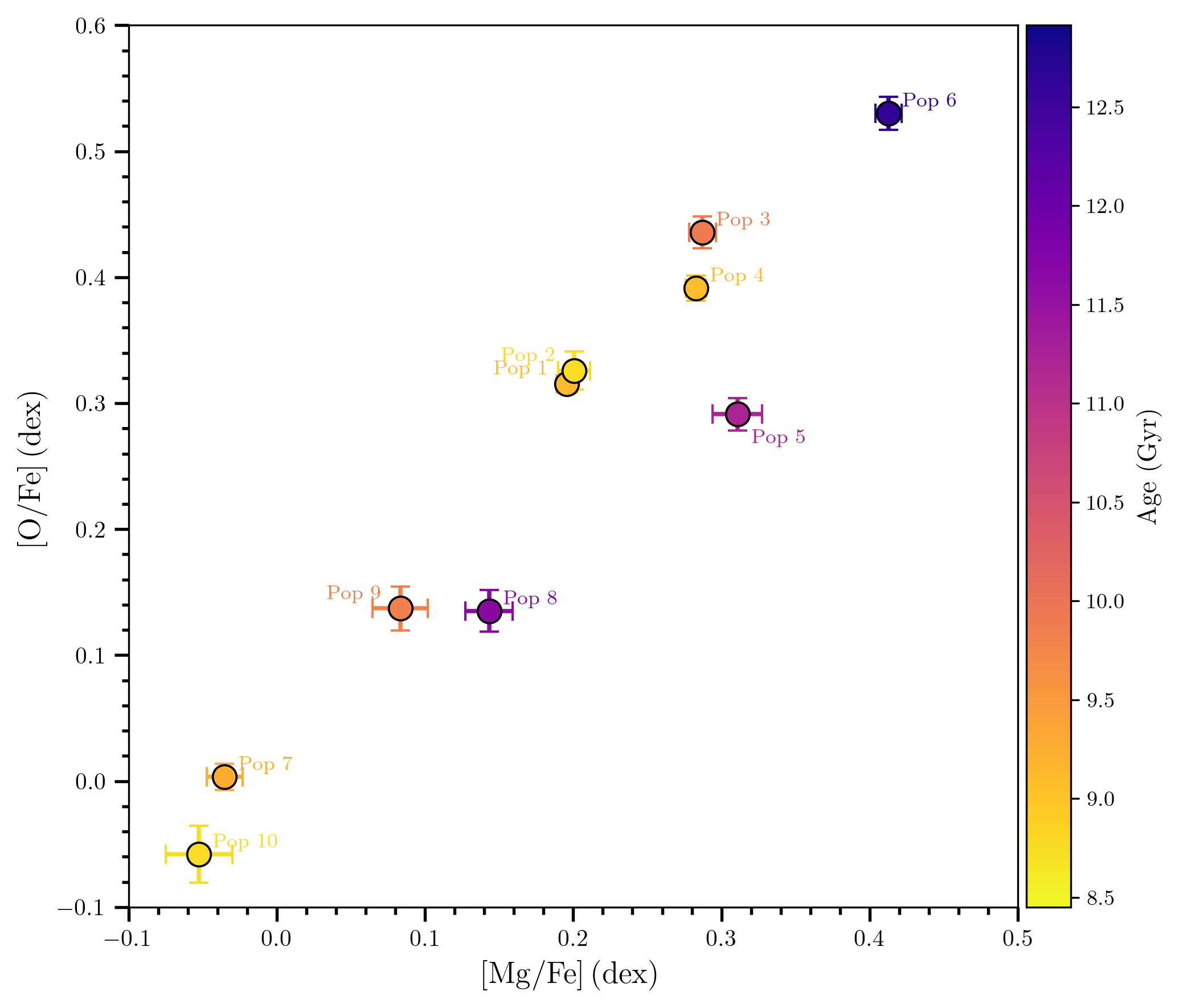}
    \caption{Distribution of the ten chemically defined populations in the $[\mathrm{Mg/Fe}]$--$[\mathrm{O/Fe}]$ plane. Each point represents the median abundance of a population, with error bars indicating the standard error in the corresponding abundance ratios. The color coding reflects the median age of each population derived from isochrone fitting.}
    \label{fig:ofe_mgfe}
\end{figure}

Several alternative scenarios for the origin of this bimodal enrichment pattern can be considered. One possibility is that 
Populations~1--4 and Populations~7--10 formed in spatially segregated regions of the progenitor dwarf galaxy, evolving largely independently before being assembled into a single system \citep{BekkiFreeman2003}. Under this picture, the chemical separation would reflect different local star-formation efficiencies rather than a strict temporal sequence. A second scenario invokes a genuine first-generation/second-generation (FG/SG) dichotomy in the classical globular cluster sense \citep{Gratton2004, Bastian2018}, where Populations~1--4 represent FG stars and Populations~7--10 represent SG stars formed from AGB ejecta diluted with pristine gas. A third possibility is that the two groups represent a continuous 
sequence in proton-capture enrichment rather than discrete episodes, with the apparent bimodality arising from the finite resolution of the Ward clustering applied to a seven-dimensional space.

The data presented here do not allow us to conclusively distinguish among these scenarios, but several lines of evidence favor the sequential self-enrichment interpretation. First, the near-constant $[\mathrm{Fe/H}]$ separation between the two groups ($\Delta [\mathrm{Fe/H}] \approx 0.3$--$0.4$~dex; Table~\ref{tab:pop_summary}) is consistent with a single additional episode of supernova enrichment between the formation of the two components, as expected in a chemically evolving system \citep{Romano2010}. Second, the $[\mathrm{O/Fe}]$--$[\mathrm{Mg/Fe}]$ diagram (Figure~\ref{fig:ofe_mgfe}) shows that the enriched populations are systematically displaced below the $[\mathrm{O/Fe}] = [\mathrm{Mg/Fe}]$ locus, indicating selective oxygen destruction via the CNO cycle without equivalent magnesium depletion — a signature expected from AGB-driven pollution rather than from spatially independent star formation \citep{Cristallo2011, Ventura2013}. Third, a phylogenetic analysis of the \ocen\ chemical tree by \citet{Jofre2025} independently identifies three primary branches in the abundance space, broadly consistent with the multi-phase structure identified here.


\subsection{Proton-Capture and CNO Processing}
\label{sec:discussion_cno}

The intermediate populations (Pop~7--10) exhibit extreme enhancements in $[\mathrm{Al/Fe}]$ and $[\mathrm{N/Fe}]$, accompanied by oxygen depletion (Figure~\ref{fig:ofe_mgfe}), indicating high-temperature hydrogen burning at $T \geq 7 \times 10^7$~K in the NeNa and Mg--Al cycles \citep{Denisenkov1990, Prantzos2007}. Population~7 represents the most extreme case, reaching $[\mathrm{Al/Fe}] = +1.0$~dex and $[\mathrm{N/Fe}] = +1.2$~dex, while the metal-poor backbone
(Pop~1--4) remains at $[\mathrm{Al/Fe}] \leq -0.05$~dex and $[\mathrm{N/Fe}] \approx +0.4$~dex (Figure~\ref{fig:al_n_plane}). The total spread of $\Delta[\mathrm{Al/Fe}] \approx 1.6$~dex across the full sample is consistent with the extreme Mg--Al anticorrelation previously identified in \textsl{APOGEE} observations of \ocen\
\citep{AlvarezGaray2024} and with the high-resolution optical measurements of \citet{Johnson2010}, who reported $[\mathrm{Al/Fe}]$ values reaching $+1.3$~dex in the most enriched giants.

\begin{figure*}
    \centering
    \includegraphics[width=\linewidth]{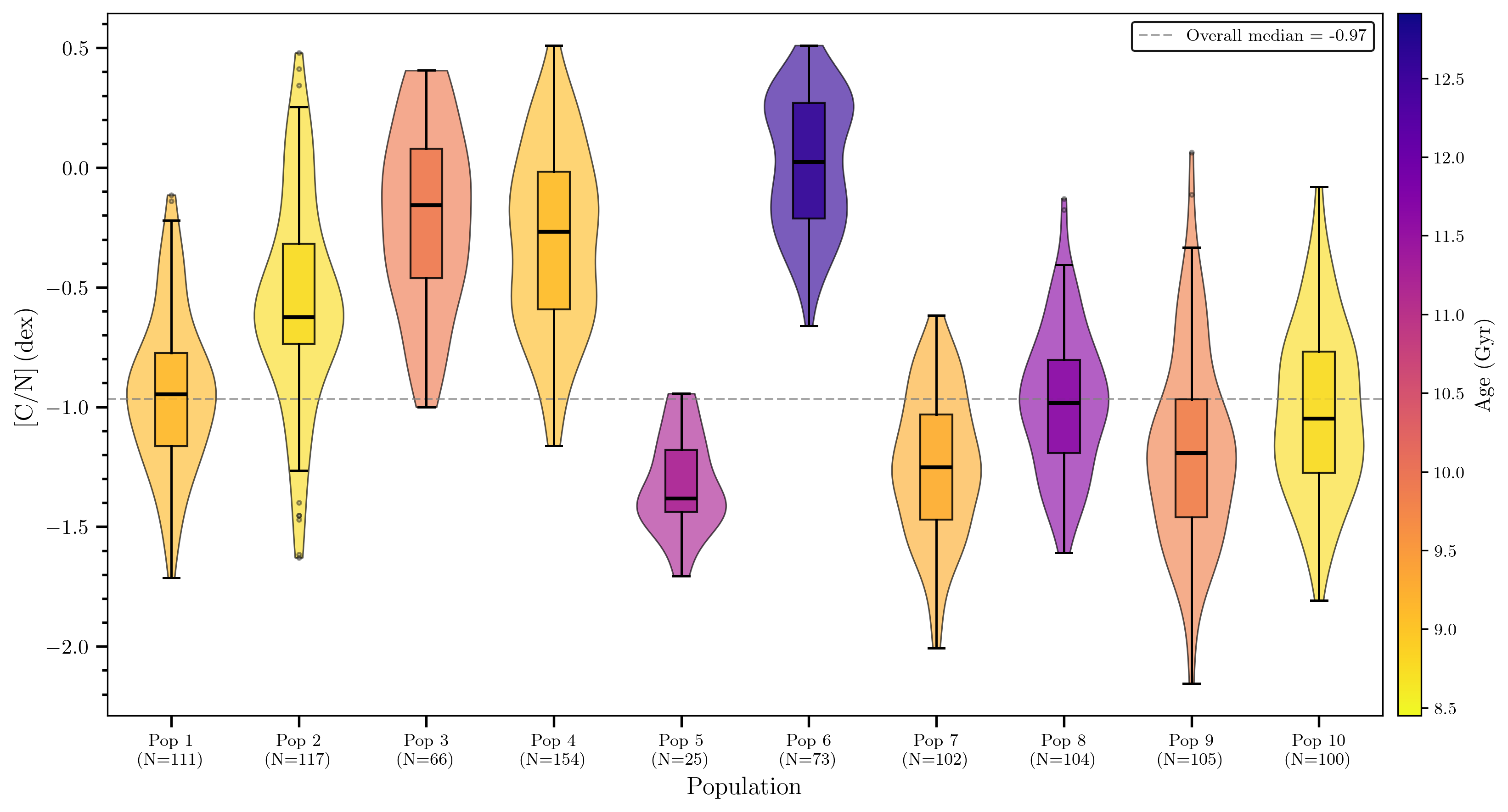}
    \caption{Violin plots of the $[\mathrm{C/N}]$ ratio distributions for the ten Ward-defined populations, color-coded by age. Boxes mark the interquartile range; horizontal lines inside boxes indicate the median value. The dashed grey line shows the overall sample median of $[\mathrm{C/N}] = -0.97$. Population~5 reaches the lowest median ($[\mathrm{C/N}] = -1.38$), followed by the intermediate proton-capture-enriched groups (Pop~7--10; medians $-1.05$ to $-1.25$), consistent with extensive CNO-cycle processing. Population~6 is a clear outlier within the intermediate-metallicity regime, with a near-solar median of $[\mathrm{C/N}] = +0.02$, confirming its chemically primordial character. Among the metal-poor populations, Populations~2 and 3 show notably elevated $[\mathrm{C/N}]$ values ($-0.62$ and $-0.16$, respectively) relative to the overall median, indicating limited internal mixing or reduced evolutionary dredge-up in these groups.}
    \label{fig:cn_violin}
\end{figure*}

The $[\mathrm{C/N}]$ distributions (Figure~\ref{fig:cn_violin}) provide complementary and quantitatively consistent evidence for this picture. The metal-poor, proton-capture-poor populations retain relatively elevated $[\mathrm{C/N}]$ ratios: Pop~3 and Pop~6 reach medians of $-0.16$ and $+0.02$~dex respectively, while Pop~1 sits near the overall sample median of $-0.97$~dex. In contrast, the intermediate enriched populations show systematically depressed values, with
Pop~7 at $[\mathrm{C/N}] = -1.25$~dex, Pop~9 at $-1.19$~dex, and Pop~10 at $-1.05$~dex. This depression reflects the conversion of carbon into nitrogen via the CN branch of the CNO cycle operating in the gas reservoir from which these populations formed, and is distinct from the internal mixing-driven $[\mathrm{C/N}]$ reduction that affects individual RGB stars at all metallicities \citep{Martell2008}.

Population~5 warrants particular attention in this context. With $[\mathrm{C/N}] = -1.38$~dex — the lowest median in the entire sample and $0.4$~dex below the overall median — and an extremely narrow distribution (Figure~\ref{fig:cn_violin}), it occupies a position clearly separated from all other populations. Its $[\mathrm{N/Fe}] = +1.39$~dex is the highest recorded, yet its $[\mathrm{Al/Fe}]= +0.66$~dex is more moderate than Pop~7. This combination places Pop~5 outside the main Al--N sequence defined by Pop~7--10 (Figure~\ref{fig:al_n_plane}), suggesting that its extreme nitrogen enhancement reflects a different balance of nucleosynthetic processes — likely a higher contribution from intermediate-mass AGB stars operating at high envelope temperatures \citep{Ventura2011, Ventura2013} — rather than simply a more advanced stage of the same enrichment sequence.

Two physical channels can in principle produce the observed proton-capture signatures: (i) early-stage pollution from the winds of fast-rotating massive stars or super-AGB stars, operating on rapid timescales (($\leq 10^8$~yr; \citealt{Decressin2007,DercoleVesperini2008}), and (ii) intermediate-mass AGB stars ($3$--$8\,M_\odot$) enriching the gas on timescales of $\sim 0.1$--$1$~Gyr \citep{Cottrell1981, Ventura2001}. Both channels predict correlated Al and N enhancements, but differ in their predicted Mg depletion: the super-AGB/massive-star scenario requires more efficient Mg--Al cycling and predicts more pronounced $[\mathrm{Mg/Fe}]$ depletion at high $[\mathrm{Al/Fe}]$ \citep{Decressin2007}. In Figure~\ref{fig:ofe_mgfe}, Pop~7 reaches $[\mathrm{Mg/Fe}] = -0.04$~dex alongside $[\mathrm{Al/Fe}] = +1.0$~dex, a level of Mg depletion consistent with contributions from both channels. The present dataset does not allow a definitive discrimination between these scenarios, but the simultaneous presence of strong $s$-process enrichment in the same populations (Section~\ref{sec:discussion_ncap}) favors a dominant AGB contribution, as massive stars are not significant $s$-process producers at the relevant metallicities \citep{Kobayashi2020}.

\subsection{Neutron-Capture and Iron-Peak Evolution}
\label{sec:discussion_ncap}

\begin{figure}
    \centering
    \includegraphics[width=1\linewidth]{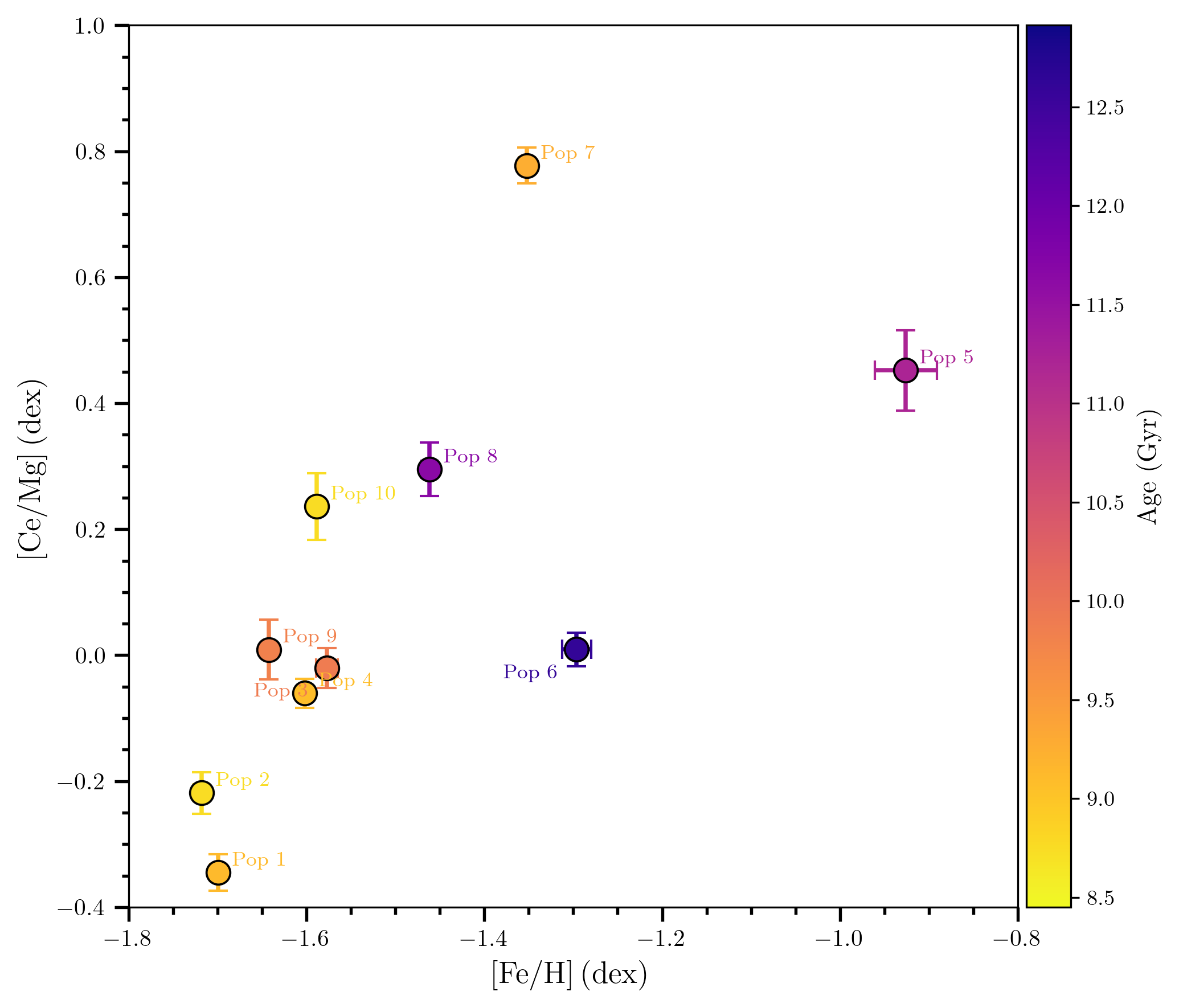}
    \caption{Distribution of the ten chemically defined populations in the $[\mathrm{Fe/H}]$--$[\mathrm{Ce/Mg}]$ plane. Each point represents the median abundance of a population, with error bars indicating the error deviation in the corresponding abundance ratios. The color coding reflects the median age of each population derived from isochrone fitting.}
    \label{fig:feh_cemg}
\end{figure}

The $[\mathrm{Ce/Mg}]$ ratio provides a sensitive diagnostic of the relative timing between $\alpha$-element production by core-collapse supernovae and $s$-process enrichment by AGB stars, since Mg is released promptly on timescales of $\lesssim 10^7$~yr while Ce accumulates on the longer AGB delay timescale of $\sim 0.3$--$1$~Gyr \citep{Busso1999, Cristallo2011}.
The systematic rise in $[\mathrm{Ce/Mg}]$ with increasing chemical enrichment level (Figure~\ref{fig:feh_cemg}) therefore traces the progressive build-up of $s$-process material in the gas reservoir of the progenitor system over time.

The metal-poor backbone populations (Pop~1--4) occupy the lowest $[\mathrm{Ce/Mg}]$ values in the sample, with Pop~1 reaching $[\mathrm{Ce/Mg}] \approx -0.35$~dex, consistent with a chemical composition dominated by core-collapse supernova ejecta and a negligible AGB contribution at this early stage. Moving to higher enrichment levels, Pop~9 and Pop~10 show intermediate values near $[\mathrm{Ce/Mg}] \approx 0.00$--$+0.23$~dex, while Pop~7 reaches the highest value among the intermediate populations at $[\mathrm{Ce/Mg}] \approx +0.78$~dex despite a modest $[\mathrm{Fe/H}] = -1.35$. This large $[\mathrm{Ce/Mg}]$ enhancement at intermediate metallicity is a clear signature of significant AGB-driven s-process enrichment decoupled from the iron-peak evolution. Although \citet{Marino2011} did not measure Ce, our results are qualitatively and quantitatively consistent with the pronounced enrichment they reported for other second-peak s-process elements, such as $[\mathrm{Ba/Fe}]$ and $[\mathrm{La/Fe}]$, in the most chemically advanced \ocen\ giants \citep[see also][]{Meszaros2020}.

A striking feature of Figure~\ref{fig:feh_cemg} is the position of Population~6. Despite sharing a similar metallicity with Pop~7 ($[\mathrm{Fe/H}] = -1.35$), its $[\mathrm{Ce/Mg}]$ ratio remains near zero, $\approx 0.78$~dex below Pop~7. This offset mirrors its primordial proton-capture composition discussed in Section~\ref{sec:discussion_cno} and reinforces the interpretation that Pop~6 formed from a gas reservoir that had not been exposed to significant AGB ejecta, despite having reached a higher iron abundance. The simultaneous absence of both proton-capture and $s$-process enrichment in Pop~6 places strong constraints on its formation environment: it must have assembled from a chemically isolated gas pocket where the AGB delay time had not yet elapsed, or where the local star-formation efficiency was too low to retain AGB ejecta \citep{DercoleVesperini2008, Conroy2012}.

Population~5 occupies a moderately elevated position at $[\mathrm{Ce/Mg}] \approx +0.45$~dex. This value is notably lower
than that of Pop~7 despite Pop~5's much higher iron abundance, suggesting that the $s$-process enrichment in the metal-rich
tail was not simply a continuation of the same AGB-driven sequence. One possibility is that Pop~5 formed on a sufficiently short timescale that only the most massive AGB stars ($\geq 4\,M_\odot$, delay time $\leq 0.3$~Gyr) had contributed to its gas reservoir \citep{Cristallo2011, Ventura2013}. Alternatively, dilution of AGB ejecta with pristine or less-enriched gas may have reduced the effective Ce yield \citep{Dercole2016}.

\begin{figure}
    \centering
    \includegraphics[width=1\linewidth]{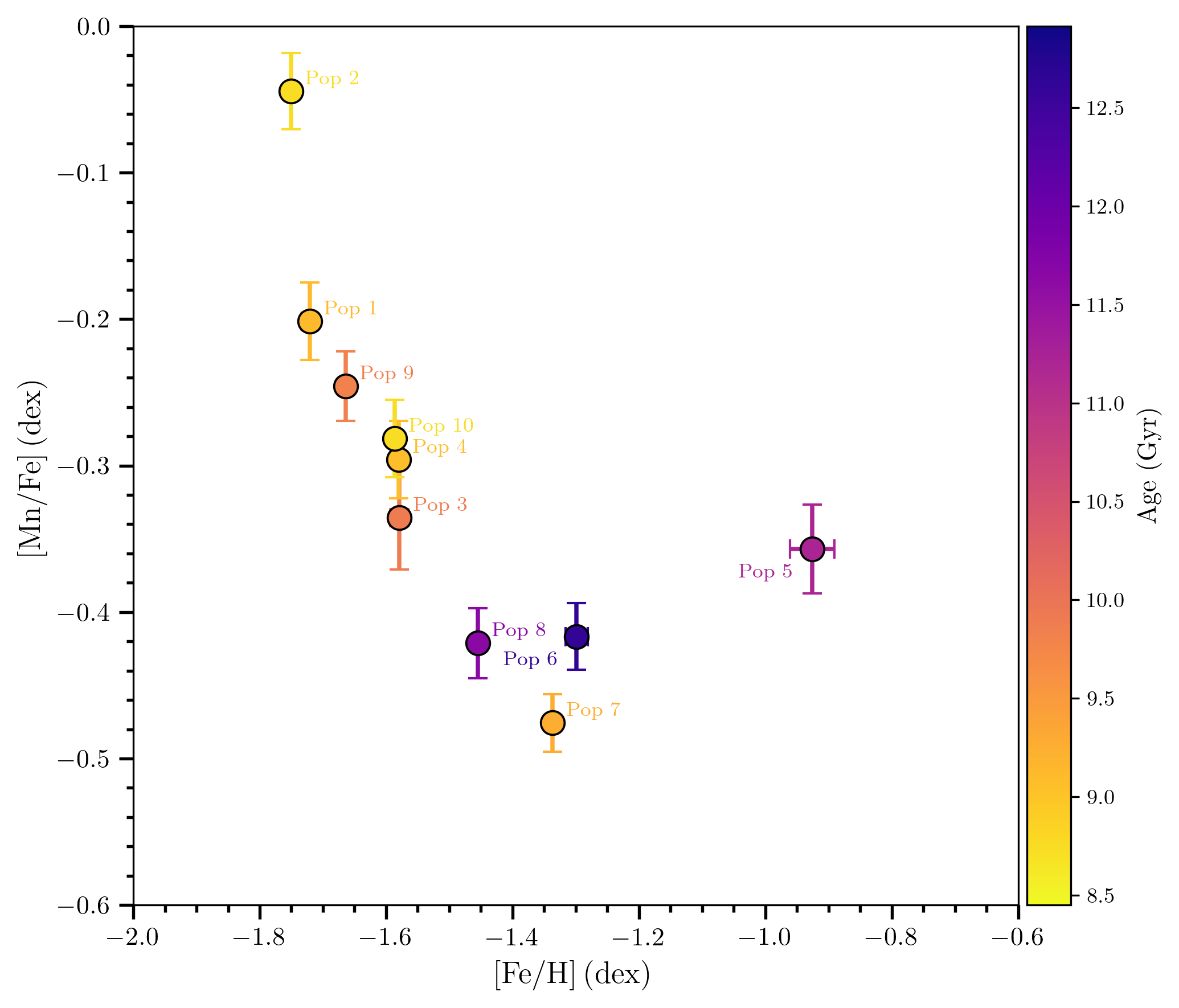}
    \caption{Distribution of the ten chemically defined populations in the $[\mathrm{Fe/H}]$--$[\mathrm{Mn/Fe}]$ plane. Each point represents the median abundance of a population, with error bars indicating the error deviation in the corresponding abundance ratios. The color coding reflects the median age of each population derived from isochrone fitting.}
    \label{fig:mnfe_feh}
\end{figure}

The $[\mathrm{Mn/Fe}]$--$[\mathrm{Fe/H}]$ plane (Figure~\ref{fig:mnfe_feh}) provides an independent constraint on
the role of Type~Ia supernovae in the chemical evolution of \ocen. In the Milky Way disc, $[\mathrm{Mn/Fe}]$ rises systematically from $\approx -0.4$~dex at $[\mathrm{Fe/H}] \approx -1.5$ to near-solar values at $[\mathrm{Fe/H}] \approx 0$, driven by the delayed contribution of Type~Ia supernovae which produce Mn with a super-solar yield relative to Fe \citep{Kobayashi2006, Nomoto2013}. No such trend is apparent in the \ocen\ populations. Instead, $[\mathrm{Mn/Fe}]$ remains persistently low across the full metallicity range, with the intermediate populations (Pop~6, 7) reaching $[\mathrm{Mn/Fe}] \approx -0.42$--$-0.47$~dex, and even the most metal-rich component (Pop~5, $[\mathrm{Fe/H}] = -0.93$) showing
$[\mathrm{Mn/Fe}] = -0.36$~dex — well below the values expected if Type~Ia supernovae had been a significant contributor \citep{McWilliam2003}.

\begin{figure}
    \centering
    \includegraphics[width=1\linewidth]{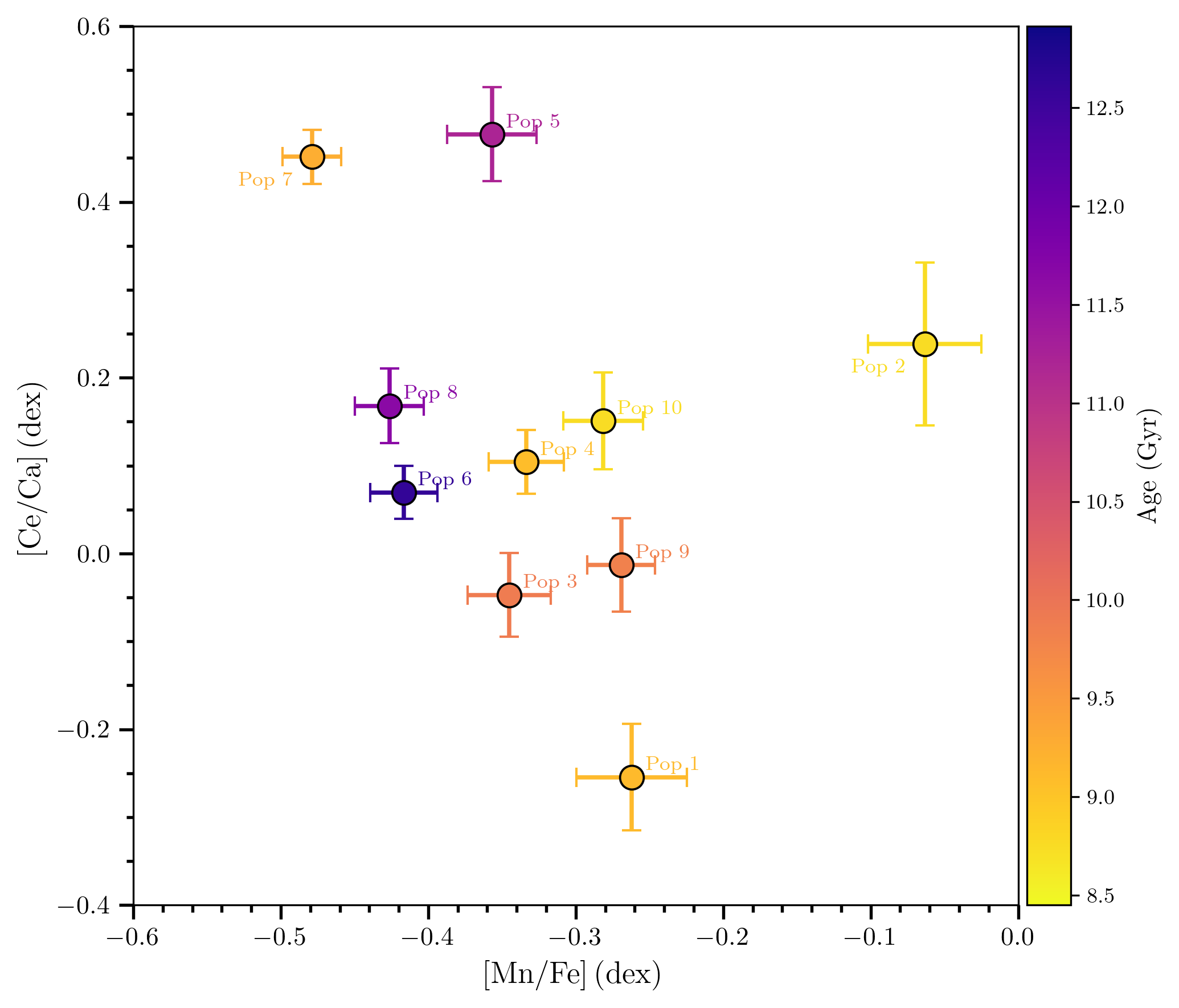}
    \caption{Distribution of the ten chemically defined populations in the $[\mathrm{Mn/Fe}]$--$[\mathrm{Ce/Ca}]$ plane. Each point represents the median abundance of a population, with error bars indicating the error deviation in the corresponding abundance ratios. The color coding reflects the median age of each population derived from isochrone fitting.}
    \label{fig:mnfe_ceca}
\end{figure}

The $[\mathrm{Ce/Ca}]$--$[\mathrm{Mn/Fe}]$ diagram (Figure~\ref{fig:mnfe_ceca}) further clarifies this picture. The
spread of $\approx 0.7$~dex in $[\mathrm{Ce/Ca}]$ — from Pop~1 at $\approx -0.27$~dex to Pop~7 and Pop~5 at
$\approx +0.45$~dex — occurs at nearly constant $[\mathrm{Mn/Fe}]$ across all populations. This orthogonality between the
$s$-process and the Type~Ia enrichment axes demonstrates that the $s$-process sequence was driven entirely by AGB stars operating within a system where Type~Ia supernovae remained sub-dominant throughout the star-formation history. The persistence of low $[\mathrm{Mn/Fe}]$ even at $[\mathrm{Fe/H}] = -0.93$ is particularly significant: it implies that the progenitor system completed the bulk of its chemical enrichment on a timescale shorter than the characteristic Type~Ia delay time of $\sim 1$--$3$~Gyr \citep{Maoz2012}, consistent with the rapid enrichment expected in the dense nuclear environment of an accreted dwarf galaxy \citep{BekkiFreeman2003, Romano2010}.

Taken together, the neutron-capture and iron-peak diagnostics paint a coherent picture: the chemical evolution of \ocen\ was
governed by a two-channel process in which core-collapse supernovae established the iron and $\alpha$-element baseline, and intermediate-mass AGB stars subsequently drove the $s$-process enrichment and proton-capture processing, all within a timescale too short for Type~Ia supernovae to make a significant contribution. This scenario is consistent with the chemical properties expected for the nuclear star cluster of a dwarf galaxy with a total stellar mass of $\sim 10^8$--$10^9\,M_\odot$ \citep{Massari2019,Pagnini2025}, and provides a natural explanation for the coexistence of extreme proton-capture and $s$-process signatures within the same system.

\subsection{Evidence for Multiple Enrichment Pathways}
\label{sec:discussion_multiple_pathways}

\begin{figure*}
    \centering
    \includegraphics[width=0.3\linewidth]{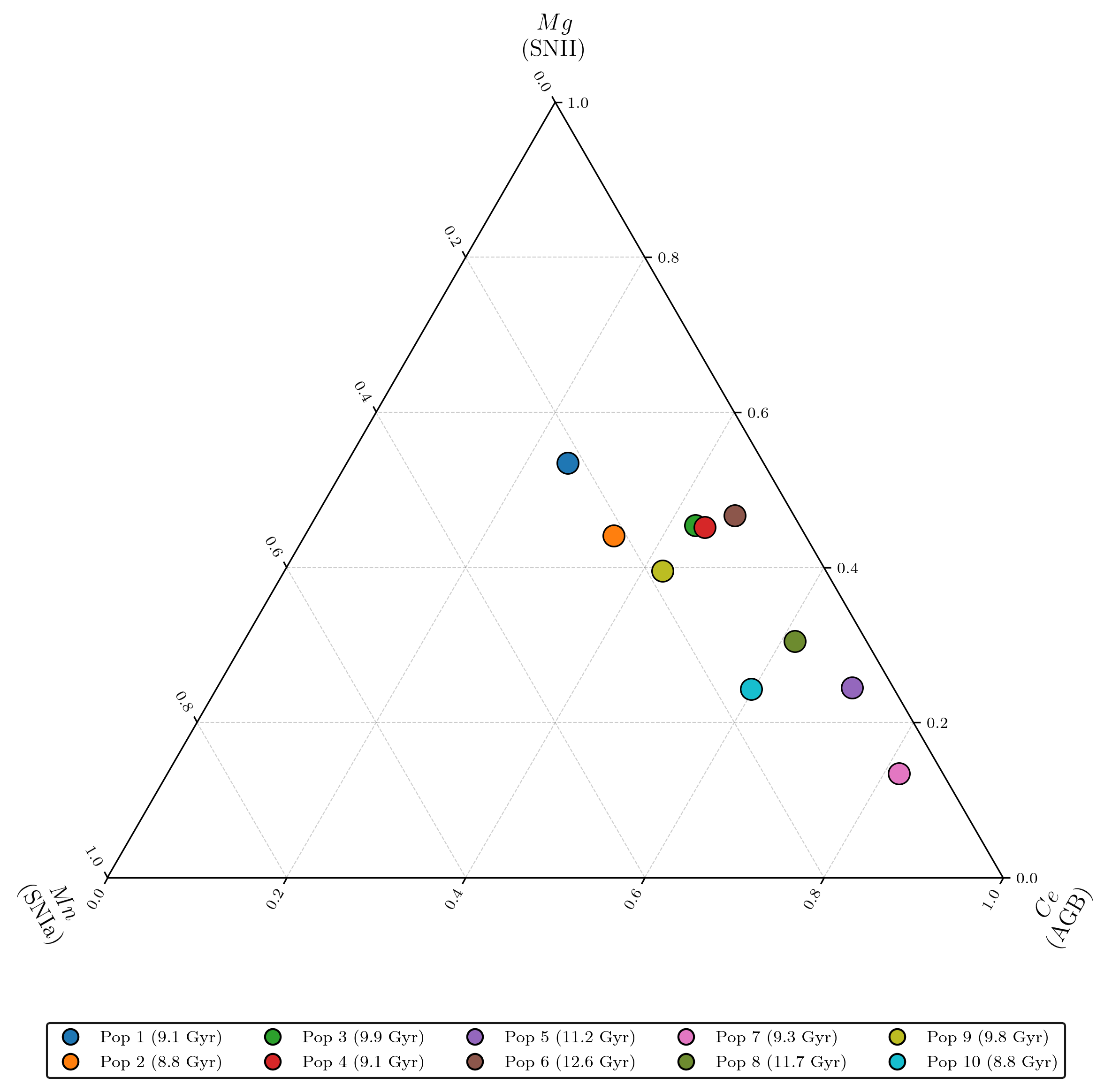}
    \includegraphics[width=0.3\linewidth]{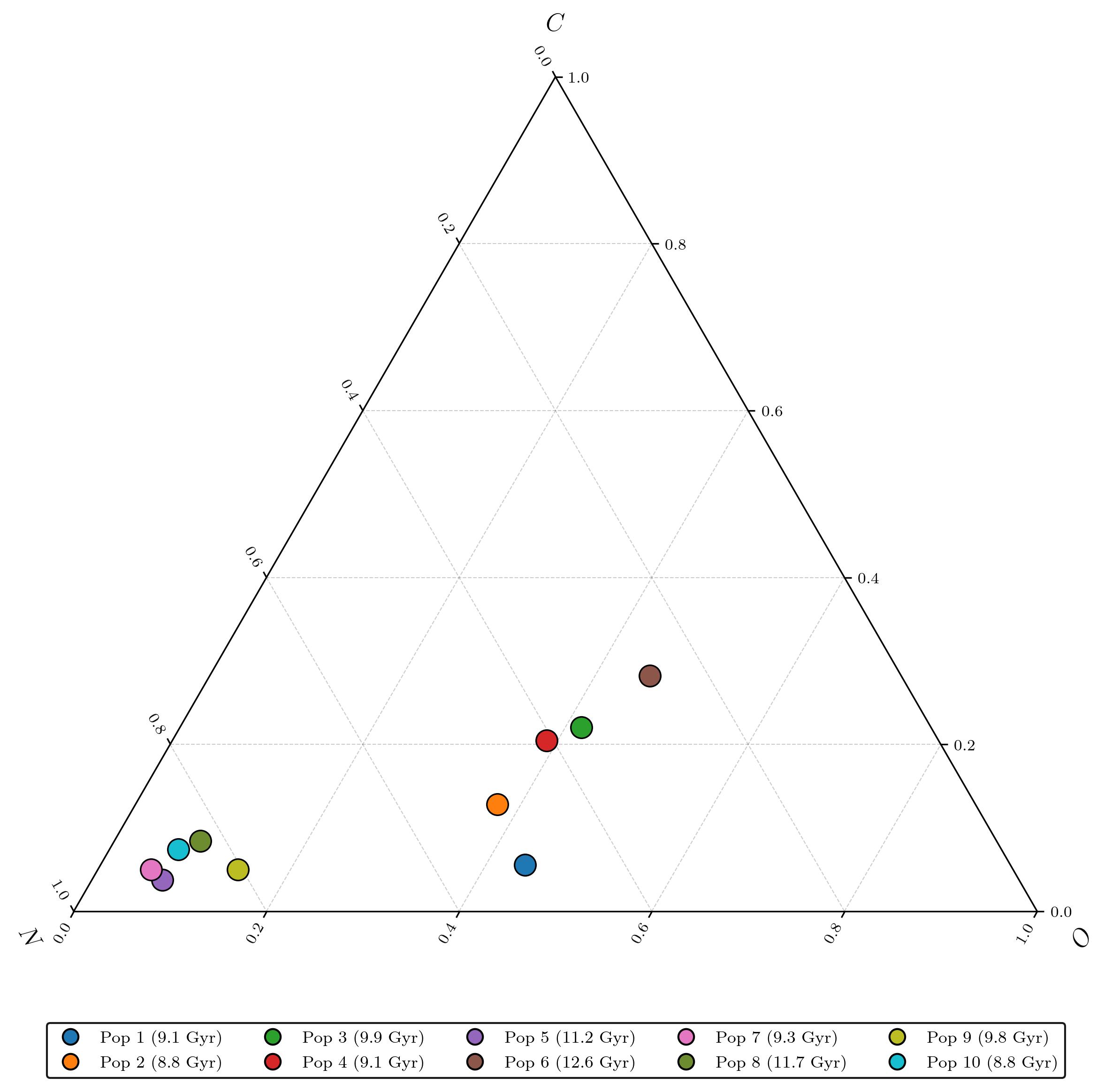}
    \includegraphics[width=0.3\linewidth]{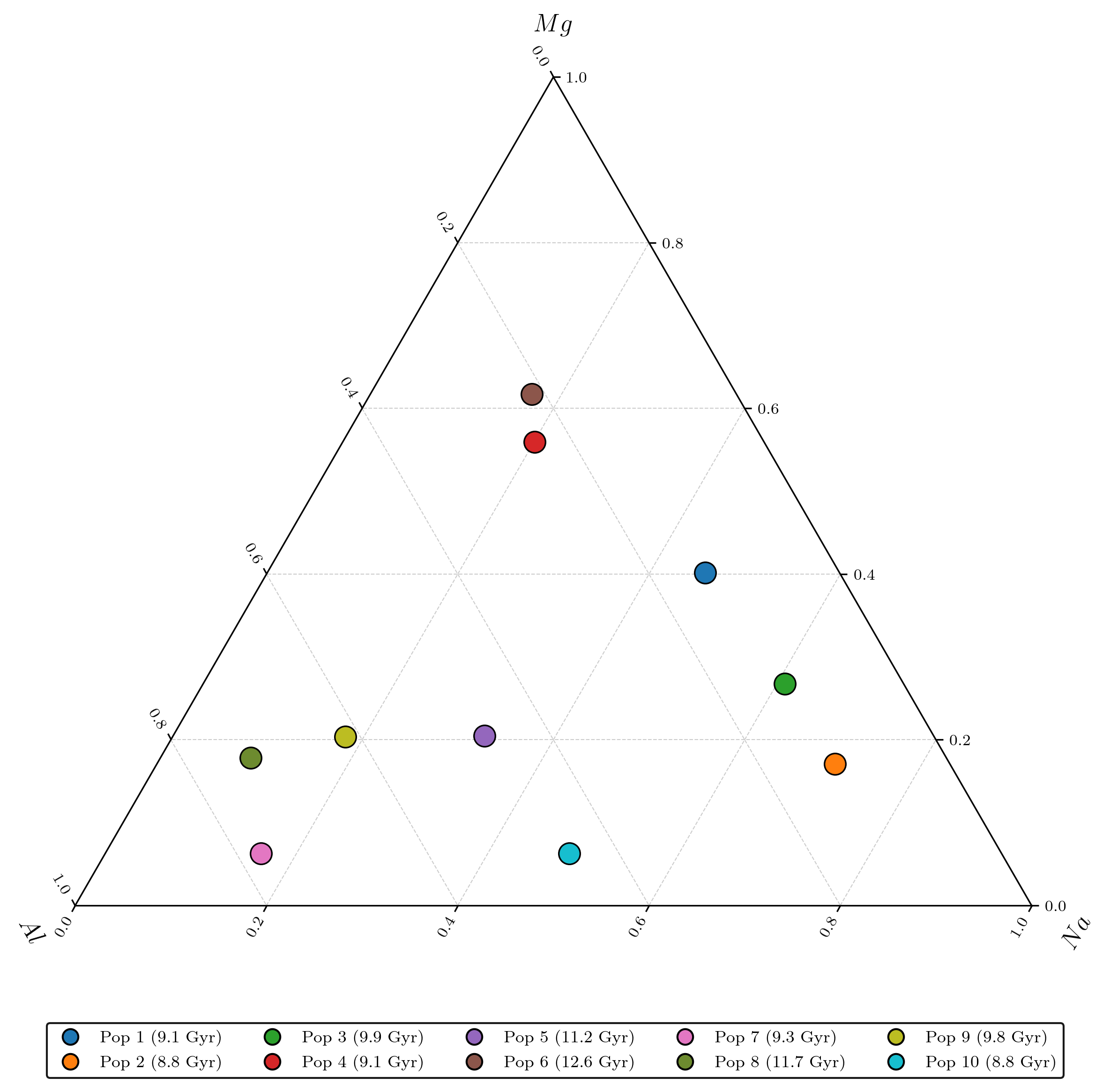}
    \caption{Global nucleosynthetic ternary diagrams for the ten $\omega$~Cen populations, mapping the relative fractional contributions of distinct enrichment channels. \textbf{Left (Mg--Mn--Ce):} The system's evolution traces a primary mixing vector from core-collapse supernovae (Mg) to AGB stars (Ce), with a universally negligible contribution from Type~Ia supernovae (Mn) across all populations. \textbf{Center (C--N--O) \& Right (Al--Mg--Na):} Tracers of high-temperature proton-capture processing. While the enriched branch (Pops~7--10) migrates continuously toward the processed (N, Al, Na) poles, Populations~6 and 5 deviate significantly from this main track, providing direct visual evidence for decoupled, multi-pathway enrichment within the progenitor dwarf galaxy.}
    \label{fig:ternary_plots}
\end{figure*}

The immense chemical diversity of $\omega$~Centauri cannot be fully explained by a simple, monotonic age-metallicity relationship or a single, well-mixed reservoir of gas. To unravel the global enrichment topology of the system, we project the ten chemically defined populations onto three nucleosynthetic ternary planes (Figure~\ref{fig:ternary_plots}): Mg--Mn--Ce, C--N--O, and Al--Mg--Na. These diagnostic planes map the relative fractional contributions of distinct physical sites across the entire cluster: core-collapse supernovae (Mg, O), Type~Ia supernovae (Mn), intermediate-mass AGB stars (Ce), and high-temperature hydrogen burning (N, Na, Al). To project the logarithmic abundance ratios onto two-dimensional barycentric coordinates, we transform the selected chemical tracers into a fractional parameter space. For a given nucleosynthetic plane defined by elements $X$, $Y$, and $Z$, we convert the standard logarithmic abundances, $[\mathrm{X/Fe}]$, into linear relative enrichment factors, $10^{[\mathrm{X/Fe}]}$. Notably, this conversion is performed strictly in the differential abundance space without scaling by the absolute solar abundances. This approach prevents intrinsically abundant elements (e.g., Oxygen) from compressing the dynamic range, ensuring that the ternary axes accurately represent the relative nucleosynthetic shifts. The fractional coordinate $f_{X}$ for element $X$ is thus computed as:

\begin{equation}
f_{X} = \frac{10^{[\mathrm{X/Fe}]}}{10^{[\mathrm{X/Fe}]} + 10^{[\mathrm{Y/Fe}]} + 10^{[\mathrm{Z/Fe}]}}.
\end{equation}

This transformation normalizes the variance across the three dimensions to unity ($f_{X} + f_{Y} + f_{Z} = 1$), allowing the system's mixing vectors to be visualized cleanly, independent of the absolute global metallicity.

The Mg--Mn--Ce ternary diagram provides a definitive visualization of the system's heavy-element evolution. The entire ten-population ensemble defines a global mixing sequence strictly between the Mg (SN~II) and Ce (AGB) poles, completely avoiding the Mn (SN~Ia) pole. The metal-poor backbone (Populations~1--4) anchors the system at the SN~II pole, reflecting their rapid, early formation. As self-enrichment proceeds, the intermediate populations (Populations~7--10) migrate systematically along the AGB vector. The universal absence of Mn-driven migration across 0.8~dex in $[\mathrm{Fe/H}]$ structurally confirms that the entire $\omega$~Cen progenitor evolved as a rapidly self-enriching, closed-box-like system that completed its star formation before the onset of significant Type~Ia supernova contributions.

Similarly, the C--N--O and Al--Mg--Na planes trace the progressive processing of the intracluster medium by proton-capture nucleosynthesis. The metal-poor populations reside solidly in the O-rich and Mg-rich domains. The enriched intermediate branch (Populations~7--10) traces a continuous evolutionary pathway toward the N and Al/Na poles, reflecting the progressive incorporation of CNO-cycled and NeNa/MgAl-processed AGB ejecta into subsequent stellar generations.

However, the ternary topology explicitly reveals that the system did not evolve along a single, monolithic track; rather, it comprises multiple, decoupled enrichment pathways. The most striking deviation from the main sequence is Population~6. Despite reaching an intermediate metallicity ($[\mathrm{Fe/H}] = -1.30$), it remains firmly anchored at the primordial (Mg, O) poles in all three diagrams, completely devoid of the Ce and N/Al/Na enhancements seen in its metallicity peer, Pop~7. If all gas at a given metallicity had been uniformly mixed, Pop~6 and Pop~7 would co-occupy the ternary space. Their stark orthogonality requires Pop~6 to have formed from a spatially or dynamically segregated gas reservoir that experienced extended SN~II enrichment but was shielded from, or diluted against, AGB pollution \citep{BekkiFreeman2003, Conroy2012}.

At the high-metallicity extreme, Population~5 ($[\mathrm{Fe/H}] = -0.93$) represents another distinctly decoupled nucleosynthetic regime. While it exhibits the most extreme migration toward the Nitrogen pole in the C--N--O plane (indicating near-complete CNO processing), its position in the Mg--Mn--Ce and Al--Mg--Na planes does not simply extrapolate the Pop~7--10 sequence. This decoupling suggests that Pop~5 formed in a unique, likely central, high-density environment where a different mass spectrum of AGB stars---or possibly super-AGB stars---dominated the chemical yields prior to the cessation of star formation \citep{Ventura2013}.

Ultimately, the ternary planes demonstrate that the $\omega$~Cen remnant is a composite structure: an amalgam of a primary, AGB-driven sequential enrichment track (Pops~1--4 transitioning to Pops~7--10) alongside distinct, physically isolated star-formation episodes (Pops~6 and 5). This multi-pathway architecture is the defining chemical signature of an accreted dwarf galaxy nucleus.

\subsection{Constraints on the Age Structure}
\label{sec:discussion_ages}

The BaSTI nested sampling posteriors (Section~\ref{sec:age_estimation}; Table~\ref{tab:nested_results}) reveal a structured age distribution across the ten Ward populations, spanning a total range of $\sim 3.4$~Gyr.

A key methodological choice underpinning these results is the adoption of $\alpha$-enhanced isochrones ($[\alpha/\mathrm{Fe}] = +0.4$, $Y = 0.30$) from the BaSTI library \citep{Pietrinferni2021} in place of solar-scaled tracks.
The MWM spectra of $\omega$\,Cen members show uniformly elevated $\alpha$-element abundances across all ten populations
(Table~\ref{tab:xfe_populations}), with $[\mathrm{Mg/Fe}] \sim +0.2$--$+0.4$~dex. Solar-scaled isochrones, which assume $[\alpha/\mathrm{Fe}] = 0$, systematically overestimate the total metallicity at a given $[\mathrm{Fe/H}]$, biasing RGB colors and hence the inferred ages and distances. The $\alpha$-enhanced BaSTI tracks correct for this by adopting the \citet{Salaris1993} metallicity scaling $[\mathrm{M/H}] = [\mathrm{Fe/H}] + \log(0.638\,f_\alpha + 0.362)$, where $f_\alpha = 10^{[\alpha/\mathrm{Fe}]}$, which shifts the effective metallicity by $+0.29$~dex relative to the iron abundance alone.
This correction is non-negligible: for a typical population at $[\mathrm{Fe/H}] = -1.6$, a solar-scaled isochrone would place the RGB $\sim 0.02$--$0.04$~mag blueward of the correct $\alpha$-enhanced locus, translating into a systematic age
underestimate of $\sim 1$--$2$~Gyr if left uncorrected \citep{Salaris1993, VandenBerg2012}. The use of $\alpha$-enhanced isochrones is therefore not a conservative choice but a physical requirement for a population as chemically distinct as $\omega$\,Cen.

The \textsc{dynesty} \citep{Speagle2020} \textsc{DynamicNestedSampler} \citep{dynamicsampling}. explores the four-dimensional parameter space $(\log\mathrm{Age},\, Z_{\mathrm{ini}},\, d,\, E(B_{\mathrm{P}}-R_{\mathrm{P}}))$ exhaustively, mapping the full posterior geometry rather than approximating it from a Markov chain. The resulting posteriors are well-behaved and unimodal for all ten populations, with effective sample sizes of $\mathrm{ESS} = 1600$--$2800$ per population and log-evidence uncertainties of $\sigma_{\ln Z} < 0.14$ (Table~\ref{tab:nested_results}), confirming that the nested
sampling runs are fully converged. The posterior credible intervals are therefore reliable representations of the age uncertainty given the observed CMD, the $\alpha$-enhanced BaSTI isochrone grid, and the spectroscopic metallicity constraints.

The six most metal-poor populations (Pop~1, 2, 3, 4, 9, 10; spectroscopic $[\mathrm{Fe/H}] = -1.50$ to $-1.76$)
yield posterior medians in the range $8.9$--$9.9$~Gyr, with a group mean of $\sim 9.2$~Gyr (Table~\ref{tab:nested_results}).
This is consistent with the old ages ($\geq 10$~Gyr) established for the dominant metal-poor component of $\omega$\,Cen
by MSTO-based analyses \citep{Sollima2005, Villanova2014}. The four best-constrained populations within this group are Pop~4 ($8.4$--$9.5$~Gyr, $\Delta = 1.1$~Gyr) and Pop~1 ($8.4$--$9.7$~Gyr, $\Delta = 1.3$~Gyr), while Pop~3 and Pop~9 show wider intervals of $\Delta = 2.6$--$2.8$~Gyr, consistent with their smaller sample sizes and broader CMD loci.

The two intermediate-metallicity, chemically anomalous populations (Pop~5 and Pop~6; spectroscopic $[\mathrm{Fe/H}] = -0.93$ and $-1.30$) are unambiguously \emph{older} than the metal-poor group: Pop~5 yields $11.3^{+0.9}_{-0.5}$~Gyr and Pop~6 yields $12.3^{+0.6}_{-0.1}$~Gyr, separated from the metal-poor group mean by $\sim 2$~Gyr and $\sim 3$~Gyr, respectively. Pop~6 is the most precisely constrained population in the sample, with a $68\%$ credible interval of only $0.7$~Gyr, driven by its large sample size ($N = 73$) and the tight, well-defined CMD locus of this group.
The finding that the most chemically evolved populations are also the oldest is consistent with a self-enrichment scenario in which a compact, rapidly star-forming core reached intermediate metallicities early in the assembly of the $\omega$\,Cen progenitor system, while the surrounding reservoir continued to form stars from less processed gas over a longer timescale
\citep{BekkiFreeman2003, Romano2007, Tsujimoto2012, deBoer2014}. The $\sim 2$--$3$~Gyr age difference matches the extended
formation timescales inferred from chemical evolution models of $\omega$\,Cen \citep{Romano2007, deBoer2014}.
However, we caution that the inferred older ages for the most chemically enriched groups may also be subject to systematic uncertainties related to helium variations. The intermediate and metal-rich populations in \ocen\ are known to host extreme helium enhancements (Y$\geq~$0.35; \citealt{Piotto2005, Clontz2025}). Because our isochrone grid adopts a fixed Y=0.30, fitting highly helium-enriched stars with standard-helium tracks can artificially drive the posterior toward older ages to reproduce the observed RGB colors. While our physical scenario of rapid core enrichment remains viable, disentangling the true age sequence from internal helium variations underscores the need for deep, multi-band MSTO photometry.

Pop~8 ($[\mathrm{Fe/H}]_{\rm spec} = -1.47$) has a posterior median of $11.2$~Gyr but a $68\%$ credible interval of
$8.6$--$14.0$~Gyr ($\Delta = 5.4$~Gyr), the widest of all ten populations. Its age is therefore consistent with both the metal-poor group and the RGB-a group, and no reliable placement within the formation sequence is possible from the present data. The broad interval reflects the combination of an intermediate metallicity --- where RGB isochrone morphology is least sensitive to age --- and the relatively diffuse CMD locus of this group \citep{vonHippel2006}.

The MWM selection function targets luminous giants, so the CMDs used here consist almost entirely of upper-RGB stars, with no coverage of the main-sequence turn-off (MSTO) or subgiant branch.

Age information in an RGB-dominated CMD enters through subtle morphological differences in the RGB slope, luminosity
function, and color dispersion, which the BaSTI nested sampling framework successfully exploits. Nevertheless, the primary age diagnostic in stellar populations, the MSTO luminosity, which shifts by $\sim 0.3$--$0.5$~mag per Gyr at these metallicities \citep{Sarajedini2007}, is absent from our data.

As a result, the age posteriors derived here, although well-sampled and internally consistent, carry systematic
uncertainties that cannot be fully quantified without MSTO photometry.

In particular, the $\sim 1$--$3$~Gyr credible intervals for the metal-poor populations (and the much wider interval for
Pop~8) reflect genuine ambiguity in the RGB morphology rather than failures of the inference framework. We therefore present the BaSTI ages as the best available constraints from spectroscopic-survey CMDs, while noting that a definitive age ranking of all ten populations will require space-based photometry reaching the MSTO ($G \geq 19$~mag at the distance of $\omega$\,Cen).

The chemical taxonomy established in this work provides a direct framework for such a follow-up analysis: the ten Ward populations define chemically homogeneous subsamples whose MSTO loci can be compared using existing \textit{HST} photometric catalogues \citep{Bellini2010, Milone2017} or future \textit{JWST} deep imaging.

\subsection{Ex-situ Origin}
\label{sec:discussion_exsitu}

The chemical properties of the \ocen\ populations provide strong and multi-dimensional evidence for an ex-situ origin,
consistent with the interpretation of the system as the remnant nucleus of an accreted dwarf galaxy \citep{BekkiFreeman2003, Massari2019, Pagnini2025}.

\begin{figure*}
    \centering
    \includegraphics[width=\linewidth]{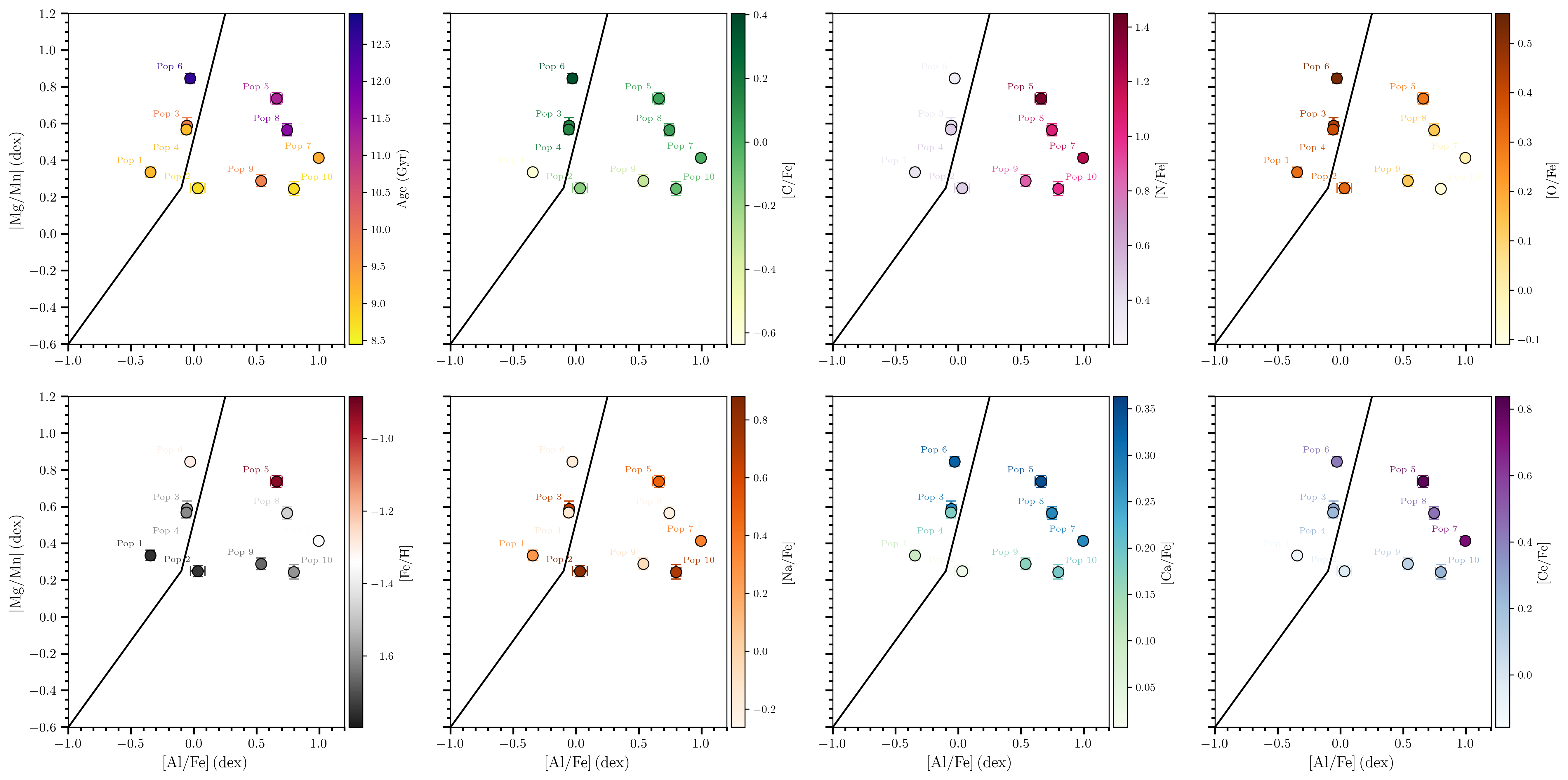}
    \caption{Distribution of the ten chemically defined populations in the $[\mathrm{Al/Fe}]$--$[\mathrm{Mg/Mn}]$ plane. The elevated $[\mathrm{Mg/Mn}]$ values across all populations indicate enrichment dominated by core-collapse supernovae, while the wide spread in $[\mathrm{Al/Fe}]$ reflects internal proton-capture processing. This combination is characteristic of chemically complex, accreted systems.}
    \label{fig:alfe_mgmn}
\end{figure*}

The most direct diagnostic is the $[\mathrm{Al/Fe}]$--$[\mathrm{Mg/Mn}]$ plane (Figure~\ref{fig:alfe_mgmn}), which has been established as a robust discriminator between in-situ and accreted stellar populations in the Milky Way halo \citep{Hawkins2015, Das2020, Horta2021}. The $[\mathrm{Mg/Mn}]$ ratio is elevated in systems that completed their chemical enrichment rapidly and were dominated by core-collapse supernovae, whereas it decreases toward solar values as the delayed iron contribution from Type~Ia supernovae becomes significant. All ten \ocen\ populations lie clearly above the Milky Way disc sequence shown in Figure~\ref{fig:alfe_mgmn}, with $[\mathrm{Mg/Mn}]$ ranging from $\approx +0.35$~dex (Pop~1) to $\approx +0.85$~dex (Pop~6), firmly in the regime associated with accreted or dwarf galaxy material \citep{Hawkins2015, Das2020}.

Crucially, this elevated $[\mathrm{Mg/Mn}]$ is maintained even at $[\mathrm{Fe/H}] = -0.93$ (Pop~5), where
in-situ Milky Way stars show markedly lower values due to the accumulated Type~Ia contribution. The persistence of
high $[\mathrm{Mg/Mn}]$ across the full $\sim 1.0$~dex metallicity baseline of \ocen\ is therefore not simply a
property of the metal-poor component, but a global characteristic of the entire system. This is consistent with the rapid
chemical evolution expected in a massive, dense nuclear star cluster, where the high stellar density and deep
potential well allow efficient self-enrichment by core-collapse supernovae while suppressing the relative contribution of
the more slowly evolving Type~Ia channel \citep{Romano2010, Recchi2015}.

The $[\mathrm{Al/Fe}]$ dimension of Figure~\ref{fig:alfe_mgmn} further reveals the internal complexity that distinguishes
\ocen\ from a simple accreted halo population. Whereas typical accreted streams and disrupted dwarf galaxies show a narrow range of $[\mathrm{Al/Fe}]$ at a given metallicity, reflecting their limited self-enrichment capacity \citep{Myeong2019, Naidu2020}, the \ocen\ populations span $\Delta[\mathrm{Al/Fe}] \approx 1.6$~dex at high and roughly constant $[\mathrm{Mg/Mn}]$. This combination — a wide proton-capture spread superimposed on a uniformly high $[\mathrm{Mg/Mn}]$ baseline — is precisely the chemical signature expected for a nuclear star cluster that experienced extended AGB self-enrichment within a dwarf galaxy potential \citep{Pfeffer2021}.

The eight-panel view of Figure~\ref{fig:alfe_mgmn} further shows that the $[\mathrm{Al/Fe}]$--$[\mathrm{Mg/Mn}]$ separation between populations is mirrored consistently across all chemical dimensions: the same populations that occupy the high-$[\mathrm{Al/Fe}]$ regime also show enhanced $[\mathrm{N/H}]$, depleted $[\mathrm{O/H}]$ and $[\mathrm{C/H}]$, and elevated $[\mathrm{Ce/H}]$, while the low-$[\mathrm{Al/Fe}]$ populations retain primordial values in each of these tracers. This coherence across eight independent abundance dimensions strongly argues against an instrumental or systematic origin for the population separation, and confirms that the Ward clustering has recovered physically meaningful chemical substructure.

At the Galactic scale, the chemical properties of \ocen\ are consistent with its proposed association with the Nephele
merger event \citep{Pagnini2025}, which is chemically and dynamically distinct from the \textsl{Gaia}--Sausage--Enceladus
accretion \citep{Massari2019, Myeong2019}. The progenitor system inferred from this association — a dwarf galaxy of initial stellar mass $\sim 10^8$--$10^9\,M_\odot$ — is massive enough to have sustained the multi-Gyr enrichment history implied by the chemical complexity of the ten populations identified here, and to have hosted an intermediate-mass black hole of the kind recently detected in the cluster core \citep{Haeberle2024}.

Taken together, the $[\mathrm{Al/Fe}]$--$[\mathrm{Mg/Mn}]$ evidence, the persistently low $[\mathrm{Mn/Fe}]$ across the
full metallicity range, the decoupled enrichment pathway of Pop~6, and the extreme proton-capture and $s$-process
signatures of Pop~7 and Pop~5 collectively paint a picture that cannot be reconciled with in-situ formation in the
Milky Way disc or halo. Instead, they point to a system that evolved as a self-contained, chemically complex entity
before being accreted and stripped to its present nuclear remnant.

An intriguing possibility raised by the $[\mathrm{Al/Fe}]$--$[\mathrm{Mg/Mn}]$ diagram is that the two-phase chemical structure of \ocen\ may reflect not only sequential enrichment within the progenitor dwarf, but also a temporal boundary coinciding with the accretion event itself. Populations~1--4 and~6, which occupy the canonical ex-situ regime of this plane, would in this picture represent the stellar populations already in place at the time of accretion. Populations~7--10, while sharing a similarly elevated $[\mathrm{Mg/Mn}]$ baseline, extend to extreme $[\mathrm{Al/Fe}]$ values that are more naturally produced in the deep potential well of a massive, self-enriching system — conditions that may have been more easily sustained after the system had settled into the Milky Way halo and its gas reservoir was stabilised by the gravitational influence of the central IMBH \citep{Haeberle2024}. Under this scenario, Population~5 would represent the ultimate endpoint of rapid, high-efficiency enrichment in which core-collapse supernova and AGB channels dominate while any Type~Ia contribution remains sub-dominant. We present this as a speculative but testable hypothesis; kinematic and spatial mapping of the ten populations using the full \citet{Vasiliev2021} proper-motion catalogue would provide a direct observational test.

\section{Conclusions}
\label{sec:conclusion}

We have presented a homogeneous spectroscopic analysis of $\omega$~Centauri using high-resolution near-infrared observations
from the Milky Way Mapper survey (MWM~DR19), cross-matched with the membership catalogue of \citet{Vasiliev2021}. Applying
Ward-linkage hierarchical clustering in a seven-dimensional chemical abundance space to 957 quality-selected members, we identify 10 chemically distinct stellar populations and characterise their abundance patterns, photometric properties,
and nucleosynthetic signatures. Our principal conclusions are as follows.

\begin{enumerate}

\item \textbf{Ten chemically distinct populations.}
Ward hierarchical clustering in the $[\mathrm{Fe/H}]$--$[\mathrm{C/H}]$--$[\mathrm{N/H}]$--$[\mathrm{O/H}]$--$[\mathrm{Al/H}]$--$[\mathrm{Na/H}]$--$[\mathrm{Ca/H}]$ space, with the partition threshold selected objectively via Davies--Bouldin and silhouette diagnostics, yields 10 populations that remain coherent in the independent \textsl{Gaia} color--magnitude diagram. The system spans $\Delta[\mathrm{Fe/H}] = 0.82$~dex ($-1.75$ to $-0.93$) and $\Delta[\mathrm{Al/Fe}] \approx 1.6$~dex across the full sample, confirming the extraordinary chemical complexity of \ocen.

\item \textbf{Two-phase chemical structure.}
The populations separate into a metal-poor backbone (Populations~1--4; $\sim 47\%$ of the sample, $-1.75 < [\mathrm{Fe/H}] < -1.58$) and an intermediate group enriched by high-temperature proton-capture nucleosynthesis (Populations~7--10; $\sim 43\%$, $-1.65 < [\mathrm{Fe/H}] < -1.35$). The large separation in light-element abundances --- $\Delta[\mathrm{Al/H}] \approx +1.76$~dex and $\Delta[\mathrm{N/H}] \approx +1.22$~dex --- at a modest $\Delta[\mathrm{Fe/H}] \approx 0.40$~dex provides a clear signature of AGB-driven self-enrichment, rather than purely supernova-driven chemical evolution \citep{Ventura2011, Bastian2018}.

\item \textbf{AGB stars as the primary enrichment driver.}
The systematic rise in $[\mathrm{Ce/Mg}]$ from $\approx -0.35$~dex (Pop~1) to $\approx +0.78$~dex (Pop~7) at nearly constant $[\mathrm{Mn/Fe}]$ across the full metallicity range suggests that $s$-process enrichment by intermediate-mass AGB stars was largely decoupled from iron-peak evolution throughout the system's history. The persistently low $[\mathrm{Mn/Fe}]$ values ($\leq -0.2$~dex even at $[\mathrm{Fe/H}] = -0.93$) indicate that Type~Ia supernovae did not become a dominant contributor before star formation ceased, implying a total enrichment timescale shorter than the characteristic Type~Ia delay time of $\sim 1$--$3$~Gyr \citep{Maoz2012}.

\item \textbf{Population~6 as evidence for decoupled enrichment pathways.}
Population~6 ($N = 73$, $[\mathrm{Fe/H}] = -1.30$) retains primordial light-element abundances ($[\mathrm{Al/Fe}] = -0.03$~dex, $[\mathrm{Ce/Mg}] \approx 0.00$~dex, $[\mathrm{C/N}] = +0.02$~dex) despite having reached a metallicity comparable to the most proton-capture-enriched populations. Its chemical contrast with the neighbouring Pop~7 ($\Delta[\mathrm{Al/Fe}] \approx 1.0$~dex, $\Delta[\mathrm{Ce/Mg}] \approx 0.8$~dex at essentially identical $[\mathrm{Fe/H}]$) strongly suggests the presence of spatially or dynamically segregated gas reservoirs within the progenitor system, in which iron enrichment proceeded without exposure to AGB ejecta \citep{DercoleVesperini2008, Conroy2012}.

\item \textbf{Population~5 as the chemically anomalous metal-rich endpoint.}                                                         
Population~5 ($N = 25$, $[\mathrm{Fe/H}]_{\rm spec} = -0.93$, $[\mathrm{Fe/H}]_{\rm iso} = -1.13^{+0.03}_{-0.06}$) represents the most chemically evolved component, combining the lowest $[\mathrm{C/N}]$ ratio in the sample ($-1.38$~dex), extreme $[\mathrm{N/Fe}] = +1.39$~dex, and elevated $[\mathrm{Ce/Mg}] \approx +0.45$~dex. Its iron abundance represents a $\sim 0.4$--$0.8$~dex jump relative to all other populations. The BaSTI isochrone fit yields an age of $11.26^{+0.85}_{-0.46}$~Gyr for this population, placing it among the oldest groups in the cluster rather than the youngest. This age inversion with respect to simple closed‑box models suggests that the most enriched gas was locked into long‑lived stars very early in a compact, high‑density core, while the surrounding, less enriched regions continued forming stars for several more gigayears. This age, combined with its extreme chemical enrichment, points to rapid early self-enrichment in a compact, high-density environment --- consistent with a nuclear star cluster origin in which a rapid initial burst of massive stars deposited large amounts of iron into a deep potential well, followed closely by s-process enrichment from intermediate-mass AGB stars, before the surrounding metal-poor populations fully assembled \citep{BekkiFreeman2003, Romano2007, Tsujimoto2012}. The high iron abundance reached at this early epoch likely reflects an exceptionally high star-formation rate and supernova~II rate density, without requiring a Type~Ia contribution \citep{Kobayashi2006, Seitenzahl2013}. We caution that Pop~5 contains only 25 stars in our sample, so its age posterior, while well-converged ($\mathrm{ESS} = 2781$), carries the largest systematic uncertainty from CMD sparsity.  

\item \textbf{All populations are consistent with an ex-situ origin.}
All ten populations occupy the elevated-$[\mathrm{Mg/Mn}]$ regime of the $[\mathrm{Al/Fe}]$--$[\mathrm{Mg/Mn}]$ plane, well above the Milky Way disc sequence and firmly in the domain of accreted or dwarf-galaxy material \citep{Hawkins2015, Das2020}. The combination of a wide proton-capture spread ($\Delta[\mathrm{Al/Fe}] \approx 1.6$~dex) superimposed on this uniformly high $[\mathrm{Mg/Mn}]$ baseline is a chemical signature broadly consistent with a nuclear star cluster that underwent extended AGB self-enrichment within the potential well of a dwarf galaxy progenitor \citep{Pfeffer2021}. This interpretation is further supported by the independent detection of an intermediate-mass black hole in the cluster core \citep{Haeberle2024}, a structural feature characteristic of nuclear star clusters rather than ordinary globular clusters, and is broadly consistent with the proposed association of \ocen\ with the Nephele accretion event \citep{Pagnini2025} and/or $\omega$ Dwarf \citep{Souza2026}.

\item \textbf{Age constraints from $\alpha$-enhanced BaSTI isochrone fitting.}
Bayesian isochrone fitting of the \textsl{Gaia} CMD with $\alpha$-enhanced BaSTI tracks ($[\alpha/\mathrm{Fe}]=+0.4$, $Y=0.30$; \citealt{Pietrinferni2021}) yields well-converged nested sampling posteriors ($\mathrm{ESS}=1600$--$2800$) and tightly constrained distances ($d = 5450$--$5523$~pc, consistent with \citealt{Haberle2025}) and extinctions $E(B_\mathrm{P}-R_\mathrm{P}) = 0.16$--$0.21$~mag, consistent with \citealt{Harris1996}. The ten populations span a significant age spread of $\sim 3.4$~Gyr, with chemically enriched populations (Pop 5 and 6) being $\approx$2‑3 Gyr older than the metal‑poor backbone: the six metal-poor populations ($[\mathrm{Fe/H}]_{\rm spec} = -1.50$ to $-1.76$) converge to $8.9$--$9.9$~Gyr, while the chemically anomalous RGB-a populations yield $11.3^{+0.9}_{-0.5}$~Gyr (Pop~5) and $12.3^{+0.6}_{-0.1}$~Gyr (Pop~6), establishing an age advantage of $\sim 2$--$3$~Gyr for the most chemically evolved component. However, we note that if these specific populations host the extreme helium enhancements ($Y\geq$~0.35) identified in previous photometric studies, the adoption of standard-helium tracks ($Y=0.30$) may artificially inflate these inferred ages via the age--helium degeneracy. The adoption of $\alpha$-enhanced over solar-scaled isochrones is a physical requirement: solar-scaled tracks would introduce a systematic color offset of $\sim 0.02$--$0.04$~mag on the RGB, translating to an age underestimate of $\sim 1$--$2$~Gyr \citep{Salaris1993, VandenBerg2012}. Because the MWM selection function excludes main-sequence turn-off stars, the posteriors rely on RGB morphology and carry intrinsic uncertainties of $1$--$3$~Gyr per population; a definitive age ranking will require space-based MSTO photometry \citep{Sarajedini2007, Bellini2010, Milone2017}.

\end{enumerate}

In conclusion, the ten chemically distinct populations identified within the seven-dimensional abundance space of $\omega$ Centauri establish the system as a premier evolutionary laboratory for galactic archaeology. The discovery of sub-structures such as Population~6—which maintains a primordial chemical signature despite its intermediate metallicity—provides compelling spectroscopic evidence for spatially segregated enrichment pathways and a complex, non-homogeneous assembly history within the progenitor. Furthermore, the coherent placement of all ten populations in the $[\mathrm{Al/Fe}]$--$[\mathrm{Mg/Mn}]$ diagnostic plane reinforces a definitive ex-situ origin, leaving little doubt regarding the cluster's extragalactic nature. Ultimately, $\omega$ Centauri stands not merely as the Milky Way’s most massive globular cluster, but as the surviving, chemically encoded nucleus of an accreted dwarf galaxy, preserving the intricate, multi-phase history of a lost member of the Galactic family.

\section*{Acknowledgements}

This study was supported by the Scientific and Technological Research Council (TÜBİTAK) MFAG-123F227, MFAG-125F196, MFAG-125F465 and 2211-C. This study was funded by the Scientific Research Projects Coordination Unit of the Istanbul University. Project number: FBA-2023-39380 and FDK-2025-41537.

\section*{Data Availability}

The spectroscopic and astrometric data underlying this article are available from public archives. MWM~DR19 spectra and derived abundances can be accessed through the SDSS Science Archive Server and CasJobs interfaces. Membership probabilities and proper-motion selections are based on the public catalogue of \citet{Vasiliev2021}. The BaSTI-IAC isochrone grids used in the fitting are publicly available from the BaSTI database. Derived population-level summary tables and plotting scripts used to generate the figures can be provided by the corresponding author upon reasonable request.



\bibliographystyle{mnras}
\bibliography{references} 




\appendix

\section{Extended Chemical Validation of the Selected Sample}
\label{sec:app_chem_validation}

\subsection{Chemical Properties of the Selected Sample}
\label{sec:chem_properties}

To verify that the MWM DR19 stars retained after kinematic and quality filtering are chemically representative of \ocen, we compare the key abundance distributions of the 1807-member sample (membership probability $\geq 0.9$; Section~\ref{sec:data_sample}) against spectroscopic benchmarks from the literature.

\subsubsection{Metallicity Distribution}
\label{sec:feh_distribution}

\begin{figure}
    \centering
    \includegraphics[width=1\columnwidth]{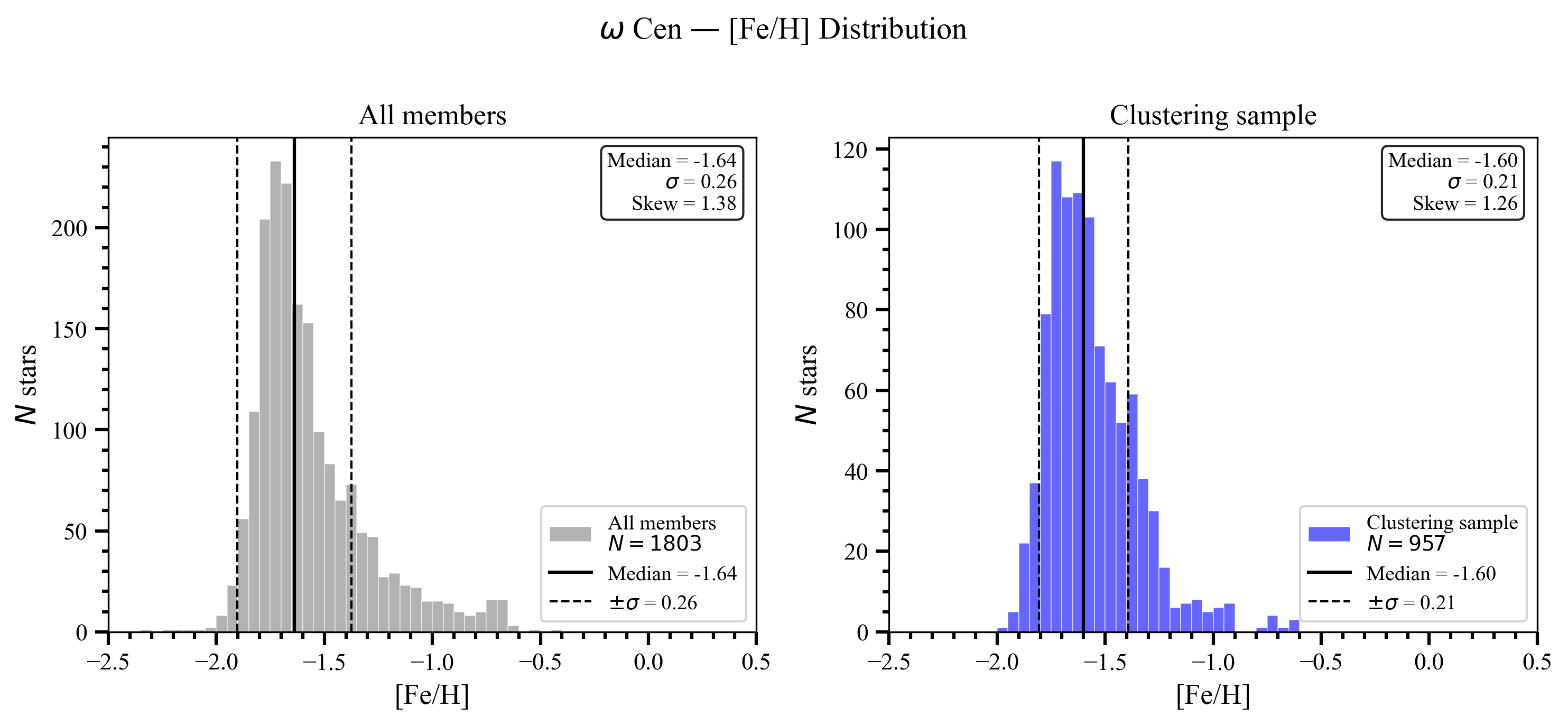}
    \caption{Metallicity distribution of the 1807 matched \ocen\ members used
    in the validation step before the final seven-element quality filtering.
    The profile is dominated by a metal-poor peak near $[\mathrm{Fe/H}]\approx-1.7$ with
    an extended metal-rich tail.}
    \label{fig:feh_distribution}
\end{figure}

The $[\mathrm{Fe/H}]$ distribution of the selected sample is shown in Figure~\ref{fig:feh_distribution}. The sample spans a range of $-2.35 \leq [\mathrm{Fe/H}] \leq -0.45$ ($\Delta[\mathrm{Fe/H}] \approx 1.9$~dex), with a median of $[\mathrm{Fe/H}]_{\mathrm{med}} = -1.64$ and a sample mean of $\langle[\mathrm{Fe/H}]\rangle = -1.56$ ($\sigma = 0.26$~dex).
The distribution is strongly skewed towards the metal-poor tail, peaking near $[\mathrm{Fe/H}] \approx -1.7$, with a secondary shoulder at $[\mathrm{Fe/H}] \approx -1.1$ and a metal-rich extension out to $[\mathrm{Fe/H}] \approx -0.5$.

This morphology is in good agreement with published high-resolution spectroscopic surveys of \ocen. \citet{Johnson2010} characterised 855 red giants over the full radial extent of the cluster and reported a metallicity range of $-2.2 \leq [\mathrm{Fe/H}] \leq -0.6$ with a dominant metal-poor peak near $[\mathrm{Fe/H}] \approx -1.75$ and progressively weaker metal-rich components. \citet{Marino2011} similarly found a peak near $-1.7$~dex and confirmed the presence of discrete sub-populations at $[\mathrm{Fe/H}] \approx -1.7$, $-1.5$, $-1.1$, and $-0.7$, consistent with the multiple star-formation episodes inferred from the cluster's broad metallicity baseline \citep{NorrisDaCosta1995}. The metallicity range, peak location, and skewed shape of our sample all fall within the bounds established by these studies, confirming that no significant metallicity bias has been introduced by the \textsl{APOGEE} targeting strategy or the membership filter.

\subsubsection{Light-Element Anticorrelations}
\label{sec:anticorrelations}

The Na--O and Al--O anticorrelations are the most direct chemical signatures of hot hydrogen-burning in globular cluster multiple populations \citep{Denisenkov1990, Prantzos2007, Carretta2009a}. These planes are examined for the quality-selected clustering subsample ($N = 957$ stars with zero quality flags on all seven tracers; Section~\ref{sec:element_selection}).

\begin{figure*}
    \centering
    \includegraphics[width=1\linewidth]{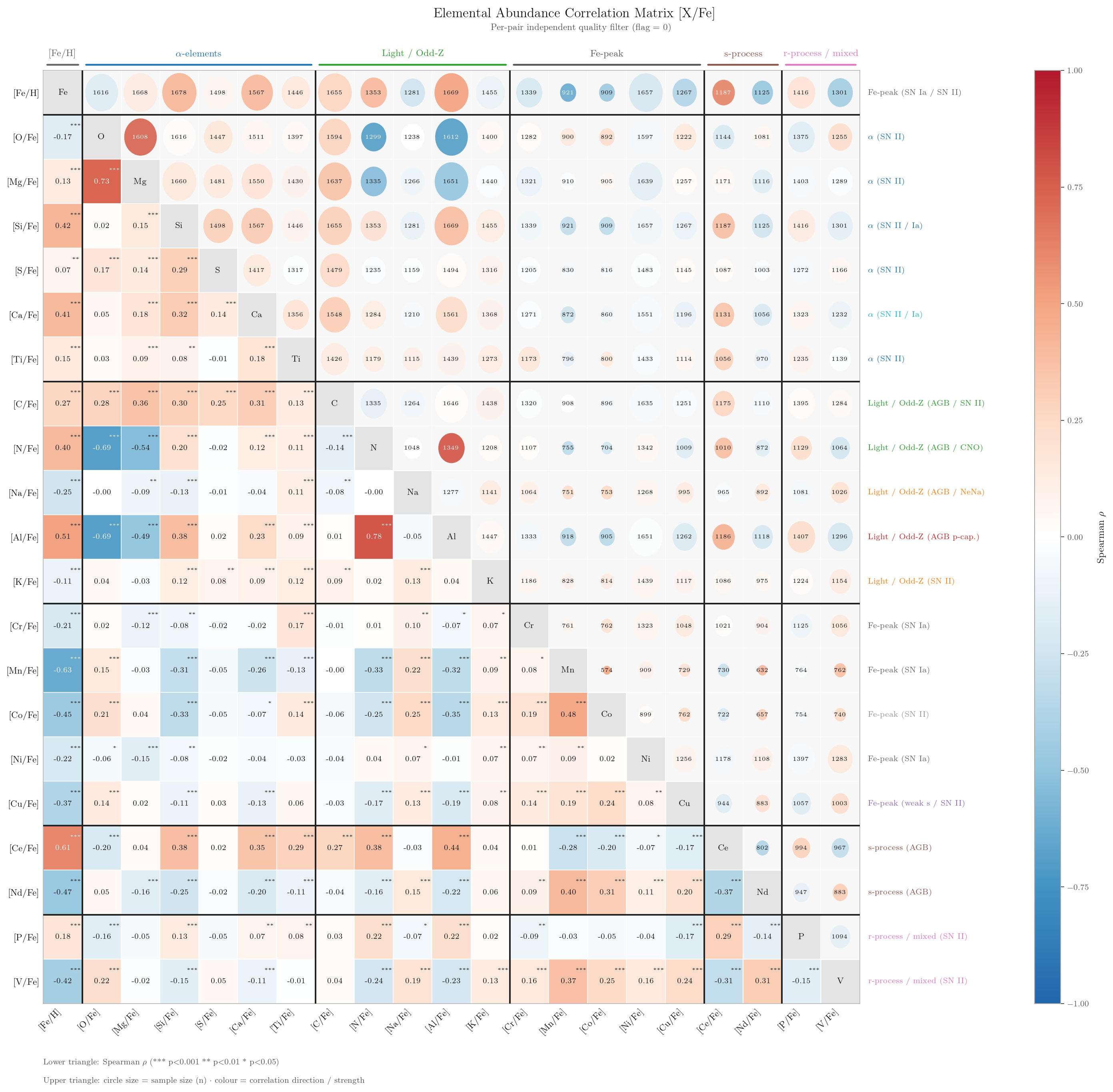}
    \caption{Spearman rank-correlation matrix for the same MWM~DR19 \ocen\ sample in the [X/Fe] frame. After removing the first-order metallicity trend, intrinsic light-element structure is more clearly visible, including proton-capture signatures such as O--Na and O--Al behavior.}
    \label{fig:corr_matrix}
\end{figure*}

The Al--O anticorrelation is detected with high significance: the Spearman rank coefficient between $[\mathrm{Al/Fe}]$ and $[\mathrm{O/Fe}]$ is $\rho = -0.76$, with $[\mathrm{Al/Fe}]$ spanning $-0.55$ to $+1.29$~dex and $[\mathrm{O/Fe}]$ ranging from $-0.55$ to $+0.92$~dex. The full $\sim 1.8$~dex spread in $[\mathrm{Al/Fe}]$ and the strong anti-correlation are hallmarks of extreme proton-capture nucleosynthesis in \ocen\ \citep{Carretta2009b} and is consistent with the range reported by \citet{Johnson2010}, who found $[\mathrm{Al/Fe}]$ values as high as $+1.3$~dex in the most chemically enriched giants. \citet{AlvarezGaray2024} also detected an extended Mg--Al anticorrelation in the \textsl{APOGEE} observations of \ocen, with the same extreme aluminium enhancement, providing direct validation that the \textsl{APOGEE}/ASPCAP pipeline
recovers the proton-capture signature reliably in this cluster; in our data, this behavior is also consistent with the [X/Fe] correlation structure in Fig.~\ref{fig:corr_matrix}.

The Na--O anticorrelation in \ocen\ is more complex than in monometallic globular clusters, because the wide metallicity baseline tends to dilute the $[\mathrm{X/Fe}]$ anti-trend when stars of all metallicities are plotted together \citep{Marino2011}. Within narrow metallicity bins, however, the anticorrelation is clearly present and follows the same locus as in other massive GCs \citep{Carretta2009a}. Our sample reproduces this behaviour: the full $[\mathrm{Na/Fe}]$
distribution spans $-0.87$ to $+1.75$~dex, consistent with the range reported by \citet{Marino2011} (down to $\approx -0.2$~dex), once the known systematic offsets between APOGEE/DR19 and optical abundance scales are accounted for
\citep[$\approx 0.3$--$0.5$~dex; see][]{Meszaros2021}.

\subsubsection{Alpha-Element Enhancement}
\label{sec:alpha_enhancement}

For a stellar population formed predominantly from core-collapse supernova (SN~II) ejecta, $\alpha$-element-to-iron ratios should be enhanced above the solar value. We also inspect $[\mathrm{Ca/Fe}]$ and $[\mathrm{Mg/Fe}]$ as a function of $[\mathrm{Fe/H}]$ for the member sample.

The sample mean values are $\langle[\mathrm{Ca/Fe}]\rangle = +0.19$ ($\sigma = 0.20$, $N = 1687$) and
$\langle[\mathrm{Mg/Fe}]\rangle = +0.17$ ($\sigma = 0.19$, $N = 1788$), both substantially above solar and consistent with the $\alpha$-enhanced plateau seen in metal-poor halo and globular cluster stars \citep{Kobayashi2006, Kobayashi2020}.
At the metal-rich end ($[\mathrm{Fe/H}] \gtrsim -1.0$), a mild decline of $[\mathrm{Ca/Fe}]$ with increasing $[\mathrm{Fe/H}]$ is visible, reflecting the onset of SN~Ia iron enrichment on the timescale of the cluster's extended star-formation history \citep{deBoer2014}. This ``knee'' feature has been previously identified in \ocen\ by \citet{Johnson2010} and its presence in our sample further confirms that the \textsl{APOGEE} data capture the full chemical history of the cluster.

\subsubsection{Radial Velocity Consistency}
\label{sec:vrad_check}

As a final consistency check, we compare the line-of-sight radial velocity distribution of the selected sample with the known systemic kinematics of \ocen. The heliocentric radial velocity distribution of the 1803 stars with valid $v_r$ measurements has a mean of $\bar{v}_r = +233.3$~km~s$^{-1}$ and a dispersion of $\sigma_{v_r} = 27.0$~km~s$^{-1}$.
This is in excellent agreement with the systemic velocity of $v_{\mathrm{sys}} = +232.2 \pm 0.5$~km~s$^{-1}$ and the
central velocity dispersion of $\sigma_0 \approx 20$--$25$~km~s$^{-1}$ reported by \citet{Meszaros2021}, confirming that the
membership filter has effectively removed field contamination while retaining a dynamically unbiased sample of cluster members.

Taken together, the metallicity distribution, light-element anticorrelations, $\alpha$-element enhancement, and radial velocity properties of the selected sample are all fully consistent with published spectroscopic and kinematic characterisations of \ocen. We therefore conclude that the MWM DR19 member sample adopted here provides a chemically representative and statistically robust basis for the hierarchical clustering analysis presented in Section~\ref{sec:method}.

\subsection{Representativeness of the Quality-Selected Subsample}
\label{sec:representativeness}

The quality cuts described in Section~\ref{sec:data_sample} reduce the 1807-member catalogue to a 957-star clustering subsample, a reduction of $\sim 47$\%. In this section we verify that this selection does not introduce systematic biases in the spatial, kinematic, or chemical properties of the retained stars.

\subsubsection{Spatial and Kinematic Representativeness}
\label{sec:spatial_kinematic}

\begin{figure}
    \centering
    \includegraphics[width=\linewidth]{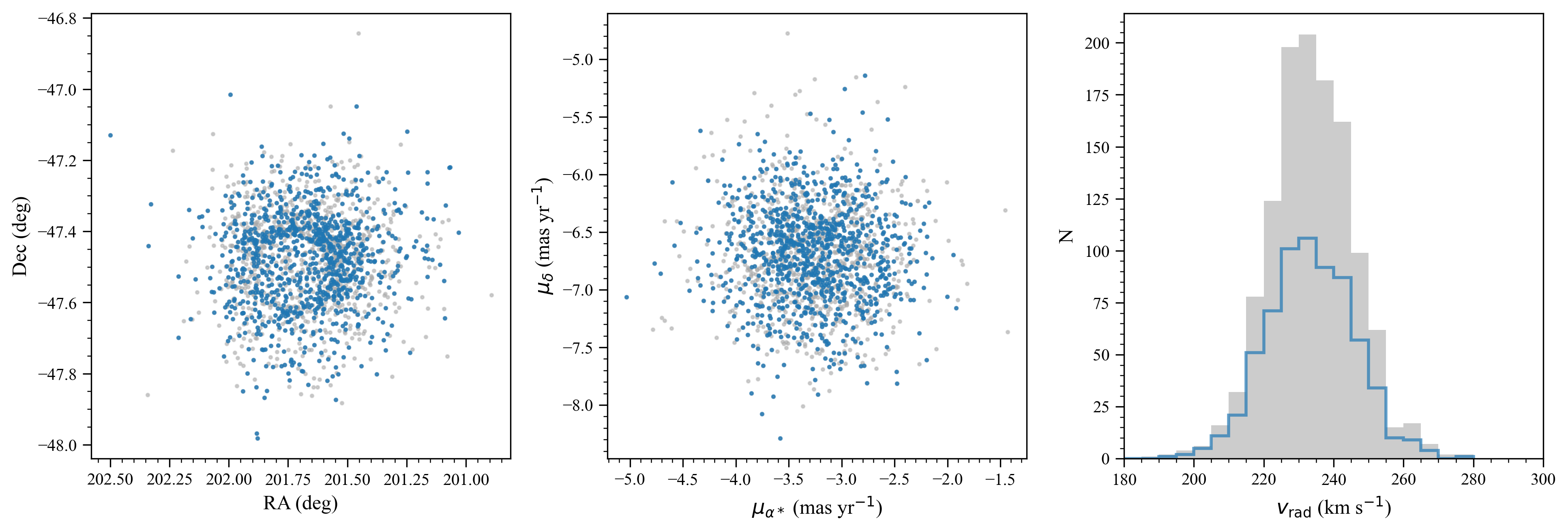}
    \caption{Spatial and kinematic comparison between the full
    1807-member sample (grey) and the 957-star quality-selected
    clustering subsample (blue). Left: on-sky distribution in
    equatorial coordinates. Centre: \textsl{Gaia}~DR3 proper-motion plane
    ($\mu_{\alpha*}$, $\mu_\delta$). Right: heliocentric radial
    velocity distributions. In all three planes the subsample
    reproduces the parent distribution without systematic offset,
    confirming the absence of spatial or kinematic selection bias
    introduced by the quality cuts.}
    \label{fig:spatial_kinematic}
\end{figure}

Figure~\ref{fig:spatial_kinematic} compares the distributions of the full 1807-member sample and the 957-star clustering subsample in three independent observational planes: the on-sky projected position (RA, Dec), the \textsl{Gaia}~DR3 proper-motion plane ($\mu_{\alpha*}$, $\mu_\delta$), and the heliocentric radial velocity distribution.

In all three planes, the clustering subsample (blue) closely follows the distribution of the full member sample (grey). The
spatial footprint covers the same radial extent and position angle range as the parent catalogue, with no evidence for a concentration toward the cluster centre or preferential sampling of any azimuthal sector. In the proper-motion plane, both samples occupy the same compact locus, confirming that the quality cuts do not preferentially remove stars at the tails of the kinematic distribution. The radial velocity histogram of the clustering subsample reproduces the peak and width of the parent distribution, consistent with the systemic velocity and velocity dispersion reported in Section~\ref{sec:vrad_check}.

These comparisons confirm that the SNR and per-element flag requirements do not introduce any detectable spatial or kinematic selection effect, and that the clustering subsample can be regarded as a kinematically and spatially unbiased representation of the $\omega$~Cen member population within the MWM~DR19 footprint.

\subsubsection{Chemical Representativeness}
\label{sec:chem_representativeness}

\begin{figure}
    \centering
    \includegraphics[width=\linewidth]{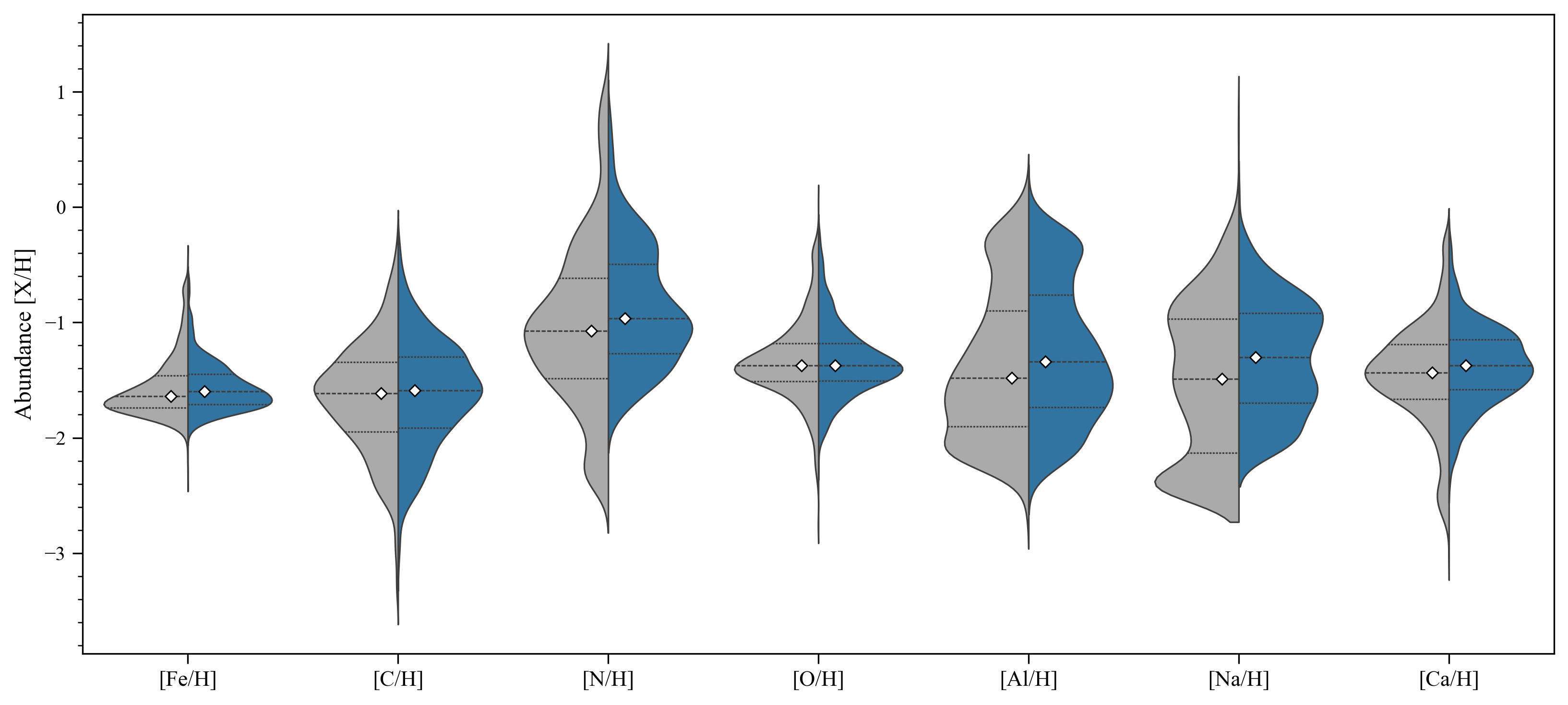}
    \caption{Split violin plots comparing the $[\mathrm{X/H}]$ abundance distributions of the full 1807-member sample (grey,
    left half) and the 957-star quality-selected clustering subsample (blue, right half) for the seven Ward clustering
    dimensions. Open diamonds mark the median; dashed lines indicate the 16th and 84th percentiles. The close agreement
    across all seven elements confirms that the quality cuts introduce no significant chemical selection bias.}
    \label{fig:violin_sample}
\end{figure}

Figure~\ref{fig:violin_sample} shows split violin plots comparing the $[\mathrm{X/H}]$ distributions of the full 1807-member sample and the 957-star clustering subsample across all seven Ward clustering dimensions.

The two distributions are in close agreement for all seven elements. Median values are consistent within measurement
uncertainties across the full abundance range, and the overall shape and width of the distributions are preserved in both
samples. The most extended tails, visible in $[\mathrm{N/H}]$ and $[\mathrm{Al/H}]$, are present in both samples, confirming
that the quality cuts do not truncate the chemically most extreme stars. The slight narrowing of the subsample distributions
reflects the removal of high-uncertainty measurements rather than the loss of chemically distinct substructure.

Taken together with Section~\ref{sec:spatial_kinematic}, these results confirm that the 957-star clustering subsample is a
faithful and unbiased representation of the full MWM~DR19 member catalogue in all relevant observational and chemical dimensions. The Ward populations identified in Section~\ref{sec:method} are therefore not an artefact of the quality selection process.

\section{Feature Selection for Ward Analysis}
\label{sec:feature_selection}

\subsection{Robustness of the Clustering Solution}
\label{sec:app_robustness}

This appendix evaluates the robustness of the Ward hierarchical clustering solution (Section~\ref{sec:method}) against both measurement uncertainties and sampling variance using two complementary tests. In both cases, we construct a stability matrix $\mathcal{S}$, where each element $S_{mn}$ represents the mean probability that stars originally assigned to population $m$ are co-assigned to population $n$ across all $N_{\rm runs}$ realizations. The diagonal elements $S_{mm}$ thus serve as internal \textbf{stability scores ($P$)} for each cluster, quantifying the robustness of the population definitions against stochastic variations.

\textbf{Monte Carlo perturbation test.} 
The abundance vector of each star was independently perturbed $N_{\rm runs} = 300$ times by drawing Gaussian noise scaled to the per-element measurement uncertainties provided by the MWM DR19 pipeline, and the clustering was re-applied at each realisation. The median internal measurement uncertainties of the seven elements range from $0.012$ to $0.050$~dex ($[\mathrm{Fe/H}]$: 0.012~dex; $[\mathrm{Na/H}]$: 0.014~dex; $[\mathrm{N/H}]$: 0.017~dex; $[\mathrm{O/H}]$: 0.018~dex; $[\mathrm{Ca/H}]$: 0.030~dex; $[\mathrm{C/H}]$: 0.034~dex; $[\mathrm{Al/H}]$: 0.050~dex).

\textbf{Bootstrap resampling test.} 
The dataset was resampled with replacement for $N_{\rm runs} = 300$ independent clustering runs. This test quantifies the sensitivity of the solution to random variations in sample size and composition.

\begin{figure*}
      \centering
      \includegraphics[width=0.48\textwidth]{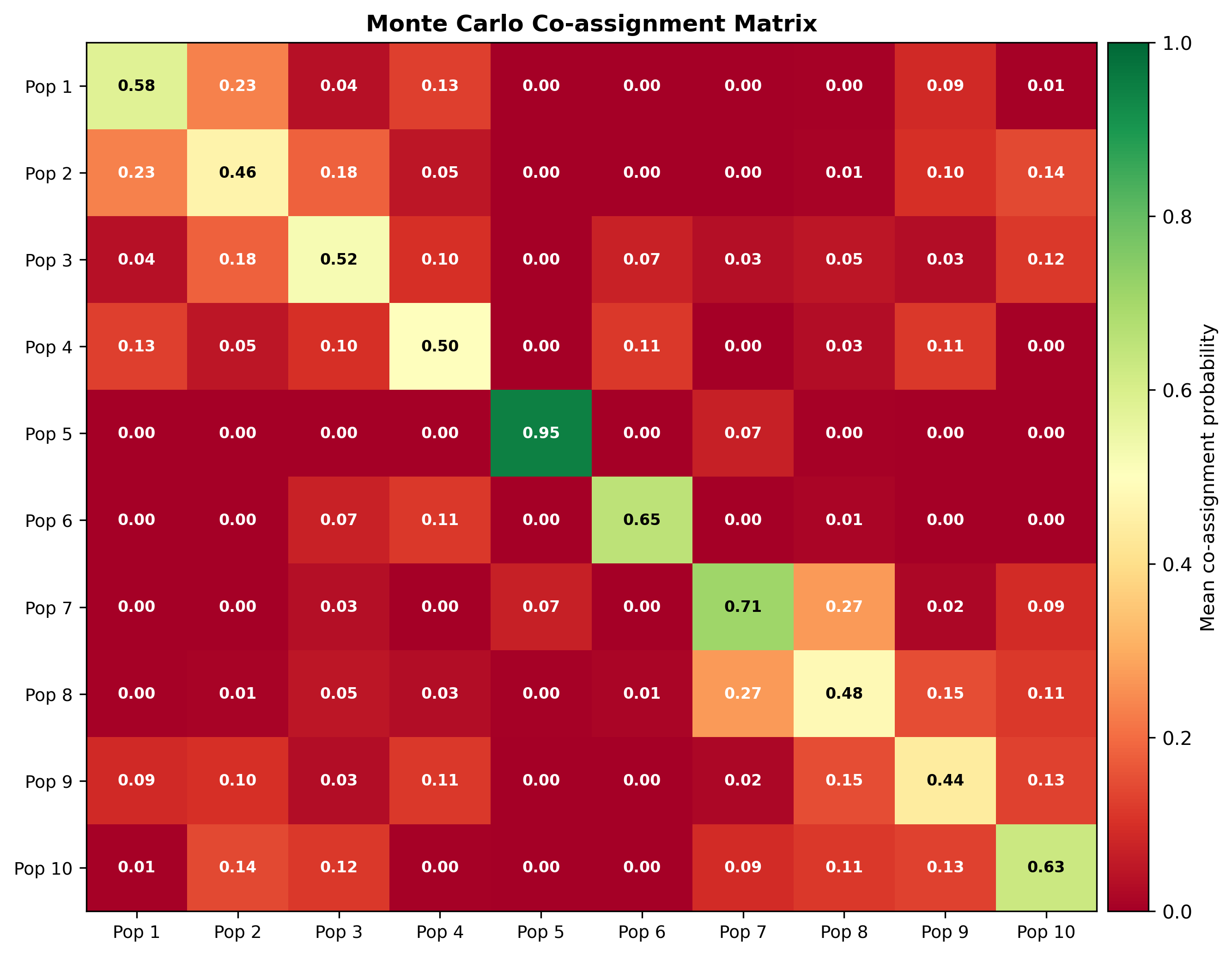}                         
      \hfill                                                                                                            
      \includegraphics[width=0.48\textwidth]{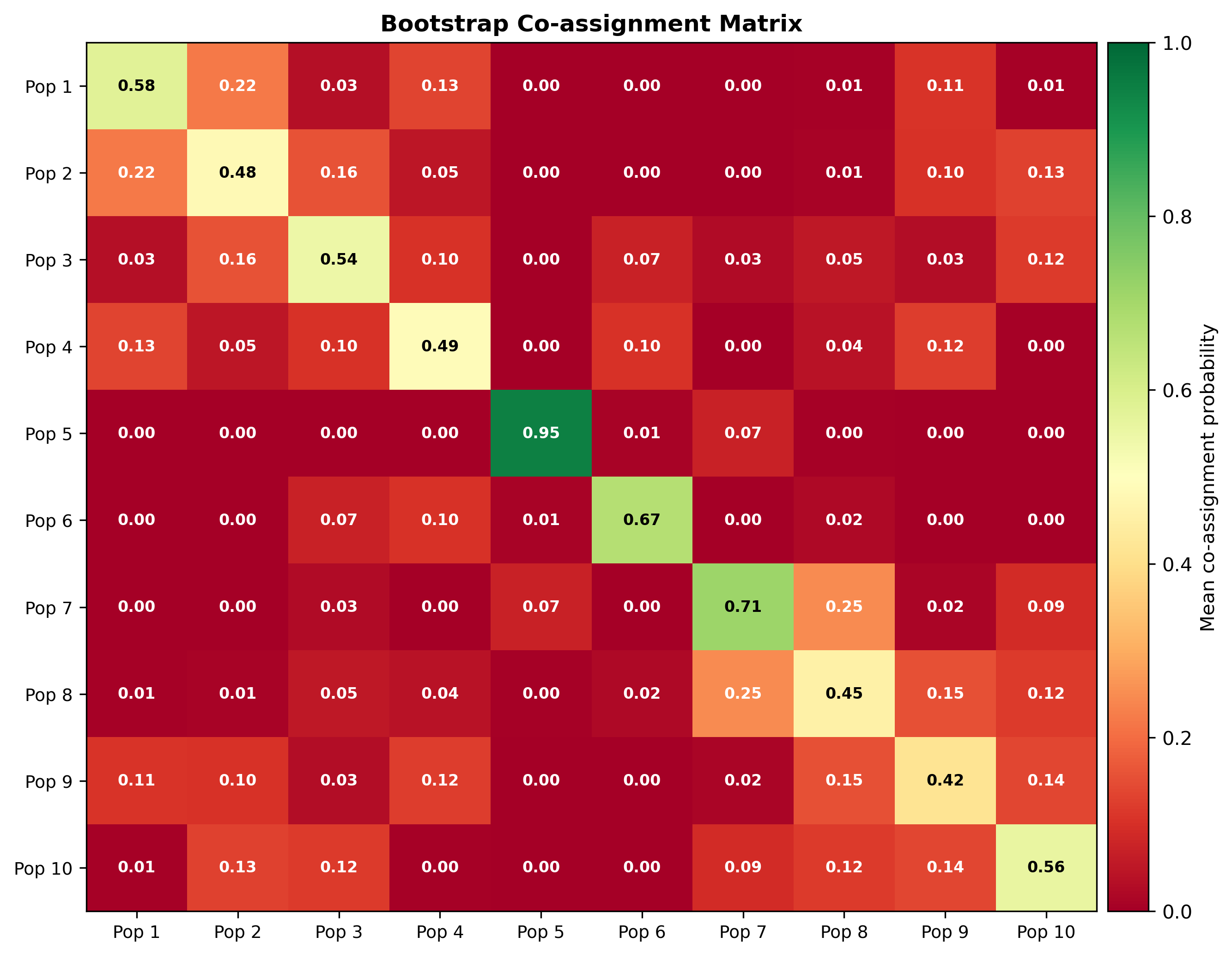}                          
      \caption{Mean pairwise co-assignment probability matrices for the ten identified populations derived from (\textit{left}) Monte Carlo perturbation ($N=300$ runs, Gaussian noise scaled to per-element measurement uncertainties) and (\textit{right}) bootstrap resampling ($N=300$ runs). Diagonal elements represent within-population stability scores; off-diagonal elements indicate the degree of cross-assignment between populations.}                                                                            
      \label{fig:stability}
\end{figure*}

The Monte Carlo and bootstrap tests yield consistent results, demonstrating that the clustering solution is robust against both measurement uncertainties and sampling fluctuations. Within-population stability scores ($P$) range from $0.42$ to $0.95$, with Pop~5 showing the highest coherence ($P = 0.95$) and Pop~9 the lowest ($P \approx 0.42$). All off-diagonal values remain $\leq 0.27$, confirming limited cross-contamination between populations under perturbation (see Figure~\ref{fig:stability}).

Importantly, the variation in stability across populations follows a clear physical trend: the most chemically extreme population (Pop~5) is highly self-consistent, while the metal-poor backbone populations (Pop~1--4) exhibit lower stability, reflecting their intrinsically continuous chemical structure. The elevated co-assignment between Pop~7 and Pop~8 ($\sim 0.27$) further indicates partial overlap in abundance space, consistent with the continuous and multi-phase chemical evolution of $\omega$~Cen rather than a set of strictly discrete populations.

\section{CMD Isochrone Fits for All Ten Populations}
\label{app:cmd_fits}

\begin{figure*}
\centering
\includegraphics[width=\textwidth]{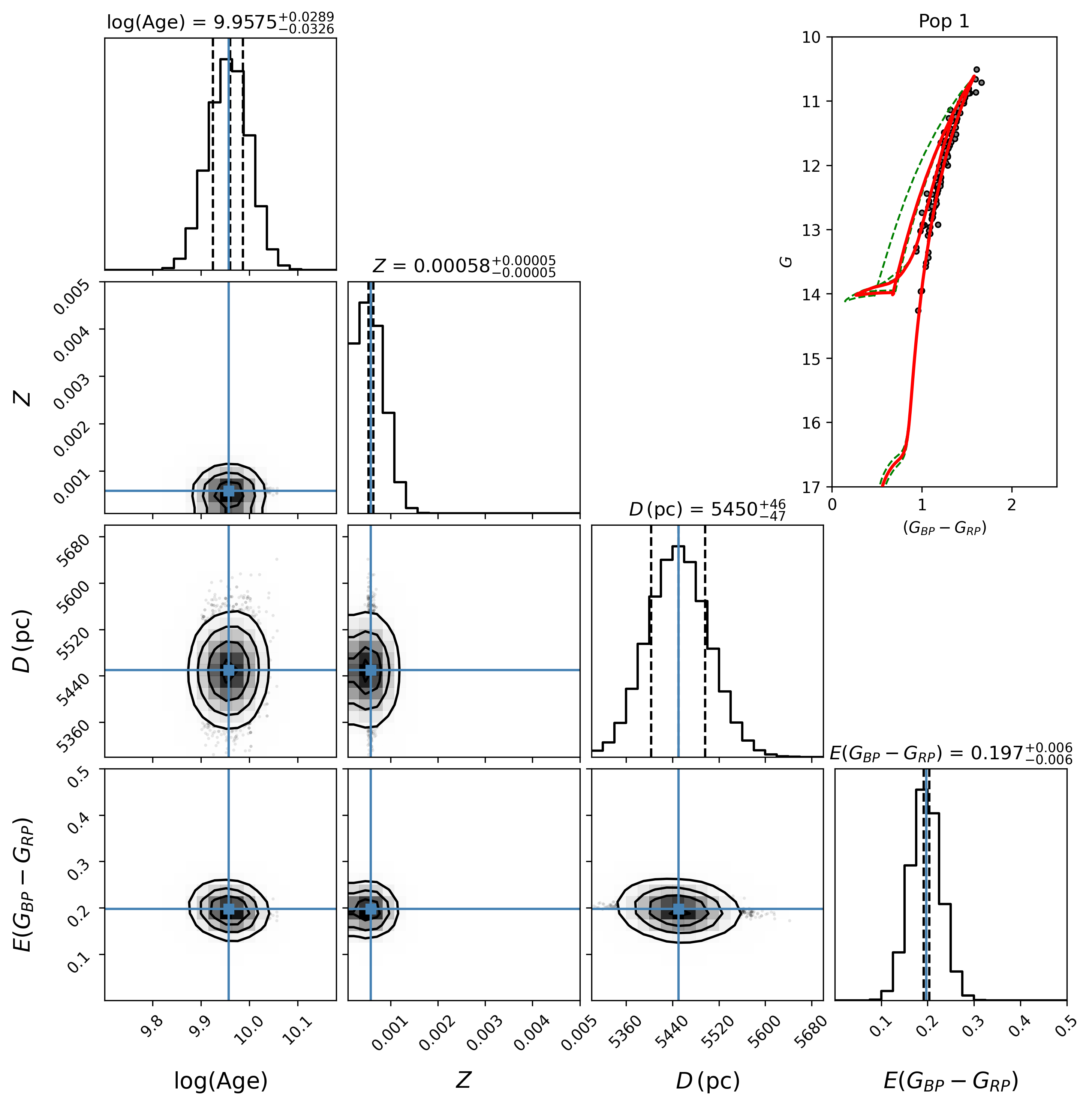}
\caption{Nested-sampling CMD fit for Population~1 ($N=111$,
$[\mathrm{Fe/H}]=-1.71$).}
\label{fig:cmd_app_pop1}
\end{figure*}

\begin{figure*}
\centering
\includegraphics[width=\textwidth]{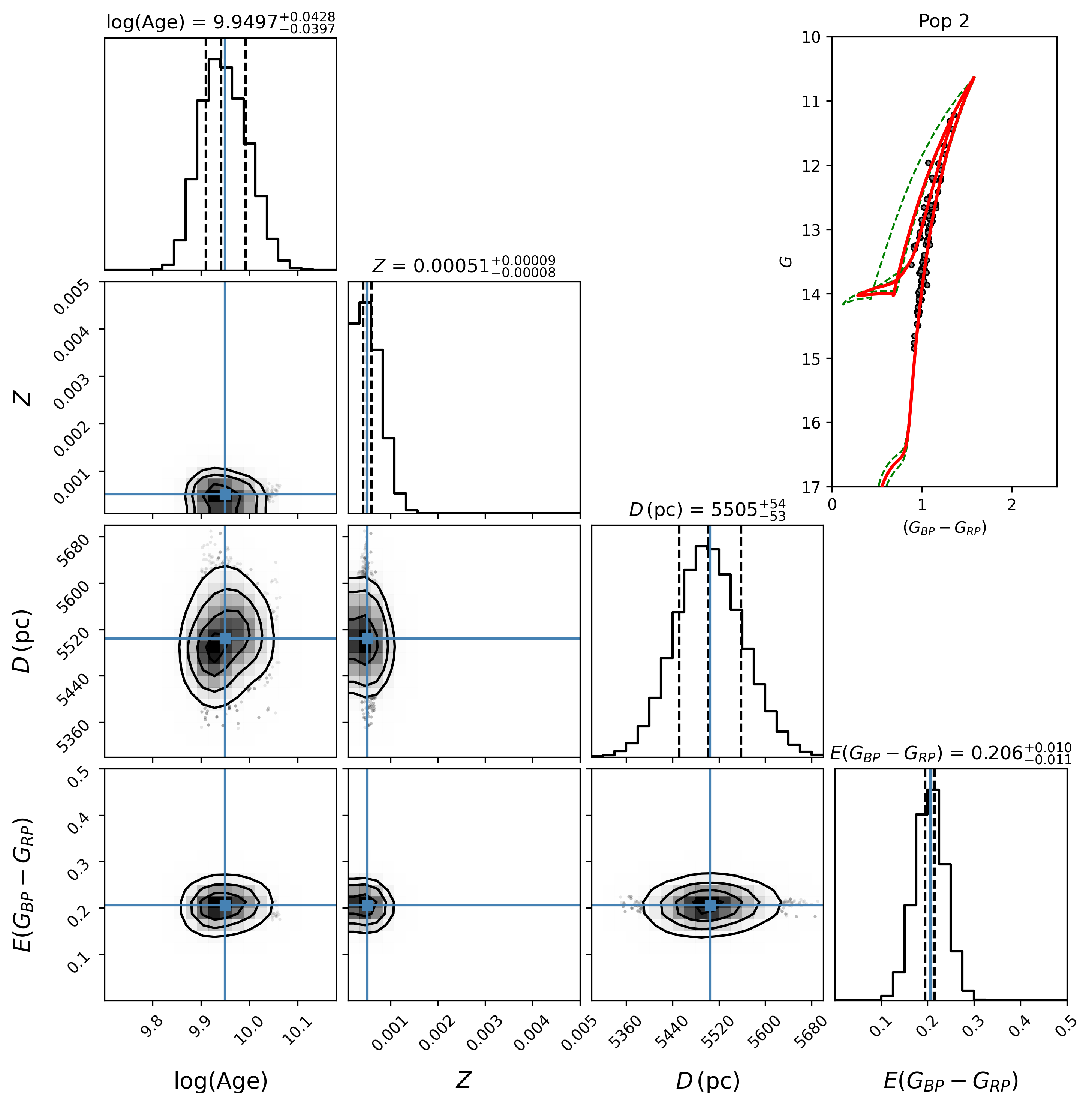}
\caption{Nested-sampling CMD fit for Population~2 ($N=117$,
$[\mathrm{Fe/H}]=-1.77$).}
\label{fig:cmd_app_pop2}
\end{figure*}

\begin{figure*}
\centering
\includegraphics[width=\textwidth]{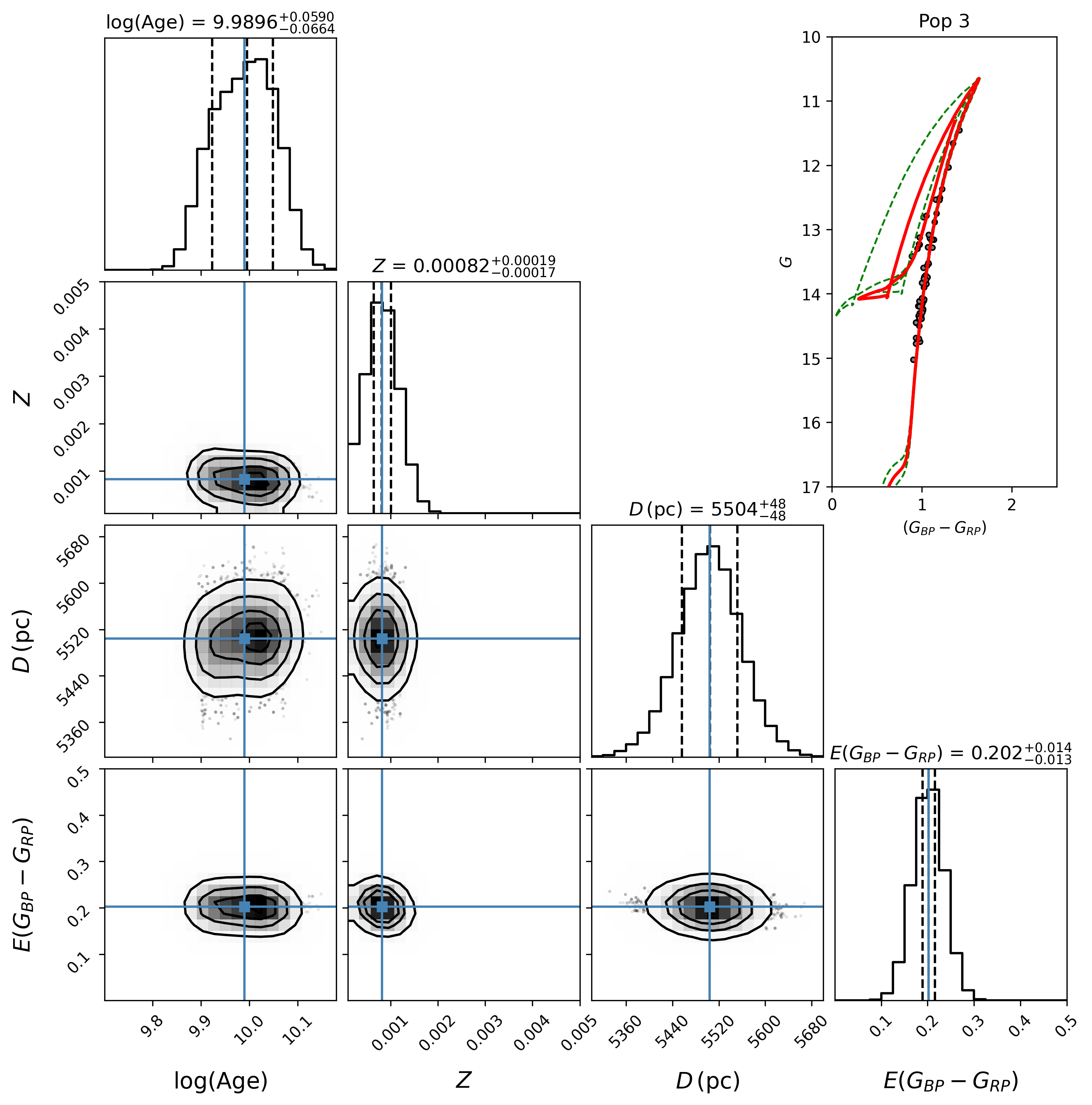}
\caption{Nested-sampling CMD fit for Population~3 ($N=66$,
$[\mathrm{Fe/H}]=-1.56$).}
\label{fig:cmd_app_pop3}
\end{figure*}

\begin{figure*}
\centering
\includegraphics[width=\textwidth]{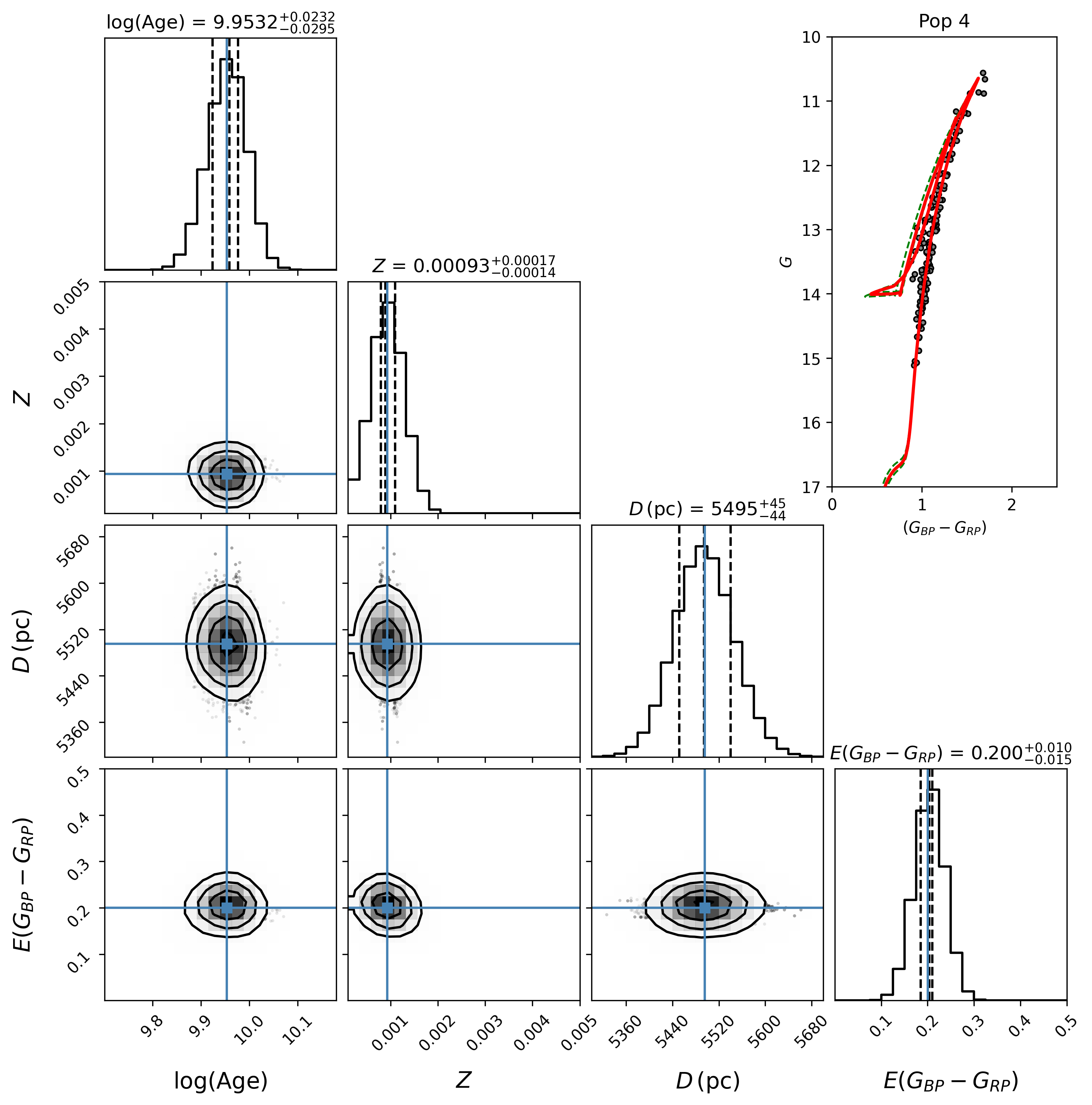}
\caption{Nested-sampling CMD fit for Population~4 ($N=154$,
$[\mathrm{Fe/H}]=-1.50$).}
\label{fig:cmd_app_pop4}
\end{figure*}

\begin{figure*}
\centering
\includegraphics[width=\textwidth]{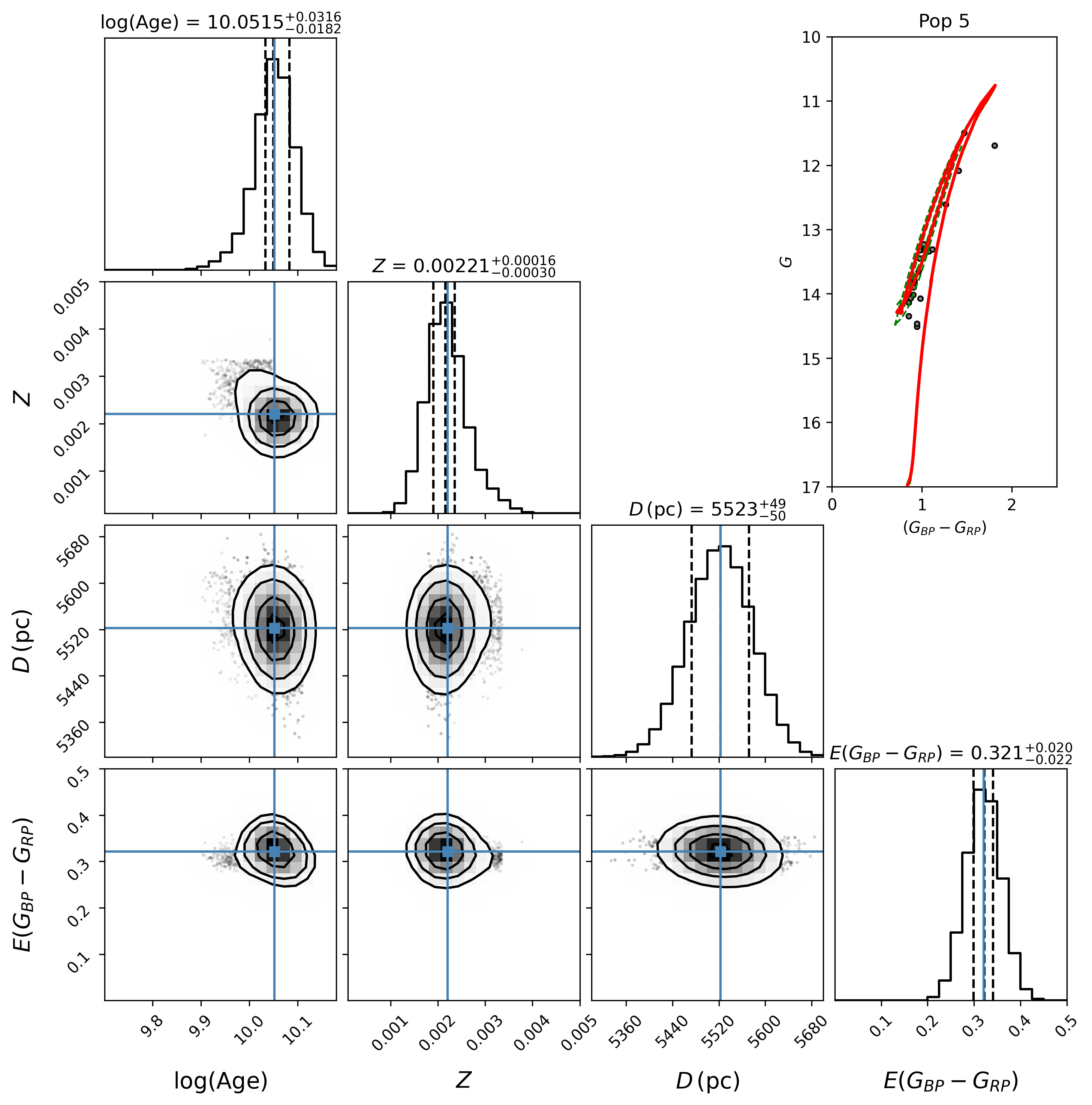}
\caption{Nested-sampling CMD fit for Population~5 ($N=25$, $[\mathrm{Fe/H}]_{\rm iso}=-1.13$). The elevated reddening estimate ($E(G_{\rm BP}-G_{\rm RP}) \approx 0.32$~mag) likely reflects the well-known age--metallicity--extinction degeneracy in photometric fitting rather than a true spatial differential reddening, particularly given the small sample size and extreme metallicity of this group.}
\label{fig:cmd_app_pop5}
\end{figure*}

\begin{figure*}
\centering
\includegraphics[width=\textwidth]{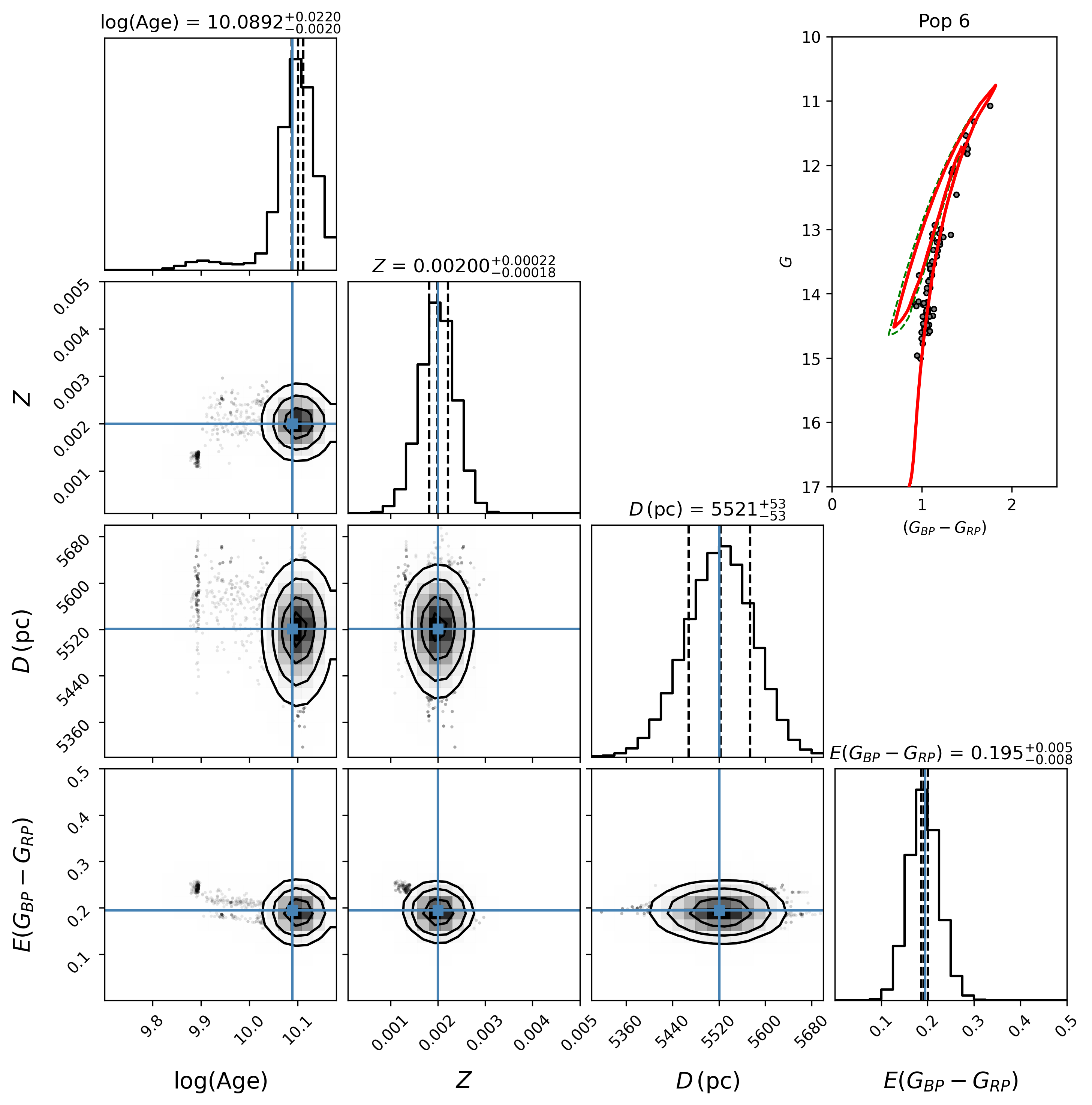}
\caption{Nested-sampling CMD fit for Population~6 ($N=73$,
$[\mathrm{Fe/H}]=-1.17$).  This population yields the oldest
mean age among the ten groups ($12.3^{+0.6}_{-0.1}$~Gyr).}
\label{fig:cmd_app_pop6}
\end{figure*}

\begin{figure*}
\centering
\includegraphics[width=\textwidth]{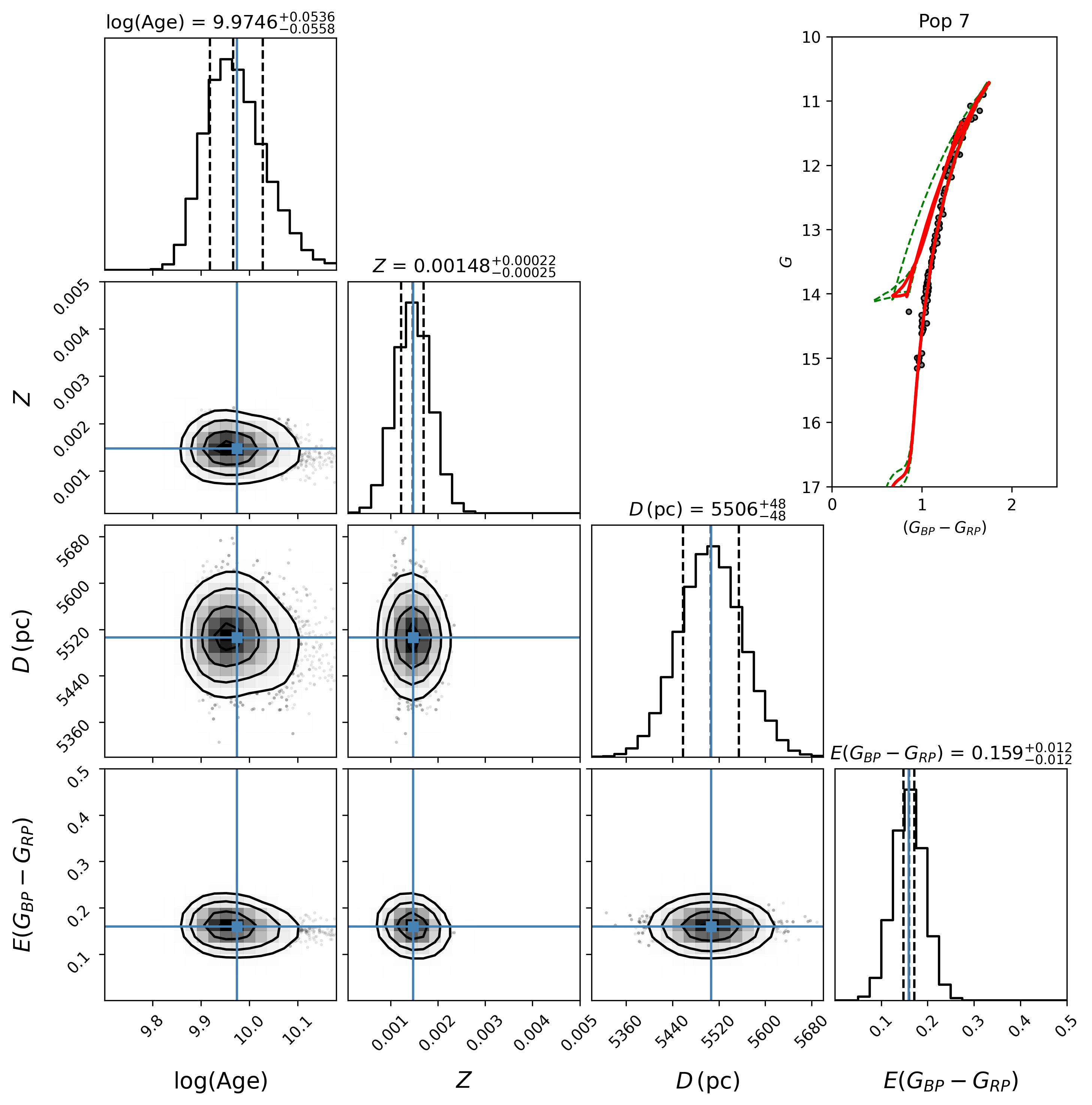}
\caption{Nested-sampling CMD fit for Population~7 ($N=102$,
$[\mathrm{Fe/H}]=-1.30$).}
\label{fig:cmd_app_pop7}
\end{figure*}

\begin{figure*}
\centering
\includegraphics[width=\textwidth]{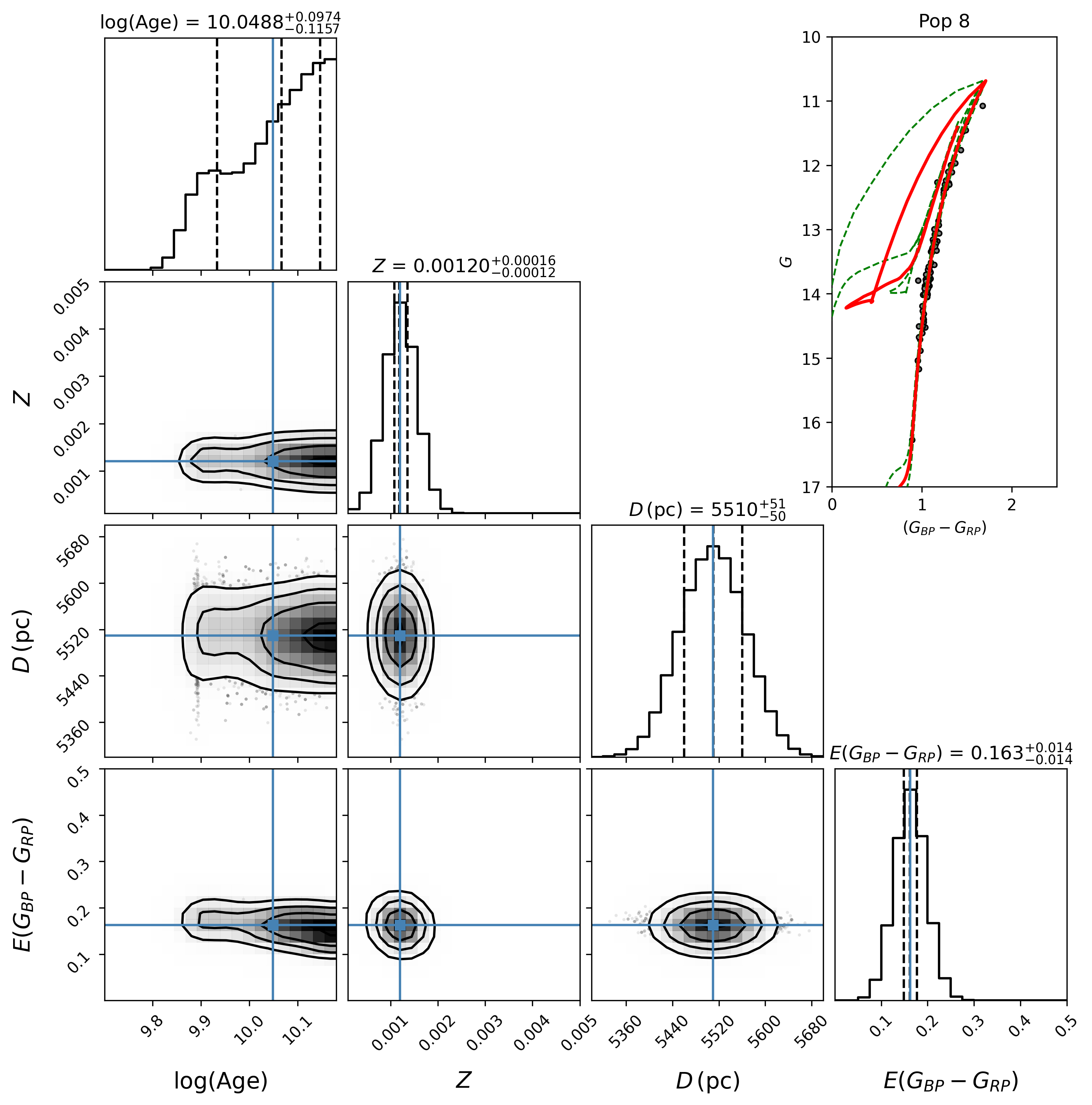}
\caption{Nested-sampling CMD fit for Population~8 ($N=104$,
$[\mathrm{Fe/H}]=-1.39$).  The broad age posterior
($\sigma_{\rm age} \approx 2.7$~Gyr) reflects the wide CMD
spread of this population.}
\label{fig:cmd_app_pop8}
\end{figure*}

\begin{figure*}
\centering
\includegraphics[width=\textwidth]{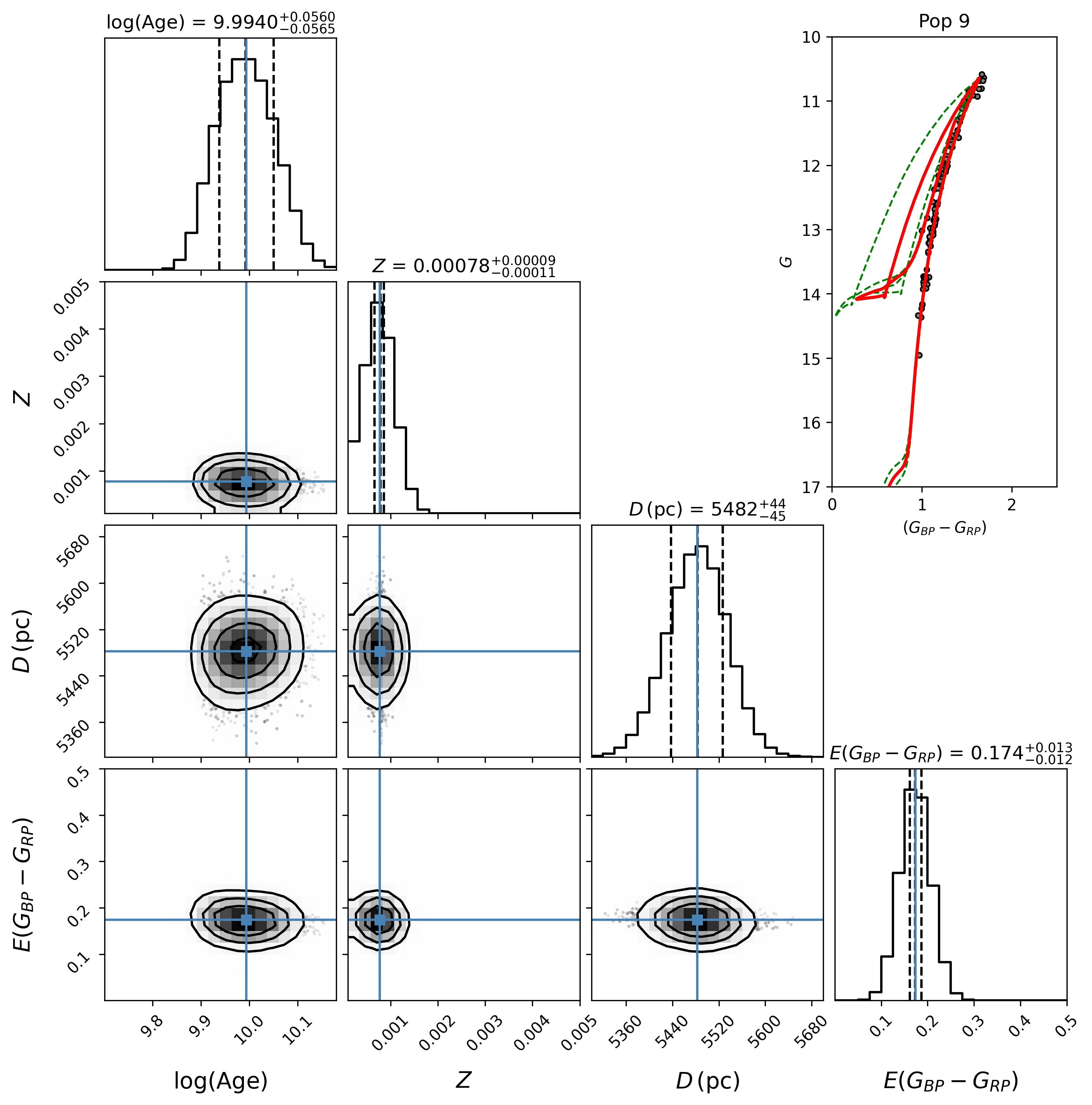}
\caption{Nested-sampling CMD fit for Population~9 ($N=105$,
$[\mathrm{Fe/H}]=-1.58$).}
\label{fig:cmd_app_pop9}
\end{figure*}

\begin{figure*}
\centering
\includegraphics[width=\textwidth]{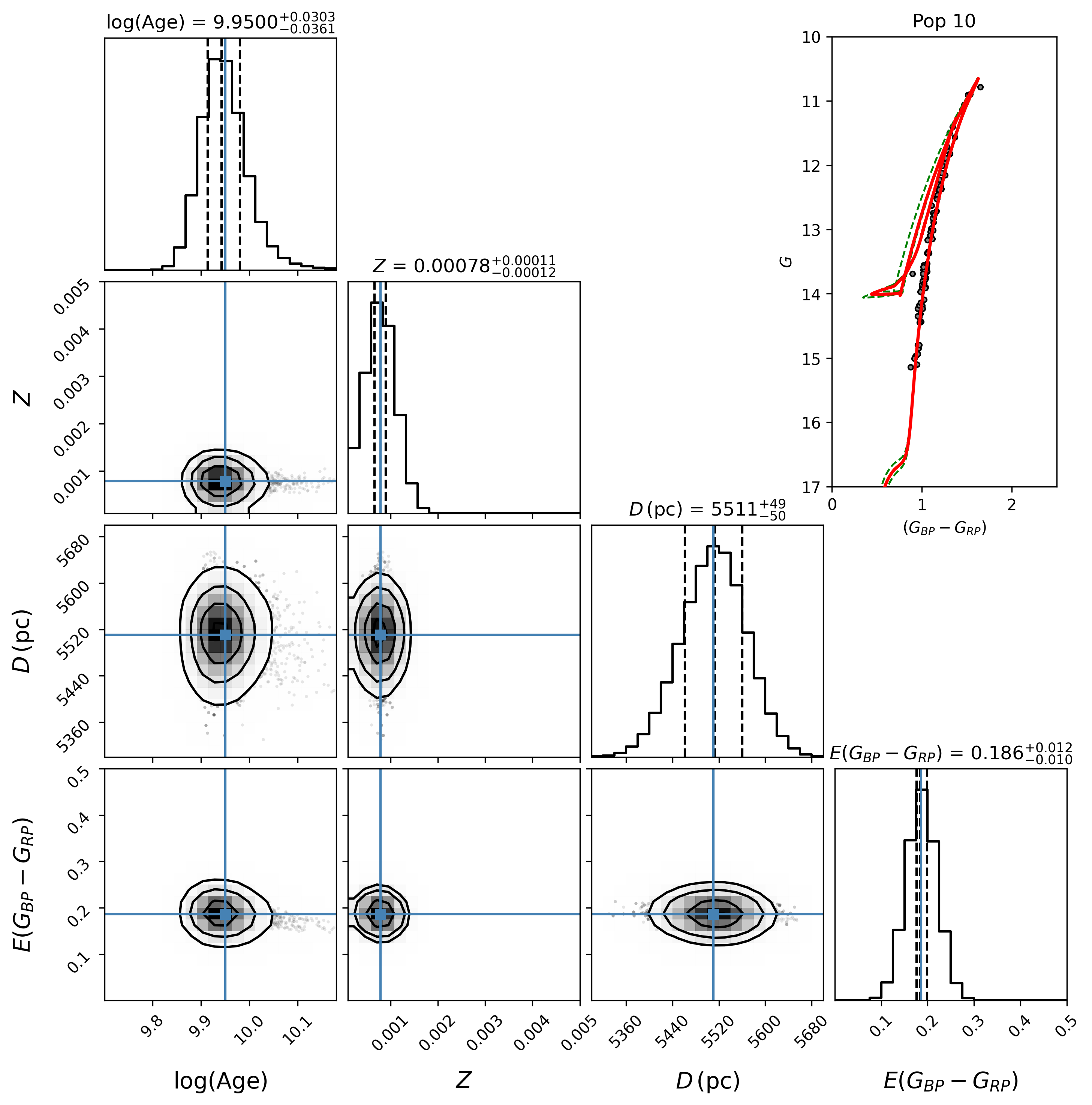}
\caption{Nested-sampling CMD fit for Population~10 ($N=100$,
$[\mathrm{Fe/H}]=-1.58$).}
\label{fig:cmd_app_pop10}
\end{figure*}

Figures~\ref{fig:cmd_app_pop1}--\ref{fig:cmd_app_pop10} show the nested-sampling results for all ten Ward populations.  Each figure combines a corner plot with a CMD inset. The top-right panel displays the \textsl{Gaia}~DR3 CMD of the population members (black points) overlaid with the best-fitting BaSTI-IAC isochrone ($[\alpha/\mathrm{Fe}]=+0.4$, $Y=0.30$; red curve) evaluated at the posterior mean parameters. The corner panels show the marginalised posterior distributions for the four free parameters $(\log\mathrm{Age},\,Z,\,D\,[\mathrm{pc}],\,E(G_{\rm BP}-G_{\rm RP}))$: diagonal panels give 1D histograms with vertical dashed lines at the 16th, 50th, and 84th percentiles (values quoted above each histogram); off-diagonal panels show 2D marginals with $1\sigma$ and $2\sigma$ contours.

The fitted distance and reddening are consistent across all ten populations (Section~\ref{sec:age_estimation}), confirming the internal coherence of the photometric calibration. Age constraints range from tight ($\sigma_{\rm age} \approx 0.5$~Gyr for Pop.~4) to broad ($\sigma_{\rm age} \approx 2.7$~Gyr for Pop.~8), reflecting the variation in sample size and in the colour--magnitude coverage of each population's RGB. Populations~6 and 5 stand out as the oldest ($\sim 12.3$ and $\sim 11.3$~Gyr, respectively).  Bayesian evidence values ($\ln\mathcal{Z}$) and effective sample sizes (ESS) for each run are listed in Table~\ref{tab:nested_results}.

\bsp	
\label{lastpage}
\end{document}